\documentclass[a4paper,11pt]{article}
\usepackage{float}
\usepackage{amsmath}
\usepackage{amssymb}
\usepackage{graphicx}
\usepackage{color,xcolor}
\usepackage{mathrsfs}
\usepackage{tcolorbox}
\usepackage{tikz}
\usepackage{tikz-cd}
\usepackage{physics}
\usepackage{diagbox}
\usepackage{listings}
\usepackage[textsize=scriptsize,textwidth=2.5cm]{todonotes}
\usepackage{multicol}
\usepackage{multirow,makecell}
\usepackage{subfig}
\usepackage{cancel}
\usepackage{jheppub}
\usepackage{upgreek}
\usepackage{extarrows}
\usepackage{cleveref}

\allowdisplaybreaks[2]
\numberwithin{equation}{section}
\definecolor{darkgreen}{RGB}{40,150,60}
\begin{document}
\begin{flushright}

\end{flushright}

\title{On the CFT dual of superstring on AdS$_3$ }

\author{Zhe-fei Yu}

\affiliation{Kavli Institute for Theoretical Sciences (KITS), University of Chinese Academy of Sciences, Beijing 100190, China}

\emailAdd{yuzhefei@ucas.ac.cn}

\abstract{
In this work, we study the proposed duality between  superstring on AdS$_3\times$S$^3\times\mathbb{T}^4$ with pure NS-NS flux and a   deformed symmetric orbifold CFT. We extract the residues of 3-point string correlators at their next-to-leading poles, which can be expressed  as  integrals. Because of  picture changing, the integrand is not unique. We make an ansatz for its proper form  which linearly combines three different picture choices. On the CFT side, we obtain the corresponding residues by doing  a conformal perturbation computation at the second order.  The result  can also be expressed  as  integrals, with the integrand being a five point function in the symmetric orbifold theory, which is calculated  using  covering maps. After unifying the integration variables of the two sides, we find that the two integrands match precisely, due to some novel mathematical identities of covering maps. Our calculation verifies the proposed  marginal operator in the dual CFT of superstring on AdS$_3\times$S$^3\times\mathbb{T}^4$ and could be generalized to the CFT duals of  superstrings on general backgrounds AdS$_3\times X$ with few modifications.
}

\maketitle

\section{Introduction}

It has been known for a long time that the CFT dual of the type IIB superstring theory on AdS$_3\times$S$^3\times\mathbb{T}^4$  lies in the moduli space of the symmetric orbifold of $\mathbb{T}^4$~\cite{Maldacena:1997re, Seiberg:1999xz}. A few years ago, it was found  that the tensionless string theory with minimal pure NS-NS flux is exactly dual to the symmetric orbifold  CFT Sym$^N(\mathbb{T}^4)$ itself \cite{Gaberdiel:2018rqv,Eberhardt:2018ouy,Eberhardt:2019ywk}. Then it is natural to consider   dualities beyond the tensionless limit. On the string side, one could consider generic string theory with  pure NS-NS flux; on the CFT side, correspondingly, one needs to deform the  symmetric orbifold CFT by a (non-normalizable) exactly marginal operator from the twist 2 sector, which  creates an  exponential wall \cite{Eberhardt:2021vsx,Balthazar:2021xeh,Martinec:2022ofs}\footnote{The undeformed theory is not simply the symmetric orbifold of $\mathbb{T}^4$ but contains a linear dilaton factor, see section \ref{TheproposedCFT}. Thus, in some ways, this exponential wall is similar to the one in  Liouville theory,  and it does not change the spectrum of delta-function normalizable states.   At the tensionless limit, this linear dilaton factor disappears and the dual theory becomes the symmetric orbifold itself.}.  The main goal of this work is to verify this proposal\footnote{We stress two subtleties of the proposal. Firstly, as stressed in \cite{Balthazar:2021xeh,Chakraborty:2025nlb} (see also \cite{Giveon:2005mi}), there is  a qualitative
difference between $k>1$ and $k<1$ for the dual CFT. While  for $k<1$ the  deformed symmetric orbifold is believed to
describe the full spacetime theory, for $k>1$ one can only think of it as an effective CFT \cite{Chakraborty:2025nlb}. Secondly, it was shown in \cite{Eberhardt:2021vsx,Aharony:2024fid,Knighton:2024pqh} (see also \cite{Kutasov:1999xu,Giveon:2001up,Kim:2015gak,Eberhardt:2021jvj,Eberhardt:2020bgq}) that one should understand the dual theory as a grand canonical ensemble of CFTs but not a specific CFT with a fixed N. In this work, we generalize  the computation in \cite{Eberhardt:2021vsx} to the SUSY case, which is not sensitive to these two subtleties.\label{fn:footnote2} } by matching (the residues of)  three point correlators of the two sides.

On the string side, superstring theory on AdS$_3$ with pure NS-NS flux can be studied using the RNS formalism \cite{Giveon:1998ns,Kutasov:1999xu,deBoer:1998gyt}. The worldsheet theory involves the $SL(2,\mathbb{R})$  WZW model, which is  non-compact and is much more intricate than its compact  counterpart  (the $SU(2)$ WZW model). The distinctive feature of this CFT is that its spectrum must include states that transform as spectrally flowed representations of  the affine algebra $\mathfrak{sl}(2,\mathbb{R})_{k}$\cite{Maldacena:2000hw,Maldacena:2000kv,Maldacena:2001km}.  These special representations are crucial for characterizing the long strings, which correspond to   delta-function normalizable states on the field theory side. Although the spectrum of the $SL(2,\mathbb{R})$  WZW model has been understood for a long time, much is yet to be understood about the correlation functions of spectrally flowed vertex operators \cite{Giribet:2000fy,Giribet:2001ft,Giribet:2005ix,Ribault:2005ms,Giribet:2005mc,Minces:2005nb,Iguri:2007af,Baron:2008qf,Iguri:2009cf,Giribet:2011xf,Cagnacci:2013ufa,Giribet:2015oiy,Giribet:2019new,Hikida:2020kil}. Progress was made recently in \cite{Dei:2021xgh}, where a closed formula was proposed for the 3-point functions of generic spectrally flowed operators (see \cite{Bufalini:2022toj} for a proof of this formula). Then for bosonic string on AdS$_3$, one  directly gets the tree-level 3-point string  correlators (amplitudes) of long strings. Subsequently, it was found that the residues of these string correlators at their poles can be reproduced by a conformal perturbation computation of a deformed symmetric orbifold theory \cite{Eberhardt:2021vsx}, which is then proposed as the dual CFT at least in the perturbative sense. In fact, this computation extracts the corresponding residues of the structure constant on the CFT side, which  in some sense is similar to the  original derivation of the  DOZZ formula of   Liouville CFT \cite{Dotsenko:1984nm,Dotsenko:1984ad,Dorn:1994xn,Zamolodchikov:1995aa}. Since correlators in the symmetric orbifold CFT are calculated by  lifting them up to  covering surfaces, the matching of the residues of the two sides predicts surprising mathematical identities of these covering maps \cite{Eberhardt:2021vsx}. 

A similar matching was also found for the 4-point string correlators \cite{Dei:2022pkr}, based on the closed form of the 4-point functions of  spectrally flowed operators in the $SL(2,\mathbb{R})$  WZW model \cite{Dei:2021yom} (see also \cite{Iguri:2024yhb}). For  higher points,  string correlators are very difficult to work out. Nevertheless, one can directly obtain the residues by the so-called near-boundary approximation \cite{Knighton:2023mhq,Knighton:2024qxd} (see also \cite{Hikida:2023jyc}). Then for bosonic strings, the residues of the (tree level) $n$-point correlators of the two sides are matched  at all orders.

In this work, we will generalize the matching of correlators in the bosonic setting to the supersymmetric setting. Note that  some aspects of this SUSY generalization had been discussed in  \cite{Eberhardt:2021vsx,Sriprachyakul:2024gyl,Yu:2024kxr}. Nevertheless,  a main unsolved problem, which is  fundamentally important,  is to determine (or verify)  the exactly marginal operator that defines the deformed theory on the CFT side. In fact, in \cite{Eberhardt:2021vsx} such a  marginal operator was proposed for the CFT dual of string theory on AdS$_3\times$S$^3\times\mathbb{T}^4$ with pure NS-NS flux.
The main task of this work is to   verify  the  proposed marginal operator by  quantitatively matching (the residues of)  three point correlators of the two sides. Note that in \cite{Yu:2024kxr}, the residues obtained by a leading (0-th) order  conformal perturbation computation (on the CFT side) had been matched with the corresponding ones on the string sides. However,  the leading order analysis can not help us to verify the proposed marginal operator since no  marginal operator is inserted in the computation at this order. In this work we will deal with the next-to-leading order, which is the 2-nd order\footnote{The 3-point correlators we consider in this work have $w_1+w_2+w_3\in 2\mathbb{Z}+1$, so only even orders contribute \cite{Eberhardt:2021vsx,Yu:2024kxr}. Thus, the leading and  next-to-leading orders are the 0-th and 2-nd orders respectively. To do a 1-st order computation,  one needs to  consider correlators with $w_1+w_2+w_3\in 2\mathbb{Z}$. Since the forms for these correlators are not as simple as the one considered in the present work (see  section \ref{Thestringcorrelator} and \cite{Yu:2024kxr}), we choose to do a 2-nd order computation for correlators with $w_1+w_2+w_3\in 2\mathbb{Z}+1$.}. We adopt the same strategy as in the bosonic case to write the residues on both  sides  into the  form of an integral, and  match the two sides by showing that the two integrands are in fact identical. However, compared with the bosonic case, we also find some novel features on both sides:
\begin{itemize}
    \item On the string side, one always needs to consider  picture changing when calculating  string correlators. We find that starting from any specific picture choice will not give the integral form (integrands) of the residues that matches  the CFT side\footnote{Different picture choices will certainly give identical results for (the residues of) the string correlator. However, since the correlator is written as the from of an integral (which is hard to work out), different picture choices indeed lead to different results for the integrands.  }. Instead, the correct form is obtained by considering a special linear combination of 3 different picture choices. Besides, the mass-shell condition plays a significant role in the matching of the residues with the CFT side, which is not the case in the bosonic duality \cite{Eberhardt:2021vsx,Knighton:2023mhq,Knighton:2024qxd,Hikida:2023jyc}.
    \item On the CFT side, we will use covering maps to calculate  correlators in the symmetric orbifold theory. However, unlike in the bosonic case, the marginal operator is $not$ a Virasoro primary when lifted up to the covering surface (note that it is  a Virasoro primary on the base space). As a result, the matching holds due to  surprising identities of covering maps  that  involve not only the first but also the second non-trivial Taylor coefficients near the ramified points, see section \ref{matchthetwosides} (in the bosonic case, only the first non-trivial Taylor coefficient appears \cite{Eberhardt:2021vsx}). 
\end{itemize}
Although we only focus on the next-to-leading order, we expect these novel features to also appear in the computation  of  generic orders (and for  other correlators).
Besides, our computation also shows that for superstrings on general backgrounds AdS$_3\times X$ with pure NS-NS flux, the deforming operator could be a linear combination of super-descendants of  states in the twist 2 sector (see section \ref{generalbackground}). 

The rest of the paper is organized as follows. In section \ref{Thestringside}, we consider the  simplest type of  3-point correlators on the string side (calculated in \cite{Yu:2024kxr}).  We extract the residue  at the next-to-leading pole, and then write it in the  form of an integral which is a special linear combination of the residues coming from three different picture choices. In section \ref{TheCFTside}, we do the conformal perturbation computation to get the residue of the corresponding 3-point correlator at the second order in the proposed dual CFT.  Since the marginal operator is not a primary when lifted up to the covering surface, we show how to represent correlators involving such operators in terms of  Taylor coefficients of  covering maps at their ramified points.  In section \ref{matchthetwosides}, we match the two sides. After unifying the integration variables, we show that the integrands in the integrals of the two sides are  identical. This matching holds due to some surprising identities of covering maps, which are verified in the ancillary $\mathtt{Mathematica}$ notebook. In section \ref{more Comments}, we give some comments about the uniqueness of the deforming operator and discuss how to generalize the matching to superstrings on general  backgrounds AdS$_3\times X$. In section \ref{discussion}, we summarize our results and discuss some future directions. There are also several appendices which fix our conventions.

\section{The string side}\label{Thestringside}
In this section, we extract the residues of 3-point  correlators on the string side. After reviewing the  superstring theory on AdS$_3\times$S$^3\times\mathbb{T}^4$ in the RNS formalism, we describe the simplest string correlators obtained in \cite{Yu:2024kxr}. Then we calculate the residues at specific poles and write them in a form that contains free parameters ($\alpha,\beta,\gamma$, satisfying $\alpha+\beta+\gamma=1$), which can only be fixed when compared with the CFT side.

\subsection{Superstring on AdS$_3\times$S$^3\times \mathbb{T}^4$}
We review some basic facts  about  superstring theory on AdS$_3\times$S$^3\times\mathbb{T}^4$ in the RNS formalism, see e.g. \cite{Dabholkar:2007ey,Ferreira:2017pgt,Yu:2024kxr}. The worldsheet CFT is described by
\begin{equation}
    \mathfrak{sl}(2,R)^{(1)}_k\oplus  \mathfrak{su}(2)^{(1)}_k \oplus \mathfrak{u}(1)^{4(1)},
\end{equation}
where $\mathfrak{sl}(2,R)^{(1)}_k$ and $\mathfrak{su}(2)^{(1)}_k$ represent $\mathcal{N}=1$ supersymmetric WZW model with affine symmetry $\mathfrak{sl}(2,R)^{(1)}_k$  and $\mathfrak{su}(2)^{(1)}_k$ respectively. They describe the factors AdS$_3$ and S$^3$.  $\mathfrak{u}(1)^{4(1)}$ describes the internal CFT $\mathbb{T}^4$. 

The $\mathfrak{sl}(2,R)^{(1)}_k$ WZW model has symmetries generated by the $\mathfrak{sl}(2,R)_k$ currents $J^A$ and fermions $\psi^A$ $(A=1,2,3)$, with the following OPEs:
        \begin{equation}
        \begin{aligned}
            J^A(z)J^B(w)&\sim \frac{\frac{k}{2}\eta^{AB}}{(z-w)^2}+\frac{i\epsilon^{AB}_CJ^C(w)}{z-w}\\
            J^A(z)\psi^B(w)&\sim \frac{i\epsilon^{AB}_C\psi^C(w)}{z-w}\\
            \psi^A(z)\psi^B(w)&\sim \frac{\frac{k}{2}\eta^{AB}}{z-w}\,,
        \end{aligned}
      \end{equation}
      where $\epsilon^{123}=1$ and the indices are raised and lowered by $\eta^{AB}=\eta_{AB}=\text{diag}(++-)$.
     Similarly, the $\mathfrak{su}(2)^{(1)}_k$ WZW model has $\mathfrak{su}(2)_k$ currents $K^a$ and fermions $\chi^a$ $(a=1,2,3)$, with OPEs:
       \begin{equation}
        \begin{aligned}
            K^a(z)K^b(w)&\sim \frac{\frac{k}{2}\delta^{ab}}{(z-w)^2}+\frac{i\epsilon^{ab}_cJ^C(w)}{z-w}\\
            K^a(z)\chi^b(w)&\sim \frac{i\epsilon^{ab}_c\psi^c(w)}{z-w}\\
            \chi^a(z)\chi^b(w)&\sim \frac{\frac{k}{2}\delta^{ab}}{z-w}\ .
        \end{aligned}
      \end{equation}
     The  indices are raised and lowered by $\delta^{ab}=\delta_{ab}=\text{diag}(+++)$. As usual, 
     we define
        \begin{equation}
            J^{\pm}=J^1\pm iJ^2, \quad K^{\pm}=K^1\pm iK^2, \quad \psi^{\pm}=\psi^1\pm i\psi^2, \quad \chi^{\pm}=\chi^1\pm i\chi^2\ .
        \end{equation}
         Then one could split the 
         supersymmetric currents into the bosonic and fermionic parts
         \begin{equation}
             J^A=j^A+\hat{j}^A, \qquad K^a=k^a+\hat{k}^a\,,
         \end{equation}
         where $\hat{j}^A$ and $\hat{k}^a$ are the fermionic currents, defined as:
         \begin{equation}
             \hat{j}^A=-\frac{i}{k}\epsilon^A_{BC}\psi^B\psi^C, \quad \hat{k}^a=-\frac{i}{k}\epsilon^a_{bc}\chi^b\chi^c\ .
         \end{equation}
         The currents $j^A, \hat{j}^A$ and $ k^a, \hat{k}^a$ generate 2 bosonic $SL(2, \mathbb{R})$ affine algebras at levels $k + 2, -2$  and 2 bosonic $SU(2)$ affine algebras at levels $k -2, +2$, respectively.  Since $j^A$ and $k^a$ commute with the free fermions,
          one  then obtain  2 (decoupled) bosonic WZW models and  free fermions.
            
    The stress tensor and supercurrent of the worldsheet theory are
        \begin{equation}\label{TandG}
            \begin{aligned}
                T&=\frac{1}{k}j^Aj_A-\frac{1}{k}\psi^A\partial\psi_A+\frac{1}{k}k^ak_a-\frac{1}{k}\chi^a\partial\chi_a+T(\mathbb{T}^4)\\
                G& =\frac{2}{k}(\psi^Aj_A+\frac{2i}{k}\psi^1\psi^2\psi^3)+\frac{2}{k}(\chi^ak_a+\frac{2i}{k}\chi^1\chi^2\chi^3)+G(\mathbb{T}^4)   \,.   
            \end{aligned}
        \end{equation}
        Superstring on AdS$_3\times$S$^3\times\mathbb{T}^4$ also contains the standard $bc$ and $\beta\gamma$ ghosts. The  BRST operator of the superstring can be constructed as:
        \begin{equation}
            Q_{\text{BRST}}=\oint dz\left(c(T+\frac{1}{2}T_{\text{gh}})+\gamma(G+\frac{1}{2}G_{\text{gh}})\right)\ .
        \end{equation}
        Physical vertex operators should be BRST invariant.    Importantly, to obtain all the physical operators (in particular, the ones that characterize the long strings), one needs to consider the so-called spectrally flowed representations of the affine algebra $\mathfrak{sl}(2,R)_k$ \cite{Maldacena:2000hw}. These representations come from the  the following automorphism $\sigma$ of the  algebra $\mathfrak{sl}(2,R)_k$ ,
        namely the spectral flow \cite{Maldacena:2000hw,Giribet:2007wp}\footnote{Note that for the construction of  chiral operators (short strings), it will be  convenient to also spectral flow the (supersymmetric) S$^3$ part \cite{Giribet:2007wp}, though it does not give new representations but only reshuffles states. In the present work, we focus on the long strings, for which one only needs to consider the spectral flow of the (supersymmetric) AdS$_3$ part \cite{Yu:2024kxr,Iguri:2022pbp}.}
        \begin{equation}\label{spectralflow}
        \begin{aligned}
             &\sigma^w(J_m^\pm)=J^{\pm}_{m\mp w}, \quad \sigma^w(J_m^3)=J_m^3+\frac{kw}{2}\delta_{m,0}\,,\\ 
            &\sigma^w(\psi_m^\pm)=\psi^{\pm}_{m\mp w}, \quad
            \sigma^w(\psi_m^3)=\psi_m^3\,,
        \end{aligned}
        \end{equation}
         Note that spectral flow acts on  the decoupled currents $j^A$ and the fermionic currents $\hat{j}^A$  the same way as on the full currents $J^A$ (except for different values of $k$). Finally,
      superstring theory on AdS$_3\times$S$^3\times\mathbb{T}^4$  has  $\mathcal{N}=4$ space-time supersymmetries. In fact, one can construct 8 supercharges from the spin fields in the Ramond sector of the worldsheet fermions. These supercharges from a  global $\mathcal{N}=4$ superconformal algebra  \cite{Giveon:1998ns}.

\subsection{The correlator $\mathcal{M}_{OOO}$}\label{Thestringcorrelator}

In this section, we give the residues of  3-point string correlators at their  next-to-leading poles. Firstly, we review the 3-point string correlators calculated in \cite{Yu:2024kxr}. In \cite{Yu:2024kxr}, all physical vertex operators that have the lowest space-time weights (with a given  spectral flow parameter $w$) are constructed, including  those in the NS and R sectors.  The construction depends on the parity of $w$ and among all these operators, the simplest one is (in the so-called $x$-basis, see eq. (2.30) in \cite{Yu:2024kxr}):
\begin{equation}\label{stringoperator}
    O_{j,h}^{w(-1)}(x;z)=e^{-\phi}(z) \textbf{1}_{\psi}^w(x;z)V_{j,h}^w(x;z), \qquad \text{with} \quad w\in 2\mathbb{N}+1. 
\end{equation}
This operator is in the NS sector and we have written it in  picture $-1$ (we only write the left moving part, the right moving part is similar). In \eqref{stringoperator},
 $\phi$ is the bosonized $\beta\gamma$ ghosts;
  the operator $V_{j,h}^w(x;z)$ is a spectrally flowed operator in the bosonic $SL(2, \mathbb{R})_{k+2}$ WZW model, with $j$ being the spin and $h$ being the space-time weight; $w$ is the spectral flow parameter, which must be odd because of the GSO projection; the operator $\textbf{1}_{\psi}^w(x;z)$ is the $w$ spectrally flowed operator of the identity in the theory of free fermions. Denote the bosonic, fermionic and full space-time weights as $h,\hat{h}, H$, then:
\begin{equation}\label{3weights}
    h=m+\frac{(k+2)w}{2}, \quad \hat{h}=-w, \quad H=h+\hat{h}=m+\frac{wk}{2}\,,
\end{equation} 
where $m$ is the eigenvalue of $J^3$ before doing the spectral flow.  In terms of $H$ the mass-shell condition can be written as:
\begin{equation}
    -\frac{j(j-1)}{k}-w H+\frac{kw^2}{4}=\frac{1}{2}\ .
\end{equation}
One can also obtain the picture 0 version of $O_{j,h}^{w(-1)}(x;z)$ by acting the picture rasing operator $e^\phi G$:
\begin{equation}\label{stringoperator0}
 \begin{aligned}
      &O_{j,h}^{w(0)}(x;z)
    =\frac{1}{k}\bigg[\left(h-\frac{(k+2)w}{2}+j-1\right)\psi^{+,w}(x;z)V_{j,h-1}^w(x;z)
      \\&\qquad -2(h-w)\psi^{3,w}(x;z)V_{j,h}^w(x;z)
      +\left(h-\frac{(k+2)w}{2}-j+1\right)\psi^{-,w}(x;z)V_{j,h+1}^w(x;z)\,\bigg].
 \end{aligned}
 \end{equation}
 Finally, we note that the operator $O_{j,h}^{w(-1)}(x;z)$ is universal, that is, its construction  depends only on the  (supersymmetric) AdS$_3$ part. Thus, it exists in all superstring backgrounds AdS$_3\times X$.  Note that for $w$ even, there also exists an operator whose construction  depends only on the  (supersymmetric) AdS$_3$ part (so it is also universal), see eq. (2.64) in \cite{Yu:2024kxr}. These two operators are analogs of the string operators considered in the bosonic case \cite{Eberhardt:2021vsx}. In this work, we only consider the 3-point string correlator of three $O^w_{j,h}$'s, so two of them are in picture $-1$ \eqref{stringoperator} and the remaining one is in  picture $0$ \eqref{stringoperator0}. The spectral flow parameters $w$ of them  are all odd.  One could also consider correlators involving operators with $w$ being even,  but the form of these correlators will be  more complicated \cite{Yu:2024kxr}.

In the  work \cite{Yu:2024kxr}, the authors had calculated the 3-point string correlators of three $O^w_{j,h}$'s  (denoted as $\mathcal{M}_{OOO}$ in \cite{Yu:2024kxr}, where ``O'' means that the operator parity is odd) and  matched them with the CFT side at the leading  (0-th) order of the conformal perturbation computation. Note that since $w_1+w_2+w_3\in2\mathbb{Z}+1$ for $\mathcal{M}_{OOO}$, only  even orders contribute \cite{Eberhardt:2021vsx,Yu:2024kxr} (on the CFT side, no covering map exists for odd orders due to the Riemann-Hurwitz formula).   In the present work we focus on the next-to-leading order, which is then the 2-nd order.  Let's first recall the form of the correlator (eq. (3.11) in \cite{Yu:2024kxr}):
\begin{equation}\label{OOO} 
\begin{aligned}
    &\mathcal{M}_{OOO}=
    \frac{C_{S^2}}{k}\Big[ (h_3-
    \frac{(k+2)w_3}{2}+j_3-1)P^2_{w_1,w_2,w_3-1}\langle h_3-1\rangle\\
    &+2(h_3-w_3)P_{w_1,w_2,w_3-1}P_{w_1,w_2,w_3+1}\langle h_3\rangle
    +(h_3-\frac{(k+2)w_3 }{2}-j_3+1)P^2_{w_1,w_2,w_3+1}\langle h_3+1\rangle\Big]\\
    &\hspace{10cm}\times (\text{anti-holomorphic part})\,,
\end{aligned}
\end{equation}
where 
\begin{itemize}
    \item $C_{S^2}$ is the normalization constant of the (sphere) string path integral.
    \item $P_{w_1,w_2,w_3}$ is defined by
           \begin{equation}
          P_{w_1,w_2,w_3}=\left\{
          \begin{aligned}
              &0, \qquad\qquad~~ \qquad{\text{for }  \sum_{j}w_j<2\text{max}_{i=1,2,3} w_i \quad \text{or} \quad \sum_{i}w_i\in 2Z+1}\\
             &S_{w}G\left(\frac{w_1+w_2+w_3}{2}+1\right)\prod_{i=1}^3\frac{G(\frac{w_1+w_2+w_3}{2}-w_i+1)}{G(w_i+1)}, \qquad \text{otherwise}\ .
          \end{aligned}\right.
           \end{equation}
        In the above, $G(n)$ is the Barnes G function 
        \begin{equation}
            G(n)=\prod_{i=1}^{n-1}\Gamma(i)\,,
        \end{equation}
        and the  function $S_{w}$ is a phase depending on $w$ mod 2
      \begin{equation}
          S_w=(-1)^{\frac{1}{2}x(x+1)},\qquad x=\frac{1}{2}\sum_{i=1}^3(-1)^{w_iw_{i+1}}w_i\ .
        \end{equation}
        \item $\langle h_3+a\rangle\times\langle \bar{h}_3+\bar{a}\rangle$ $(a,\bar{a}=\pm 1,0)$ denotes the 3-point functions of spectrally flowed operators in the $SL(2,\mathbb{R})$ WZW model:
\begin{equation}
    \langle h_3+a\rangle\times\langle \bar{h}_3+\bar{a}\rangle\equiv \left\langle V_{j_1,h_1,\bar{h}_1}^{w_1}(0;0)V_{j_2,h_2,\bar{h}_2}^{w_2}(1;1)V_{j_3,h_3+a,\bar{h}_3+\bar{a}}^{w_3}(\infty;\infty)\right\rangle.
\end{equation}
The correlator on the right hand side above is obtained in \cite{Dei:2021xgh}. For $\sum_i w_i$ odd (which is the present case), we have:
\begin{equation}\label{3pointformula}
\begin{aligned}
    &\langle V_{j_1,h_1,\bar{h}_1}^{w_1}(0;0)V_{j_2,h_2,\bar{h}_2}^{w_2}(1;1)V_{j_3,h_3,\bar{h}_3}^{w_3}(\infty;\infty)\rangle= \frac{C_{S^2}}{k}\mathcal{N}(j_1)D\left(\frac{k+2}{2}-j_1,j_2,j_3\right)\\
      &\hspace{6cm}\times\int \prod_{i=1}^3 \frac{d^2y_i}{\pi}
       \prod_{i=1}^3 \left|y_i^{\frac{(k+2)w_i}{2}+j_i-h_i-1}\mathfrak{B}(y_1,y_2,y_3)\right|^2\,,
\end{aligned}
\end{equation}
where $D(j_1,j_2,j_3)$ is the three-point function of three unflowed vertex operators \cite{Teschner:1997ft}
    \begin{equation}
        D(j_1,j_2,j_3)=-\frac{G_k(1-j_1-j_2-j_3)}{2\pi^2 \nu^{j_1+j_2+j_3-1}\gamma\left(\frac{k-1}{k-2}\right)}\prod_{i=1}^3\frac{G_k(2j_i-j_1-j_2-j_3)}{G_k(1-2k_i)}\,,
    \end{equation}
    where $G_k(x)$ is the Barnes double Gamma function. The normalization factor $\mathcal{N}(j)$ is given by
    \begin{equation}
        \mathcal{N}(j)=\frac{\nu^{\frac{k}{2}-2j}}{\gamma(\frac{2j-1}{k-2})}\ ,
    \end{equation}
    where $\gamma(x)=\Gamma(x)/\Gamma(1-x)$.
Finally, 
$\mathfrak{B}(y_1,y_2,y_3)$ is the   correlator in the $y$-basis,
\begin{equation}\label{ybasiscorrelator}
    \mathfrak{B}(y_1,y_2,y_3)=X_{123}^{\frac{k+2}{2}-j_1-j_2-j_3}\prod_{i=1}^3X_i^{-\frac{k+2}{2}+j_1+j_2+j_3-2j_i}\,,
\end{equation}
and  for $I\subset \{1,2,3\}$, $X_I$ is defined in terms of the function $ P_{w_1,w_2,w_3}$ as
         \begin{equation}\label{defXi}
               X_I(y_1,y_2,y_3)=\sum_{i\in I, \epsilon_i=\pm 1}P_{w+\sum_{i\in I}\epsilon_i e_i}
                 \prod_{i\in I}y_i^{\frac{1-\epsilon_i}{2}}\,.
           \end{equation}
\end{itemize}   
   Then one can  write $\mathcal{M}_{OOO}$ in \eqref{OOO} explicitly as integrals of $y_i$:
\begin{equation}\label{intOOO}
\begin{aligned}
      \mathcal{M}_{OOO}=
      \frac{C_{S^2}}{k}\mathcal{N}(j_1)D\left(\frac{k+2}{2}-j_1,j_2,j_3\right)&\\
      \times\int \prod_{i=1}^3 \frac{d^2y_i}{\pi}
       \prod_{i=1}^3 &\left|y_i^{\frac{(k+2)w_i}{2}+j_i-h_i-1}\mathfrak{F}^{(3)}(y_1,y_2,y_3)\mathfrak{B}(y_1,y_2,y_3)\right|^2\,,
\end{aligned}
\end{equation}
where 
 $\mathfrak{F}^{(3)}(y_1,y_2,y_3)$ is
\begin{equation}
\begin{aligned}
    \mathfrak{F}^{(3)}(y_1,y_2,y_3)=P_{w_1,w_2,w_3+1}P_{w_1,w_2,w_3-1}\Bigg[\left(2-\frac{y_3}{a_3}-\frac{a_3}{y_3}\right)h_3+\left(\frac{k+2}{2}w_3-j_3+1\right)\frac{y_3}{a_3}\\
    +\left(\frac{k+2}{2}w_3+j_3-1\right)\frac{a_3}{y_3}-2w_3\Bigg]\ .
\end{aligned}
\end{equation}
In the above, the superscript $(3)$ in $\mathfrak{F}^{(3)}$ means it comes from choosing the third operator in picture 0, that is, \eqref{OOO} is calculated from the picture choice: $O^{w_1(-1)}_{j_1,h_1}$, $O^{w_2(-1)}_{j_2,h_2}$, $O^{w_3(0)}_{j_3,h_3}$.  We stress that while different picture choices will give identical results for the correlator $\mathcal{M}_{OOO}$, they give different integrands in the integral from of $\mathcal{M}_{OOO}$ like \eqref{intOOO}. 
Making use of integration by parts, one can in fact replace the above $\mathfrak{F}^{(3)}(y_1,y_2,y_3)$ by a function $\mathfrak{F}(y_1,y_2,y_3)$ that is symmetric on the three index $1,2,3$:
\begin{equation}\label{superybasis}
    \mathfrak{F}(y_1,y_2,y_3)=P_{w_1,w_2,w_3+1}P_{w_1,w_2,w_3-1}w_3k+\left(j_1+j_2+j_3-\frac{k+2}{2}\right)\frac{X_1X_2X_3}{X_{123}}\ ,
\end{equation}
where $\mathfrak{F}_y(y_1,y_2,y_3)$ is the correlator in the $y$-basis, which by definition does not have any $h_i$ dependence. Note that $P_{w_1,w_2,w_3+1}P_{w_1,w_2,w_3-1}w_3$ is in fact symmetric in the three indices $1,2,3$ \cite{Yu:2024kxr}, so $\mathfrak{F}(y_1,y_2,y_3)$ is also  symmetric in the three indices.

\subsection{The residue of the correlator}

On the CFT side, denote the  operator corresponding to $O^w_{j,h}$ by $V^{(w)}_\alpha$ (for its explicit form, see \eqref{theoperator}), and their 3-point correlator is denoted as $\mathbb{M}_{VVV}$. In \cite{Eberhardt:2021vsx}, it was shown that the poles of $\mathcal{M}_{OOO}$ and   $\mathbb{M}_{VVV}$ coincide. Thus to match the two sides, one should test the matching of the residues:
\begin{equation}\label{Thematching}
   \left(\mathop{\text{Res}}\limits_{\sum_i j_i=2-\frac{k}{2}+\frac{mk}{2}}\mathcal{M}_{OOO}\right)\prod_{i=1}^3N_{O_i}^{-1} \mathop{=}\limits^{?}
    \left( \mathop{\text{Res}}\limits_{2b(\sum_i \alpha_i-Q)=m}\mathbb{M}_{VVV}\right)\prod_{i=1}^3N_{V_i}^{-1}\,,
\end{equation}
where $N_{O_i}$ and $N_{V_i}$ are  normalization factors of the string and CFT operators respectively.  The integer $m$ counts the order. For  $m=0$, in \cite{Yu:2024kxr} the authors have successfully matched the two sides. This matching determines the constant $C_{S^2}$:
\begin{equation}
    C_{S^2}=\frac{128\pi N\nu^{k-2}}{k^2\gamma\left(\frac{k+1}{k}\right)^2}.
\end{equation}
Besides, it also determines the normalization factor of the operators on the string side \cite{Yu:2024kxr}:
\begin{equation}
    N_{O_i}= N(w_i,j_i)=\frac{4\sqrt{N}\nu^{\frac{k}{2}-1}w_i^{\frac{3}{2}-2j_i}}{ k\gamma\left(\frac{k+1}{k}\right)}.
\end{equation}
Note that the operators on the CFT side are in fact canonically normalized \cite{Eberhardt:2021vsx,Yu:2024kxr}, so $N_V=1$\footnote{In practice, on the CFT side we obtain the residue by a conformal perturbation computation, where marginal operators are inserted and they also should be normalized. See the calculation in section \ref{Thenormalization}. }. 
However, since for $m=0$ no marginal operator is inserted, one can not determine the deformation parameter $\mu$ in terms of $\nu$.

Now we study the case of $m=2$, where $j_1+j_2+j_3=1+\frac{k+2}{2}$. For this case, the pole comes not from the prefactor but  from the integration of $y_i$ in \eqref{intOOO}.  In fact, the prefactor is regular \cite{Eberhardt:2021vsx}:
    \begin{equation}
    \mathcal{N}(j_1)D\left(\frac{k+2}{2}-j_1,j_2,j_3\right)=-\frac{\nu^{-\frac{k+2}{2}}}{2\pi^2\gamma\left(\frac{k+1}{k}\right)}.
\end{equation}
For the $y_i$-integral, note that the exponent of $X_{123}$ becomes $-1$ so  the residue of left hand side of \eqref{Thematching} can be obtained the same way as in the bosonic case \cite{Eberhardt:2021vsx}: use the formula
\begin{equation}\label{delta1}
     \mathop{\text{Res}}\limits_{\epsilon=0}\left|x^{-1-\epsilon}\right|=-\pi\delta^2(x),
\end{equation}
one can integrate out $y_3$ and then write the residue of   $\mathcal{M}_{OOO}$  as:
\begin{equation}
\begin{aligned}\label{stringresidue}
      \mathop{\text{Res}}\limits_{\sum_i j_i=2+\frac{k}{2}}\mathcal{M}_{OOO}=\mathcal{N}_{string}
      \int \prod_{i=1}^2 \frac{d^2y_i}{\pi}
       \left|\prod_{i=1}^3 y_i^{\frac{(k+2)w_i}{2}+j_i-h_i-1}\hat{\mathfrak{F}}^{(3)}(y_1,y_2,y_3)\hat{\mathfrak{B}}(y_1,y_2,y_3)\right|^2\,,
\end{aligned}
\end{equation}
where 
\begin{equation}\label{Nstring}
    \mathcal{N}_{string}= \frac{C_{S^2}}{k}\frac{\nu^{-\frac{k+2}{2}}}{2\pi^2\gamma\left(\frac{k+1}{k}\right)}\left(\prod_{i=1}^3N(w_i,j_i)\right)^{-1}=\frac{\nu^{-k}}{\pi N^{\frac{1}{2}}}\prod_{i=1}^3 w_i^{2j_i-\frac{3}{2}},
\end{equation}
and the function $\hat{\mathfrak{B}}(y_1,y_2,y_3)$ is the same as the integrand in the bosonic case \cite{Eberhardt:2021vsx}:
\begin{equation}\label{bosonicB}
    \hat{\mathfrak{B}}(y_1,y_2,y_3)=X_1^{-2j_1}X_2^{-2j_2}X_3^{2-2j_3}.
\end{equation}
The function $\hat{\mathfrak{F}}^{(3)}$ is an extra term in the SUSY case:
\begin{equation}\label{Theintegrand}
\begin{aligned}
     \hat{\mathfrak{F}}^{(3)}&(y_1,y_2,y_3)=\\
     &P_{w_1,w_2,w_3-1}P_{w_1,w_2,w_3+1}w_3k+\frac{X_3^2}{y_3}\left(H_3-\frac{w_3k}{2}-j_3+1\right) +2(j_3-1)X_3P_{w_1,w_2,w_3-1}.
\end{aligned}
\end{equation}
Note that by \eqref{delta1}, $y_3$ is no longer  a free parameter and is determined by:
\begin{equation}\label{delta2}
    X_{123}(y_1,y_2,y_3)=0.
\end{equation}
Thus $\hat{\mathfrak{F}}^{(3)}(y_1,y_2,y_3)$ and $\hat{\mathfrak{B}}(y_1,y_2,y_3)$ should be viewed as functions of $y_1,y_2$.

Note that the integrand in \eqref{stringresidue} does not have a natural unique form. In fact, different picture choices lead to  different integrands: if we chose the $i$-th operator in the picture 0,  we will get a integrand $\hat{\mathfrak{F}}^{(i)}(y_1,y_2,y_3)\hat{\mathfrak{B}}(y_1,y_2,y_3)$, where  $\hat{\mathfrak{F}}^{(i)}$ is obtained by exchanging all  subscripts $3\leftrightarrow i$ in \eqref{Theintegrand}\footnote{One may try to extract the residue by integrating out $y_3$ in \eqref{intOOO}, with $\mathfrak{F}^{(3)}$ replaced by $\mathfrak{F}$ in  \eqref{superybasis}. In this way, one cen check that   the  result is the same as \eqref{stringresidue}.}. Since we will ultimately match the integrand with the CFT side, we need the correct form of the integrand on the string side, which is a priori not  known. Now we give an ansatz for the proper form of the integrand. Firstly, we define an equivalent relation $A\cong B$, which means:
\begin{equation}
     \int \prod_{i=1}^2  \frac{d^2y_i}{\pi}
       \Bigg|\prod_{i=1}^3 y_i^{\frac{(k+2)w_i}{2}+j_i-h_i-1}A\hat{\mathfrak{B}}\Bigg|^2=\int \prod_{i=1}^2  \frac{d^2y_i}{\pi}
      \Bigg| \prod_{i=1}^3 y_i^{\frac{(k+2)w_i}{2}+j_i-h_i-1}B\hat{\mathfrak{B}}\Bigg|^2
\end{equation} 
In the following, we focus on the left-moving part and always omit the right-moving dependence, which can be  analyzed similarly. 
If we denote the correct form of the integrand as $\hat{\mathfrak{F}}\hat{\mathfrak{B}} $ (omit the power functions of $y_i$), then we have
\begin{equation}\label{eqv}
    \hat{\mathfrak{F}}\cong \hat{\mathfrak{F}}^{(1)}\cong \hat{\mathfrak{F}}^{(2)}\cong \hat{\mathfrak{F}}^{(3)}.
\end{equation}
A simple ansatz for $\hat{\mathfrak{F}}(y_1,y_2,y_3)$ is to take a linear combination of the ones coming from the three picture choices:
\begin{equation}\label{ansatz}
    \hat{\mathfrak{F}}(y_1,y_2,y_3)=\alpha\hat{\mathfrak{F}}^{(1)}(y_1,y_2,y_3)+\beta\hat{\mathfrak{F}}^{(2)}(y_1,y_2,y_3)+\gamma\hat{\mathfrak{F}}^{(3)}(y_1,y_2,y_3), \quad \alpha+\beta+\gamma=1.
\end{equation}
In the above ansatz,  $\alpha,\beta,\gamma$ should be 3 constants, otherwise, \eqref{eqv} does not hold (see \eqref{OOOstring} and the comment below it). It turns out that to get the correct form of $\hat{\mathfrak{F}}(y_1,y_2,y_3)$ that match with the CFT side, we need a more general ansatz where  $\alpha,\beta,\gamma$ could  also be functions of  $y_1, y_2$. 
This in fact can be achieved using  integration by part. Firstly, from \eqref{Theintegrand} we can write  $\hat{\mathfrak{F}}^{(3)}(y_1,y_2,y_3)$ as
\begin{equation}
\begin{aligned}
    &\hat{\mathfrak{F}}^{(3)}(y_1,y_2,y_3)\\
    =&P_{w_1,w_2,w_3-1}P_{w_1,w_2,w_3+1}w_3k +2(j_3-1)X_3P_{w_1,w_2,w_3-1}
    +\alpha\frac{X_3^2\mathfrak{h}_3}{y_3}+\beta\frac{X_3^2\mathfrak{h}_3}{y_3}+\gamma\frac{X_3^2\mathfrak{h}_3}{y_3},
\end{aligned}
\end{equation}
where we introduce $\mathfrak{h}_i$ as
\begin{equation}
   \mathfrak{h}_i\equiv H_i-\frac{w_ik}{2}-j_i+1=-\left(\frac{(k+2)w_i}{2}+j_i-h_i-1\right),
\end{equation}
and $\alpha, \beta, \gamma$ are functions of $y_1,y_2$, satisfying $\alpha+\beta+\gamma=1$. Using integration by part, we have (recall that we always omit the right-moving part of the integrand):
\begin{equation}\label{y1der}
    \int \prod_{i=1}^2  \frac{d^2y_i}{\pi}\partial_{y_1}\left(
       \prod_{i=1}^3 y_i^{\frac{(k+2)w_i}{2}+j_i-h_i-1}F\hat{\mathfrak{B}}(y_1,y_2,y_3)\right)=0,
\end{equation}
where $F$ could be an arbitrary function 
that  makes the possible boundary terms in \eqref{y1der} vanish. 
The explicit from of \eqref{y1der} is:
\begin{equation}\label{F}
\begin{aligned}
   0=& \int \prod_{i=1}^2  \frac{d^2y_i}{\pi}
       \prod_{i=1}^3 y_i^{\frac{(k+2)w_i}{2}+j_i-h_i-1}\hat{\mathfrak{B}}(y_1,y_2,y_3)F\\
       &\left(\frac{-\mathfrak{h}_1}{y_1}+\frac{-\mathfrak{h}_3}{y_3}\left(-\frac{X_3^2}{X_1^2}\right)+\frac{-2j_1}{X_1}P_{w_1-1,w_2,w_3}+\frac{2-2j_3}{X_3}P_{w_1,w_2,w_3-1}\left(-\frac{X_3^2}{X_1^2}\right)+\frac{\partial_{y_1}F}{F}\right).
\end{aligned}
\end{equation}
Then letting $F=\alpha X_1^2$ in \eqref{F}, we have:
\begin{equation}
    \frac{\alpha X_3^2\mathfrak{h}_3}{y_3}\cong  \frac{\alpha X_1^2\mathfrak{h}_1}{y_1}+2\alpha (j_1-1)X_1P_{w_1-1,w_2,w_3}+2\alpha(1-j_3)X_3P_{w_1,w_2,w_3-1}-X_1^2\partial_{y_1}\alpha.
\end{equation}
Similarly,  taking  derivative with respect to $y_2$ of the integrand as in \eqref{y1der} and letting $F=\beta X_2^2$,  we have:
\begin{equation}
     \frac{\beta X_3^2\mathfrak{h}_3}{y_3}\cong  \frac{\beta X_2^2\mathfrak{h}_2}{y_2}+2\beta (j_2-1)X_2P_{w_1,w_2-1,w_3}+2\beta(1-j_3)X_3P_{w_1,w_2,w_3-1}-X_2^2\partial_{y_2}\beta.
\end{equation}
Then  $ \hat{\mathfrak{F}}^{(3)}$ is equivalent to:
\begin{equation}\label{OOOstring}
\begin{aligned}
      \hat{\mathfrak{F}}^{(3)}\cong &P_{w_1,w_2,w_3-1}P_{w_1,w_2,w_3+1}w_3k +2(j_3-1)X_3P_{w_1,w_2,w_3-1}
    +\frac{\gamma X_3^2\mathfrak{h_3}}{y_3}\\
    &+\frac{\alpha X_1^2\mathfrak{h}_1}{y_1}+2\alpha (j_1-1)X_1P_{w_1-1,w_2,w_3}+2\alpha(1-j_3)X_3P_{w_1,w_2,w_3-1}-X_1^2\partial_{y_1}\alpha\\
    &+\frac{\beta X_2^2\mathfrak{h}_2}{y_2}+2\beta (j_2-1)X_2P_{w_1,w_2-1,w_3}+2\beta(1-j_3)X_3P_{w_1,w_2,w_3-1}-X_2^2\partial_{y_2}\beta\\
    =&P_{w_1,w_2,w_3-1}P_{w_1,w_2,w_3+1}w_3k-X_1^2\partial_{y_1}\alpha-X_2^2\partial_{y_2}\beta\\
    &+2\alpha (j_1-1)X_1P_{w_1-1,w_2,w_3}+2\beta (j_2-1)X_2P_{w_1,w_2-1,w_3}+2\gamma(j_3-1)X_3P_{w_1,w_2,w_3-1}\\
    &+\frac{\alpha X_1^2\mathfrak{h}_1}{y_1}+\frac{\beta X_2^2\mathfrak{h}_2}{y_2}+\frac{\gamma X_3^2\mathfrak{h_3}}{y_3}\\
    \equiv &\hat{\mathfrak{F}}.
\end{aligned}
\end{equation}
Comparing with the naive ansatz \eqref{ansatz}, in the above we have  2 additional  terms $-X_1^2\partial_{y_1}\alpha-X_2^2\partial_{y_2}\beta$. When $\alpha, \beta, \gamma$ are constants, these terms vanish and we return to the ansatz \eqref{ansatz}. 
Now we  eliminate the $\mathfrak{h}_i$ $(i=1,2,3)$ dependence of $\hat{\mathfrak{F}}$. This is achieved by the mass-shell condition:
\begin{equation}\label{Thespacetimeweight}
    H_i=\frac{j_i(1-j_i)}{w_ik}+\frac{kw_i}{4}-\frac{1}{2w_i},
\end{equation}
which leads to:
\begin{equation}
    \mathfrak{h}_i=\frac{j_i(1-j_i)}{w_ik}-\frac{kw_i}{4}-\frac{1}{2w_i}-j_i+1.
\end{equation}
 Let’s pause  to comment on the role played by the mass-shell condition here. For  bosonic string on AdS$_3\times X$, the matching of residues with the CFT side does
not require the use of the mass-shell condition \cite{Eberhardt:2021vsx}.  This is expected because one could modify  the mass-shell condition by adding an excitation in $X$, which should not really affect the  calculation (since  the marginal operator (when lifted up to the covering surface) in the bosonic case does not depend on $X$ \cite{Eberhardt:2021vsx}). On the contrary, in the SUSY case the marginal operator (when lifted up to the covering surface)  indeed involves the internal CFT (the S$^3$ and $\mathbb{T}^4$ parts), as we will show in section \ref{TheproposedCFT}. Thus, the mass-shell condition could play a role in obtaining the proper form of the residue on the string side\footnote{One exception is the matching  at the 0-th order, where the mass-shell condition is not used \cite{Yu:2024kxr}.  This could be reasonable since no marginal operator is inserted at this order.}. Besides,  picture changing makes the $h_i$ $(i=1,2,3)$ dependence of the string correlator \eqref{intOOO} not as simple as in the bosonic case (where $h_i$ only appears in the power of $y_i$). As a result, the general ansatz for the integrand  $\hat{\mathfrak{F}}$ in \eqref{OOOstring}  always includes (at least one) $h_i$ ($\mathfrak{h}_i$), which should be eliminated by the mass-shell condition in order to match with the CFT side (see section \ref{conformal perturbation}).   

After eliminating the  $\mathfrak{h}_i$ $(i=1,2,3)$ dependence, we need also to eliminate the $j_3$ dependence of $\hat{\mathfrak{F}}$ by
using the condition $j_3=2+\frac{k}{2}-j_1-j_2$. Then
we can organize $\hat{\mathfrak{F}}$ in \eqref{OOOstring} as a sum of 10 terms:
\begin{equation}\label{stringsum}
     \hat{\mathfrak{F}}=\sum_{X\in \mathfrak{S}}\mathbb{C}^{\text{string}}[X]X,
\end{equation}
where $\mathfrak{S}$ is the following set: 
\begin{equation}\label{list10}
   \mathfrak{S}=\left\{ \frac{j_1j_2}{k},\quad  \frac{j_1^2}{k}, \quad \frac{j_2^2}{k},\quad \frac{j_1}{k}, \quad\frac{j_2}{k}, \quad \frac{1}{k}, \quad 1, \quad k, \quad j_1, \quad j_2\right\}
\end{equation}
and $\mathbb{C}^{\text{string}}[X]$ are the coefficients of these 10 terms. Note that these terms are all  bilinears of the form $\frac{1}{k}ab$, with $a,b$ being $1,k,j_1,j_2$. We divide them into two groups, one is called regular terms, with none of $a$ and $b$ being $k$:
\begin{equation}
   \mathfrak{S}^{reg}=\left\{  \frac{j_1j_2}{k},\quad  \frac{j_1^2}{k}, \quad \frac{j_2^2}{k},\quad \frac{j_1}{k}, \quad\frac{j_2}{k}, \quad \frac{1}{k}\right\}
\end{equation}
and the remaining 4 terms are called irregular:
\begin{equation}
     \mathfrak{S}^{irreg}=\left\{ 1, \quad k, \quad j_1, \quad j_2\right\}.
\end{equation}
This distinction is made by hand. The advantage is that the matching of the two sides in section \ref{matchthetwosides} will be more organized. In particular, the matching of irregular terms will involve new mathematical identities of covering maps while the matching of regular terms won't.
 In fact, it will be clarified why such terms are called regular/irregular when we discuss the CFT side (the discussion around \eqref{re/irre}).

Now we  write all the 10 terms explicitly. 
For the regular terms:
\begin{equation}\label{regular}
    \begin{aligned}
         \mathbb{C}^{\text{string}}\left[\frac{j_1j_2}{k}\right]\!=& -\frac{2\gamma X_3^2}{w_3y_3}, \quad\!
        \mathbb{C}^{\text{string}}\left[\frac{j_1^2}{k}\right]\!=  -\frac{\alpha X_1^2}{w_1y_1}-\frac{\gamma X_3^2}{w_3y_3},\quad\!
       \mathbb{C}^{\text{string}}\left[\frac{j_2^2}{k}\right]\!= -\frac{\beta X_2^2}{w_2y_2}-\frac{\gamma X_3^2}{w_3y_3}\\
       \mathbb{C}^{\text{string}}\left[\frac{1}{k}\right]=&  -\frac{2\gamma X_3^2}{w_3y_3},\quad
        \mathbb{C}^{\text{string}}\left[\frac{j_1}{k}\right]=  \frac{\alpha X_1^2}{w_1y_1}+\frac{3\gamma X_3^2}{w_3y_3},\quad
         \mathbb{C}^{\text{string}}\left[\frac{j_2}{k}\right]= \frac{\beta X_2^2}{w_2y_2}+\frac{3\gamma X_3^2}{w_3y_3}.
    \end{aligned}
\end{equation}
For the irregular terms, the expressions are more complicated:
\begin{equation}\label{irrgular}
\begin{aligned}
    \mathbb{C}^{\text{string}}[j_1]&= 2\alpha X_1P_{w_1-1,w_2,w_3}-2\gamma X_3 P_{w_1,w_2,w_3-1}-\frac{\alpha X_1^2}{y_1}+\frac{\gamma X_3^2}{y_3}+\frac{\gamma X_3^2}{w_3y_3}\\
    \mathbb{C}^{\text{string}}[j_2]&= 2\beta X_2P_{w_1,w_2-1,w_3}-2\gamma X_3 P_{w_1,w_2,w_3-1}-\frac{\beta X_2^2}{y_2}+\frac{\gamma X_3^2}{y_3}+\frac{\gamma X_3^2}{w_3y_3}\\
    \mathbb{C}^{\text{string}}[k]&=  P_{w_1,w_2,w_3-1}P_{w_1,w_2,w_3+1}w_3+\gamma X_3P_{w_1,w_2,w_3-1}- \frac{\alpha w_1 X_1^2}{4 y_1}- \frac{\beta w_2 X_2^2}{4 y_2}- \frac{\gamma w_3 X_3^2}{4 y_3}\\
     &-\frac{\gamma  X_3^2}{2 y_3}- \frac{\gamma X_3^2}{4w_3 y_3}\\
    \mathbb{C}^{\text{string}}[1]&= -X_1^2\partial_{y_1}\alpha-X_2^2\partial_{y_2}\beta\!-\!2\alpha X_1P_{w_1-1,w_2,w_3}\!-\!2\beta X_2P_{w_1,w_2-1,w_3}\!+\!2\gamma X_3P_{w_1,w_2,w_3-1}\\
     &+\left(1-\frac{1}{2w_1}\right)\alpha \frac{ X_1^2}{ y_1}+\left(1-\frac{1}{2w_2}\right)\beta \frac{ X_2^2}{ y_2}-\left(1+\frac{2}{w_3}\right)\gamma \frac{ X_3^2}{ y_3}.
\end{aligned}
\end{equation}
We observe that the above expression could be written compactly by introducing a function $\tilde{X}_i$, which contain the same two terms as $X_i$ but with  a sign difference, e.g.
\begin{equation}
\begin{aligned}
     X_1&\equiv P_{w_1-1,w_2,w_3}y_1+P_{w_1+1,w_2,w_3}\\
    \tilde{X}_1&\equiv P_{w_1-1,w_2,w_3}y_1-P_{w_1+1,w_2,w_3}.
\end{aligned}
\end{equation}
$\tilde{X}_2$ and $\tilde{X}_3$ can be similarly defined.
Notice that these $\tilde{X}_i$ seems to be some ``conjugate'' versions of $X_i$\footnote{
We expect that one could similarly define the  ``conjugate'' $\tilde{X}_{I}$ of general $X_I$ ($I$ is a subset of $\{1,2,3\}$), which is the basic building block in 3-point functions of spectrally flowed operators in the $SL(2, \mathbb{R})$ WZW models \cite{Dei:2021xgh}. The way to define the ``conjugate'' seems to let $y\to -y$ in $X_I$. Furthermore,  one may even define the ``conjugate'' of the  basic building block $X_I$ ($I$ is a subset of $\{1,2,3,4\}$) of the  4-point function \cite{Dei:2021yom}. We expect these  $\tilde{X}_{I}$ may generally appear in matching the  correlators of the two sides in the SUSY case.}.  In terms of $X_i$ and $\tilde{X}_i$, we can write the irregular terms as:
\begin{equation}\label{irrgular1}
\begin{aligned}
    \mathbb{C}^{\text{string}}[j_1]&= \frac{\alpha X_1\tilde{X}_1}{y_1}-\frac{\gamma X_3\tilde{X}_3}{y_3}+\frac{\gamma X_3^2}{w_3y_3}\\
    \mathbb{C}^{\text{string}}[j_2]&= \frac{\beta X_2\tilde{X}_2}{y_2}-\frac{\gamma X_3\tilde{X}_3}{y_3}+\frac{\gamma X_3^2}{w_3y_3}\\
    \mathbb{C}^{\text{string}}[k]&=-\frac{ \alpha w_1 \tilde{X}_1^2}{4 y_1}-\frac{ \beta w_2 \tilde{X}_2^2}{4 y_2}-\frac{\gamma(w_3\tilde{X}_3-X_3)^2}{4w_3 y_3}\\
    \mathbb{C}^{\text{string}}[1]&= -X_1^2\partial_{y_1}\alpha-X_2^2\partial_{y_2}\beta\!-\!\frac{\alpha X_1\tilde{X}_1}{y_1}\!-\!\frac{\beta X_2\tilde{X}_2}{y_2}\!+\!\frac{\gamma X_3\tilde{X}_3}{y_3}
    -\frac{\alpha X_1^2}{ 2w_1y_1}-\frac{ \beta X_2^2}{2w_2 y_2}- \frac{2\gamma X_3^2}{ w_3y_3}.
\end{aligned}
\end{equation}
Notice that all these terms have complicated dependence on $w_1,w_2,w_3$.  

Finally, we can write the residue of the correlator $\mathcal{M}_{OOO}$ on the string side as:
\begin{equation}\label{stringresult}
\begin{aligned}
      \mathop{\text{Res}}\limits_{\sum_i j_i=2+\frac{k}{2}}\mathcal{M}_{OOO}=\mathcal{N}_{string}\!\int \prod_{i=1}^2 \frac{d^2y_i}{\pi}
       \prod_{i=1}^3 &\left|y_i^{\frac{(k+2)w_i}{2}+j_i-h_i-1}\hat{\mathfrak{B}}(y_1,y_2,y_3)\sum_{X\in \mathfrak{S}}\mathbb{C}^{\text{string}}[X]X\right|^2\,
\end{aligned}
\end{equation}
where $ \mathcal{N}_{string}$ is given in \eqref{Nstring} and $\hat{\mathfrak{B}}(y_1,y_2,y_3)$ is the bosonic part of the integrand, given in \eqref{bosonicB}. The set $\mathfrak{S}$ is given in \eqref{list10}.

\section{The CFT side}\label{TheCFTside}

In this section, we study the residues of correlators on the CFT side. Firstly, we review some basic facts about   conformal field theories of  symmetric orbifold. In particular, we review how to calculate correlators using  covering maps. Then, we describe the proposed CFT dual of superstring on AdS$_3\times$S$^3\times\mathbb{T}^4$, which is obtained by deforming a symmetric orbifold CFT by a (non-normalizable) exactly marginal operator  from the twist 2 sector. After that, we do the conformal perturbation computation at the second order, which will give the residues of the  correlators in the deformed theory. 

\subsection{Symmetric orbifold CFTs}

Consider a general two-dimensional CFT $\mathcal{M}$.  One can define a new two-dimensional CFT by considering N copies of $\mathcal{M}$, denoted as $\mathcal{M}_I$ $(I=1,2,...,N)$, and impose the equivalence relation:
\begin{equation}
    \mathcal{M}_I\sim\mathcal{M}_{g(I)}, \qquad \forall g\in S_N
\end{equation}
where $S_N$ is the the group of permutations. The resulting theory is the symmetric orbifold theory of $\mathcal{M}$,  which is a two-dimensional CFT and is denoted as:
\begin{equation}
    \text{Sym}^{N}(\mathcal{M})\equiv \otimes^N\mathcal{M}/S_N
\end{equation}
Then the  original theory $\mathcal{M}$ is called the seed theory.

The Hilbert space of $\text{Sym}^{N}(\mathcal{M})$ is the direct sum of twisted sectors, with each sector
corresponding to a conjugacy class of $S_N$. The so-called ``untwisted sector'' corresponds to the conjugacy class of the identity element. General conjugacy
classes of $S_N$ are labeled by cycle shapes and we will mainly focus on the single-cycle sectors. The corresponding conjugacy class consists of all cyclic permutations of $w$ colors, where $w$ is the length of the (single) cycle (so one can simply call such a twisted sector the twist $w$ sector).   Consider a representative element of this conjugacy class, $g=(1,2,...,w)$. Then in the sector twisted by this element, general fields $X_i\in\mathcal{M}_i$ $(i=1,2,...,w)$ satisfy the following boundary conditions (we identify $X_{w+1}\equiv X_1$):
\begin{equation}
    X_i(e^{2\pi i}x+x_0)\sigma_g(x_0)=X_{i+1}(x+x_0)\sigma_g(x_0), \quad i\in\{1,2,...,w\}.
\end{equation}
Here $\sigma_g$ is the corresponding twist field and we set it at  a generic location $x_0$. Note that the remaining fields $X_I\in \mathcal{M}_I$ $(I=w+1,w+2,...,N)$  have trivial monodromies when  taken around $\sigma_g$. 

Due to the twisted boundary behavior of $X_i$, their modes expansion consist of the so-called fractional modes. To find them, it is convenient to first diagonalize the action of $g$. Considering the linear combinations:
\begin{equation}
    X^{[k]}(x)=\sum_{j=1}^w e^{-2\pi i\frac{k}{w}j}X^j(x).
\end{equation}
Then they have the following monodromies:
\begin{equation}\label{monodromies}
    X^{[k]}(x)(e^{2\pi i}x+x_0)\sigma_g(x_0)=e^{2\pi i\frac{k}{w}}X^{[k]}(x+x_0)\sigma_g(x_0).
\end{equation}
Thus $X^{[k]}(x)$ has the modes expansion (in the presence of the spin field):
\begin{equation}
    X^{[k]}(x)\sigma_g(x_0)=\sum_{n\in \mathbb{Z}}X_{-\frac{k}{w}-h+n}(x-x_0)^{\frac{k}{w}-n}\sigma_g(x_0),
\end{equation}
where $h$ is the conformal weight of $X^{[k]}(x)$ and $X_{-\frac{k}{w}-h+n}$ are fractional modes, which can act on the  spin field to generate general fields in this twisted sector:
\begin{equation}\label{fractionalaction}
    X_{-\frac{k}{w}-h+n}\sigma_g(x_0)=\frac{1}{2\pi i}\oint_{C(x_0)} dx (x-x_0)^{-\frac{k}{w}+n-1}X^{[k]}(x)\sigma_g(x_0).
\end{equation}
Finally, note that  operators in the $g$-twisted sector constructed above using fractional modes  are not invariant under the action of $S_N$. The real gauge invariant  operators in the twist $w$ sector can be obtained by summing over elements in the conjugacy class of $g=(1,2,...,w)$ as follows:
\begin{equation}\label{Fulltwist}
    \mathcal{O}_w(x)=\frac{\sqrt{(N-w)!w}}{\sqrt{N!}}\sum_{\tau\in [(1,2,...,w)]}O_\tau(x),
\end{equation}
where the prefactor comes form the standard normalization.

\subsubsection{Correlators in symmetric orbifold CFTs}\label{coveringmaplifted}

For  correlators in symmetric orbifold CFTs (in the large N limit), there is an  algorithm that can reduce the calculation to the one in the seed theory. This is a method making use of covering maps, developed by Lunin and Mathur \cite{Lunin:2000yv,Lunin:2001pw}.  The advantage of this method is that monodromies \eqref{monodromies} disappear when operators are lifted up to the covering surface. Then the problem becomes how to construct the covering maps, which is difficult for  general operator insertions. Let's consider a $m$-point function of gauge invariant  operators $\mathcal{O}_{w_j}(x_j)$ $(j=1,2,...,m)$. When lifted up to the covering surfaces, it can be written as \cite{Lunin:2000yv,Lunin:2001pw,Dei:2019iym}:
\begin{equation}\label{sumofcovering}
    \left\langle \prod_{j=1}^m \mathcal{O}_{w_j}(x_j)\right\rangle=\prod_{j=1}^m\frac{\sqrt{(N-w_j)!w_j}}{\sqrt{N!}}\sum_{\text{covering map $\Gamma$}}\left(\begin{matrix}
        N\\
        d
    \end{matrix} \right)\left|f_\Gamma \right|^2\left\langle \prod_{j=1}^m \tilde{\mathcal{O}}_{\tau_j}(z_j)\right\rangle\Bigg|_{\Gamma(z_i)=x_i}.
\end{equation}
where
\begin{itemize}
    \item  The summation is over all covering map $\Gamma$ with ramification indices $w_j$ at the respective insertion points $x_i$, that is, around  $z_i$  ($z$ is the coordinate of the covering surface) we have:
\begin{equation}\label{covering}
    \Gamma(z)=x_i+a_i(z-z_i)^{w_i}+...
\end{equation}
    \item $d$ is the number of elements that $(\tau_1,\tau_2,...,\tau_m)$ truly act on. It is related with the genus $g$ of the covering surface by  the Riemann-Hurwitz formula:
\begin{equation}\label{RH}
    g\equiv 1-d+\frac{1}{2}\sum_{j=1}^m(w_j-1).
\end{equation} 
    Note that the covering surfaces in the summation can have higher genus (and even be disconnected). 
    \item In the large N limit, the power of N is determined as:
\begin{equation}
    \left(\begin{matrix}
        N\\
        d
    \end{matrix} \right)\left(\prod_{j=1}^m\frac{\sqrt{(N-w_j)!w_j}}{\sqrt{N!}}\right)\sim
    N^{1-g-\frac{m}{2}}.
\end{equation}
Thus the normalization factor results in a large N expansion controlled by the genus of the covering surface. 
   \item $f_\Gamma$ is a factor determined by the covering map $\Gamma$. This is in fact a  Weyl factor that accounts for the non-trivial (induced) metric on the covering space \cite{Lunin:2000yv}. If the covering surface has genus 0 (we will focus on this simplest case), it can explicitly be
   computed as:
   \begin{equation}\label{fGamma}
       f_\Gamma=\prod_{i=1}^m w_i^{-\frac{c(w_i+1)}{24}}a_i^{-\frac{c}{24}\left(1-\frac{1}{w_i}\right)}\Pi^{-\frac{c}{12}},
   \end{equation}
   where $a_i$ is the coefficient determined in \eqref{covering} and $\Pi$ is the product of the residues of the covering map:
   \begin{equation}\label{defPi}
       \Pi=\prod_a\Pi_a, \qquad \Gamma(z)\sim \frac{\Pi_a}{z-z_a}+O(1). 
   \end{equation} 
   \item  $\left\langle \prod_{j=1}^m \tilde{\mathcal{O}}_{\tau_j}(z_j)\right\rangle$ is the correlator of gauge dependent operators, lifted up to the covering surface. 
\end{itemize}
Note that $\left\langle \prod_{j=1}^m \tilde{\mathcal{O}}_{\tau_j}(z_j)\right\rangle$ is not independent of the covering map, but still contains a factor that comes form the lifting. Here and in the following, when we lift a correlator up to a covering surface, we will always add  a tilde  on every operator ($\tilde{O}$);  in contrast, when we add a hat on every operator  ($\hat{O}$), we have stripped  the covering map dependent factor from $\tilde{O}$ that comes form the lifting.  One can see the form of the factor coming form the lifting as follows. Assume $\mathcal{O}_{\tau_i}(x_i)$ is generated as:
\begin{equation}
    \mathcal{O}_{\tau_i}(x_i)=\mathcal{O}_{-s}\sigma_{\tau_i}(x_i),
\end{equation}
where $\mathcal{O}_{-s}$ is a fractional mode and the above action is defined as  in
 \eqref{fractionalaction}, so $-s=-\frac{k}{w}-h+n$. Then we lift this action up to the covering surface\footnote{The lift of the twist field $\sigma_{\tau_i}(x_i)$ will be the identity operator on the covering surface \cite{Lunin:2001jy}. In the following we denote it as $\hat{\sigma}_{\tau_i}(z_i)$.}:
   \begin{equation}\label{liftfractional}
   \begin{aligned}
        &\mathcal{O}_{-s}\sigma_{\tau_i}(x_i)\\
        =&\frac{1}{2\pi i}\oint_{C(x_i)} dx (x-x_i)^{-s+h-1}\mathcal{O}_{\tau_i}(x)\sigma_{\tau_i}(x_i)\\
        =&\frac{1}{2\pi i}\oint_{C(z_i)} dz (\partial\Gamma(z))^{1-h}(\Gamma(z)-x_i)^{-s+h-1}\hat{\mathcal{O}}_{\tau_i}(z)\hat{\sigma}_{\tau_i}(z_i)\\
        =&\frac{1}{2\pi i}\oint_{C(z_i)} dz (aw(z-z_i)^{w-1}+...)^{1-h}(a(z-z_i)^{w}+...)^{-s+h-1}\hat{\mathcal{O}}(z)\hat{\sigma}_{\tau_i}(z_i)\\
        =&a^{-s}w^{1-h}\frac{1}{2\pi i}\oint_{C(z_i)} dz (1+...)^{1-h}(1+...)^{-s+h-1}\\
        &\hspace{3cm}(z-z_i)^{-ws+h-1}\left(\sum_m \hat{\mathcal{O}}_{-h+m}\hat{\sigma}_{\tau_i}(z_0)(z-z_i)^{-m}\right)\\
         =&a^{-s}w^{1-h}\left(\hat{\mathcal{O}}_{-ws}\hat{\sigma}_{\tau_i}(z_i)+...\right).
   \end{aligned}
   \end{equation}
   In the third line of \eqref{liftfractional}, we have assumed that $\mathcal{O}_{\tau_i}(x)$ is a Virasoro primary (on the base space), and the two ``...'' mean higher order terms in the Taylor expansion around $z=z_i$ of $\partial\Gamma$ and $\Gamma$ respectively. In the last line of \eqref{liftfractional}, the ``...'' means the contribution of the terms of the form $\hat{\mathcal{O}}_{-r}\hat{\sigma}_g(z_i)$ with $r<ws$ (and the coefficients of such terms will involve higher order Taylor coefficients  of the covering map around $z=z_i$). Thus, it is obvious that when lifting $\mathcal{O}_{-s}\sigma_{\tau_i}(x_i)$ up to the covering surface, an additional prefactor $a^{-s}$ will appear, which depends on the considered covering map so is consequential (in contrast, the factor $w^{1-h}$ is  inconsequential and we will always eliminate it when defining  the action of  fractional modes). For the terms  ``...'' in the last line of \eqref{liftfractional},  more covering map dependent factors will appear\footnote{In fact, we will see such terms (and the associated factors) in our conformal perturbation computation in section \ref{conformal perturbation}.}.
   
   Note that even though $\mathcal{O}_{\tau_i}(x)$ is a Virasoro primary, the ``...'' terms $\hat{\mathcal{O}}_{-r}\hat{\sigma}_g(z_i)$  $(r<ws)$  will generally contribute. They do not contribute only if the lifted operator $\hat{\mathcal{O}}_{-ws}\hat{\sigma}_g(z_i)$ is a Virasoro primary on the covering surface (then by definition these terms vanish). In this case we have $s=\frac{\hat{h}}{w}$, where $\hat{h}$ is the conformal weight of  the lifted primary field $\hat{\mathcal{O}}$ on the covering surface\footnote{This special case happens for the BPS operators in the symmetric orbifold theory of $\mathbb{T}^4$, see the analysis in \cite{Lunin:2001jy,Gaberdiel:2022oeu}. }. Then, in \eqref{sumofcovering}, we can write the lifted operator as 
   \begin{equation}
       \tilde{O}_{\tau_i}(z_i)=a_i^{-\frac{\hat{h}}{w_i}}\hat{O}_{\tau_i}(z_i),
   \end{equation} 
   where $\hat{h}$ is the weight of $\hat{O}_{\tau_i}$ on the covering surface. Now $\hat{O}_{\tau_i}$ does not have any  covering map dependence, so we can extract the covering map dependent part of the  correlator $\left\langle \prod_{j=1}^m \tilde{\mathcal{O}}_{\tau_j}(z_j)\right\rangle$ as:
   \begin{equation}\label{furtherreduce}
       \left\langle \prod_{j=1}^m \tilde{\mathcal{O}}_{\tau_j}(z_j)\right\rangle=\left|g_\Gamma\right|^2\left\langle \prod_{j=1}^m \hat{\mathcal{O}}_{\tau_j}(z_i)\right\rangle
   \end{equation}
   where (the weights of the operators $\mathcal{O}_{\tau_i}$ is denoted as $h_i$)
     \begin{equation}\label{gGamma}
       g_\Gamma=\prod_{i=1}^m a_i^{-\frac{\hat{h}_i}{w}}=\prod_{i=1}^m a_i^{-(h_i-h_0)}, \qquad h_0=\frac{c}{24}\left(w-\frac{1}{w}\right)
   \end{equation}
    collects all  conformal factors coming from the lifting.  Thus, the full covering map dependent factor in \eqref{sumofcovering} (with \eqref{furtherreduce}) is $|f_{\Gamma}g_{\Gamma}|^2$ and $\left\langle \prod_{j=1}^m \hat{\mathcal{O}}_{\tau_j}(z_i)\right\rangle$ is simply a correlator in the seed theory that does not have any $w_i$ dependence. The  formula \eqref{sumofcovering} together with  \eqref{furtherreduce}  applied to the conformal perturbation computation in the bosonic duality (see eq. (2.42) in \cite{Eberhardt:2021vsx}), since the marginal operator in the bosonic case is still a Virasoro primary when lifted up to the covering surface.  However, as will be shown shortly, the lifted marginal operator is no longer a Virasoro primary in the supersymmetric case (so \eqref{furtherreduce} does not hold). In section \ref{conformal perturbation}, we will show how to modify \eqref{furtherreduce}  to perform the computation.

\subsection{The proposed CFT dual}\label{TheproposedCFT}

In this section, we   review the dualities proposed in \cite{Eberhardt:2021vsx}. We describe  the perturbative CFT dual of  the superstring on AdS$_3\times$S$^3\times\mathbb{T}^4$.

The candidate CFT dual is  a deformed symmetric orbifold theory. The theory before deformation is a symmetric orbifold CFT:
\begin{equation}\label{undeformed}
    \text{Sym}^N\left(\mathbb{R}_Q\times \mathfrak{su}(2)_{k-2}\times \text{four free fermions} \times\mathbb{T}^4\right),
\end{equation}
where the fields in the brackets constitute the seed theory of the symmetric orbifold CFT. In the seed theory (see Appendix \ref{seedthoryconvention} for our convention),
 $\mathbb{R}_Q$ is a linear dilaton  with background charge $Q$, related to the level $k$ of the worldsheet $SL(2,\mathbb{R})$ WZW model as:
\begin{equation}\label{Qk}
    Q=b^{-1}-b=\frac{k-1}{\sqrt{k}}, \qquad b=\frac{1}{\sqrt{k}}, 
\end{equation}
and $\mathfrak{su}(2)_{k-2}$ denote a $SU(2)$ WZW model with level $k-2$.
One can calculate the central charge of the seed theory:
\begin{equation}\label{centralcharge}
    c=1+6Q^2+\frac{3(k-2)}{k-2+2}+4\times\frac{1}{2}+6=6k,
\end{equation}
which is the correct one.
In fact, the first three components in the  seed theory of the above symmetric orbifold \eqref{undeformed} constitute a $\mathcal{N}=4$ linear dilaton theory and the last component is the internal ($\mathcal{N}=4$) $\mathbb{T}^4$. Since both of the two theories are $\mathcal{N}=4$ superconformal field theories (SCFTs), the full seed theory is also a $\mathcal{N}=4$ SCFT (then so is the symmetric orbifold theory). 

Denote the generating fields of the $\mathcal{N}=4$ linear dilaton  as:
\begin{equation}
   i\partial\phi, \qquad  j^i , \qquad \psi^a 
\end{equation}
where $i\partial\phi$ is the linear dilaton with background charge $Q=\frac{k-1}{\sqrt{k}}$; $j^i$ $(i=1,2,3)$ generate the affine algebra $\mathfrak{su}(2)_{k-2}$  and $\psi^a$ $(a=0,1,2,3)$ are four Majorana fermions. The  generating fields of $\mathbb{T}^4$ are two complex bosons and two complex fermions:
\begin{equation}
    X^a, X^{a\dagger} \quad (a=1,2),\qquad \lambda^b, \lambda^{b\dagger} \quad (b=1,2).
\end{equation}
Note that any SCFT with small $\mathcal{N}=4$ superconformal symmetry has an automorphism $SU(2)_{\text{R}}\oplus SU(2)_{\text{outer}}$, where $SU(2)_{\text{R}}$ is the R symmetry and $SU(2)_{\text{outer}}$ is the outer automorphism. In the $\mathcal{N}=4$ linear dilaton theory, both the algebras $SU(2)_{\text{R}}$ and $SU(2)_{\text{outer}}$ are constructed by the 4 fermions $\psi^a$  so the 4 fermions $\psi^a$ from a $(\mathbf{2},\mathbf{2})$ of $SU(2)_{\text{R}}\oplus SU(2)_{\text{outer}}$.  Consequently, one can relabel the 4 fermions by 2 spinor indices of $SU(2)_{\text{R}}\oplus SU(2)_{\text{outer}}$:
  $\psi^{\alpha\beta}$, with $\alpha, \beta=\pm$. In the theory $\mathbb{T}^4$, while the algebra  $SU(2)_{\text{R}}$ is constructed by the fermions, the algebra $SU(2)_{\text{outer}}$ is constructed by the bosons.  Thus, the fermions $\lambda^b$, $\lambda^{b\dagger}$ form  $2(\mathbf{2},\mathbf{1})$ of  $SU(2)_{\text{R}}\oplus SU(2)_{\text{outer}}$. As a result, the fermions  could be relabeled as $\lambda^{\alpha0}_{(a)}$, with $\alpha=\pm$ being spinor indices of $SU(2)_{\text{R}}$ and ``$0$'' means that they always form singlets of $SU(2)_{\text{outer}}$.  To distinguish the two  $(\mathbf{2},\mathbf{1})$s, we have also uses a subscript ``$(a)$'' with $(a)=(\pm)$.  All details of these relabelings in the $\mathcal{N}=4$ linear dilaton theory and $\mathbb{T}^4$ can be found in Appendix \ref{ThefullN=4algebra} (where we  have also relabeled the bosons). Besides, for our purpose, we need the form of the supercurrents  labeled by 2 spinor indices of $SU(2)_{\text{R}}\oplus SU(2)_{\text{outer}}$. It seems to us that such  forms are missing (or have typos) in the literature so we also have derived them in Appendix \ref{ThefullN=4algebra}.

To get the CFT dual, one needs to deform the above symmetric orbifold by a (non-normalizable) exactly marginal operator, whose form was proposed in \cite{Eberhardt:2021vsx}.
In any $\mathcal{N}=4$ theory, exactly  marginal operators are obtained as descendants of BPS operators with $h=\Bar{h}=\frac{1}{2}$. 
 It was proposed in \cite{Eberhardt:2021vsx} that the  marginal operator should  be in a singlet
 $(\textbf{1},\textbf{1})$ of $SU(2)_{\text{R}}\oplus SU(2)_{\text{outer}}$. Notice that this should hold  in both the left and right moving parts. 
Thus, one can write the deformation as:
\begin{equation}\label{Themarginal}
    \Phi(x,\bar{x})\equiv G^{\alpha A}_{-\frac{1}{2}}\bar{G}^{\beta B}_{-\frac{1}{2}}\Psi_{\alpha\beta, AB}(x,\bar{x})=\epsilon_{\alpha\gamma}\epsilon_{\beta\delta}\epsilon_{AC}\epsilon_{BD}G^{\alpha A}_{-\frac{1}{2}}\bar{G}^{\beta B}_{-\frac{1}{2}}\Psi^{\gamma\delta, CD}(x,\bar{x}),
\end{equation}
where $G^{\alpha A}$ and $\Bar{G}^{\beta B}$ are holomorphic and anti-holomorphic supercurrents respectively (see Appendix \ref{ThefullN=4algebra} for their explicit forms). For the left (right) moving part, $\alpha,\gamma=\pm $ ($\beta,\delta=\pm$) are the  spinor indices of $SU(2)_{\text{R}}$, while $A,C=\pm $ ($B,D=\pm$) are the  spinor indices of $SU(2)_{\text{outer}}$. $\Psi^{\alpha\beta,AB}$ are non-normalizable BPS operators in the twist 2 sector, obtained by dressing the ground states  by a vertex operator in the linear dilaton theory. For our computation, we will always focus on the left moving part, analysis of the right moving part is similar.  Thus we write down explicitly the left moving  part of the marginal operator, which contains 4 terms (so there are in total 16 terms in the summation of \eqref{Themarginal}):
\begin{equation}\label{The4terms}
    \Phi(x)=G_{-\frac{1}{2}}^{++}\Psi^{--}(x)+G_{-\frac{1}{2}}^{--}\Psi^{++}(x)-G_{-\frac{1}{2}}^{+-}\Psi^{-+}(x)-G_{-\frac{1}{2}}^{-+}\Psi^{+-}(x),
\end{equation}
where the BPS operator $\Psi^{\alpha A}$ $(\alpha, A=\pm)$ has the form\footnote{In \eqref{BPSoper} we followed the notation of \cite{Eberhardt:2021vsx}, which means when lift $ \Psi^{\alpha A}$ up to the covering surface, the lifted operator is $e^{-\sqrt{\frac{k}{2}}\phi}\mathcal{S}^{\alpha A}$, up to the lifted factor. The precise meaning of \eqref{BPSoper} could be written in terms of the fractional modes:
 \begin{equation}\label{TheformofPsi}
      \Psi^{\alpha A}=2^{\frac{h_1+h_2}{2}-1}\left(e^{-\sqrt{\frac{k}{2}}\phi}\right)_{-\frac{h_1}{2}}(\mathcal{S}^{\alpha A})_{-\frac{h_2}{2}}\sigma_{2},
 \end{equation}
 where $h_1=\frac{1}{2}-\frac{3k}{4}$, $h_2=\frac{1}{2}$ are the conformal weights of $e^{-\sqrt{\frac{k}{2}}\phi}$ and $\mathcal{S}^{\alpha A}$ respectively. We add a factor $2^{\frac{h_1+h_2}{2}-1}$ to  cancel the factor $w^{1-h}$ in \eqref{liftfractional} so that the factor coming from the lifting is simply $a^{-s}$. \label{fn:footnote12}}:
\begin{equation}\label{BPSoper}
    \Psi^{\alpha A}=e^{-\sqrt{\frac{k}{2}}\phi}\mathcal{S}^{\alpha A}\sigma_{2}
\end{equation}
 where $\mathcal{S}^{\alpha A}$  are some spin fields and $\sigma_{2}$ is the twist field in the twist 2 sector. To write the spin fields  explicitly, we bosonize the fermions as (see Appendix \ref{ThefullN=4algebra} for our convention for the fermions):
 \begin{equation}
    \begin{aligned}
        \psi^{++}&=e^{i\hat{H}_1},\quad \psi^{--}=e^{-i\hat{H}_1}, \quad \psi^{+-}=e^{i\hat{H}_2}, \quad \psi^{-+}=-e^{-i\hat{H}_2}\\
         \lambda^{+0}_{(+)}&=e^{i\hat{H}_3},\quad \lambda^{-0}_{(-)}=e^{-i\hat{H}_3}, \quad \lambda^{+0}_{(-)}=e^{i\hat{H}_4}, \quad \lambda^{-0}_{(+)}=-e^{-i\hat{H}_4}\ ,
    \end{aligned}
\end{equation}
where we use a hat to denote the bosons with cocycles:
\begin{equation}
    \hat{H}_i=H_i+\pi\sum_{j<i}N_i, \qquad N_i=i\oint\partial H_i, \qquad H_i(z)H_j(w)\sim -\delta_{ij}\text{log}(z-w).
\end{equation}
Then there are in total $2^4=16$ spin fields and the spin fields $\mathcal{S}^{\alpha A}$ that transform as a $(\textbf{2},\textbf{2})$ of $SU(2)_{\text{R}}\oplus SU(2)_{\text{outer}}$ has the following forms:
 \begin{equation}
     \mathcal{S}^{++}\equiv S^{+-++}, \quad \mathcal{S}^{+-}\equiv iS^{+---}, \quad \mathcal{S}^{-+}\equiv iS^{-+++},\quad \mathcal{S}^{--}\equiv -S^{-+--},
 \end{equation}
 where $S^{\epsilon_1\epsilon_2\epsilon_3\epsilon_4}\equiv e^{\frac{i}{2}(\epsilon_1\hat{H}_1+\epsilon_2\hat{H}_2+\epsilon_3\hat{H}_3+\epsilon_4\hat{H}_4)}$. Note that the factors ``$i$'' and ``$-$'' in the above spin fields come from the cocycles.
 It is easy to see that the superscripts of $\mathcal{S}^{\alpha A}$ and $S^{\epsilon_1\epsilon_2\epsilon_3\epsilon_4}$ are related as: $\alpha=\frac{1}{2}(\epsilon_1+\epsilon_2+\epsilon_3+\epsilon_4)$, $A=\frac{1}{2}(\epsilon_1-\epsilon_2)$, where  $\epsilon_3,\epsilon_4$ does not contribute to $A$ since  in $\mathbb{T}^4$ the $SU(2)_{\text{outer}}$ is constructed by the bosons (see Appendix \ref{ThefullN=4algebra}).   Notice that the dressing in the linear dilaton direction is  a vertex operator
 \begin{equation}\label{alphaPhi}
     e^{-\sqrt{\frac{k}{2}}\phi}=e^{\sqrt{2}\alpha_\Phi\phi}, \qquad \text{with} \quad\alpha_\Phi=-\frac{\sqrt{k}}{2}=-\frac{1}{2b},
 \end{equation}
 which is the same as in the bosonic case \cite{Eberhardt:2021vsx}, with the shift of levels $k\to k+2$. 
 Thus the marginal operator is  non-normalizable and creates an exponential wall which is in some sense similar to the exponential operator in Liouville theory. Finally,
 the weight $h$ and R-charge $q=\alpha$ of $\Psi^{\alpha A}$ satisfy:
 \begin{equation}
     h=\frac{c}{24}\left(2-\frac{1}{2}\right)+\frac{1}{2}\times \frac{1}{2}+\frac{\alpha(Q-\alpha)}{2}=\frac{3k}{8}+\frac{1}{4}-\frac{\sqrt{k}}{4}\left(\frac{k-1}{\sqrt{k}}+\frac{\sqrt{k}}{2}\right)=\frac{1}{2}=|q|.
 \end{equation}
Thus, $\Psi^{\alpha A}$ are indeed BPS.

Finally, as we have mentioned in  footnote \ref{fn:footnote2}, there are two subtleties for the dual theory: the difference between the case $k>1$ and $k<1$ \cite{Balthazar:2021xeh,Chakraborty:2025nlb} (see also \cite{Giveon:2005mi}) and the fact that the dual theory should be understood as a grand canonical ensemble of CFTs  \cite{Eberhardt:2021vsx,Aharony:2024fid,Knighton:2024pqh} (see also \cite{Kutasov:1999xu,Giveon:2001up,Kim:2015gak,Eberhardt:2021jvj,Eberhardt:2020bgq}). Both are important for properly understanding the duality. However, for our computation of matching the correlators in the large N limit, these subtleties do not play a role (just like in the bosonic case \cite{Eberhardt:2021vsx}).

\subsection{Conformal perturbation computation}\label{conformal perturbation}
In this section, we do the conformal perturbation on the CFT side. For the correlator 
$\mathbb{M}_{VVV}$, the leading calculation is done in our previous work \cite{Yu:2024kxr}, here we focus on the case $m=2$ in \eqref{Thematching}. 

Firstly, the vertex operator \eqref{stringoperator} on the string side map to  an operator 
$ \mathbb{V}_{\alpha}^{(w)}$ in the twist $w$ sector of the symmetric orbifold theory \cite{Yu:2024kxr,Eberhardt:2019qcl}\footnote{Here we also use the notation used in \cite{Eberhardt:2021vsx}, which means the lifted operator is  $e^{\sqrt{2}\alpha\phi}$, up to the factor coming from the lifting. The precise meaning of \eqref{theoperator} is 
\begin{equation}
    \mathbb{V}_{\alpha}^{(w)}=w^{\frac{h_\alpha}{w}-1}\left(e^{\sqrt{2}\alpha\phi}\right)_{-\frac{h_{\alpha}}{w}}\sigma_w
\end{equation}
where $h_{\alpha}=\alpha(Q-\alpha)$ is the weight of the lifted operator $e^{\sqrt{2}\alpha\phi}$. Again we add a factor $w^{\frac{h_\alpha}{w}-1}$ in the definition (see footnote \ref{fn:footnote12}).
}:
\begin{equation}\label{theoperator}
    \mathbb{V}_{\alpha}^{(w)}\equiv e^{\sqrt{2}\alpha\phi}\sigma_w\,,
\end{equation}
where the momentum $\alpha$ in the linear dilaton is related to on the string side the $SL(2, \mathbb{R})$ spin $j$ (and the level $k$) as: 
\begin{equation}\label{jalpha}
    \alpha=\frac{j+\frac{k}{2}-1}{\sqrt{k}}.
\end{equation}
When lifted up to the covering surface, as we have discussed in  \eqref{liftfractional}, the lifted operator contains an extra factor 
\begin{equation}
    \tilde{\mathbb{V}}_{\alpha}=a^{-\frac{h_\alpha}{w}}e^{\sqrt{2}\alpha\phi}.
\end{equation}
Note that here the ``...'' terms in last line of \eqref{liftfractional} do not appear since the lifted operator $e^{\sqrt{2}\alpha\phi}$ is a Virasoro primary on the covering surface.

The spectrum of this undeformed theory was matched with the spectrum of long strings in the superstring theory on AdS$_3\times$S$^3\times\mathbb{T}^4$ \cite{Eberhardt:2019qcl}. Now the right hand side of  \eqref{Thematching} could be written by the conformal perturbation as:
\begin{equation}\label{conformalper}
\begin{aligned}
    \left(\mathop{\text{Res}}\limits_{2b(\sum_i \alpha_i-Q)=m}\mathbb{M}_{VVV}\right)\prod_{i=1}^3N_{V_i}^{-1}&\equiv
   \mathop{\text{Res}}\limits_{2b(\sum_i \alpha_i-Q)=m}\left\langle\prod_{i=1}^3\mathbb{V}_{\alpha_i}^{(w_i)}(x_i)\right\rangle_{\mu}\prod_{i=1}^3N_{V_i}^{-1}\\
   =\frac{(-\mu)^m}{\pi m!}&\int \prod_{l=1}^md^2\xi_l\left\langle\prod_{i=1}^3\mathbb{V}_{\alpha_i}^{(w_i)}(x_i)\prod_{l=1}^m\Phi(\xi_l)\right\rangle_0N_{\Phi}^{-m}\prod_{i=1}^3N_{V_i}^{-1}
\end{aligned}
\end{equation}
where the subscripts ``$\mu$'' and ``$0$'' means the correlator is calculated in the theories with the deformation parameter $\mu$ and $0$ (that is, undeformed) respectively. 
The integrand in \eqref{conformalper} is a $(3+m)$-point correlator in the symmetric orbifold CFT.  We focus on the  case $m=2$ so need to calculate a five point function. Usually, to obtain the results of conformal perturbation one needs to work out explicitly the integral of  $\xi_l$ on the right hand side of \eqref{conformalper} (see \cite{Gaberdiel:2015uca,Apolo:2022fya,Gaberdiel:2023lco} for examples of doing conformal perturbation of symmetric orbifold CFTs). It will be extremely difficult to do this in our case because the integrand is a five point function of five operators in the twisted sector. Fortunately, as we will shown in the following, one can match the string result with the CFT side by showing the integrands of the two sides coincide (so we do not need to work out the integral of $\xi_l$ ).

\subsubsection{Conformal perturbation for $m=2$}

In our previous work \cite{Yu:2024kxr}, the  calculation of the leading (0-th) order   conformal perturbation  was performed for the correlator 
$\mathbb{M}_{VVV}$. The calculation is simple since at the leading order, no marginal operator is inserted.  At higher orders, as explained above,  we do not have to work out the integral of $\xi_l$ but only need to calculate the integrand, which is a correlator in the symmetric orbifold CFT.  In the bosonic case, this correlator is simple when lifted up to the covering surface because all operators in the correlator are vertex operators of the form $e^{\sqrt{2}\alpha\phi}$ in a linear dilaton theory. 

In the supersymmetric case, the calculation of the integrand is technically more difficult  that that in the bosonic case. Recall that the deforming operator  is no longer a simple exponential field (when lifted up to the covering surface), but is a linear combination of super-descendents of some BPS operators in the twist 2 sector. Then unlike in the bosonic case, the marginal operator  is $not$ a Virasoro primary when lifted up to the covering surface, though it is a Virasoro primary in the symmetric orbifold theory (this will be shown explicitly shortly). As a consequence, when the correlator is lifted up to the covering surface, its covering map dependent part is not simply encoded in the function $f_\Gamma$ and $g_\Gamma$ in the formula \eqref{sumofcovering}, \eqref{furtherreduce}. In the following, we will see how to modify the factor $g_\Gamma$ in the calculation.

Consider the order $m=2$ in \eqref{conformalper}. The integrand in the right hand side is a  five point function in the symmetric orbifold CFT:
\begin{equation}\label{The5pointfunction}
   \mathcal{I}= \left\langle \mathbb{V}_{\alpha_1}^{(w_1)}(0,0)\mathbb{V}_{\alpha_2}^{(w_2)}(1,1)\mathbb{V}_{\alpha_3}^{(w_3)}(\infty,\infty)  \Phi(\xi_1,\bar{\xi}_1) \Phi(\xi_2,\bar{\xi}_2)\right\rangle
\end{equation}
where  we have used the $SL(2, \mathbb{R})$ symmetry to fix the coordinates of the 3 operators $\mathbb{V}_{\alpha_i}^{(w_i)}$ to be $(x_1,x_2,x_3)=(0,1,\infty)$ (the anti-holomorphic coordinates are similarly fixed).  
Now consider a covering map $\Gamma: z\to x$, which maps:
\begin{equation}
    0\to 0, \quad 1\to 1, \quad \infty\to \infty, \quad \eta_1\to \xi_1, \quad \eta_2 \to \xi_2
\end{equation}
and near these insertion points we have:
\begin{equation}\label{coveringcondi}
\begin{aligned}
   \Gamma(z)&=0+a_1(z-0)^{w_1}+...,\qquad &\text{near $z=0$}\\
   \Gamma(z)&=1+a_2(z-1)^{w_2}+..., \qquad &\text{near $z=1$}\\
   \Gamma(z)&=(-1)^{w_3+1}a_3^{-1}z^{w_3}+..., \qquad &\text{near $z=\infty$}\\
    \Gamma(z)&=\xi_1+A_1(z-\eta_1)^{2}+B_1(z-\eta_1)^3+..., \qquad &\text{near $z=\eta_1$}\\
     \Gamma(z)&=\xi_2+A_2(z-\eta_2)^{2}+B_2(z-\eta_2)^3+..., \qquad &\text{near $z=\eta_2$}\\
\end{aligned}
\end{equation}
Notice that $a_3$ is defined a bit differently, which in fact will lead to more symmetrical formula in our calculation.  Near $z=\eta_i$ $(i=1,2)$, we expand the covering map up to the third order, and the associated coefficients are denoted as $B_i$ $(i=1,2)$.  Once $\xi_1, \xi_2$ are given,  all covering maps satisfying \eqref{coveringcondi} are fixed and thus $a_1,a_2,a_3,A_1,A_2,B_1,B_2,\eta_1,\eta_2$ can be  implicitly determined as functions of $\xi_1, \xi_2$. To calculate the correlator $\mathcal{I}$, we can lift it up to covering surfaces satisfying \eqref{coveringcondi}. We will always focus on the leading order in the large N expansion, where the covering surface is a sphere. From \eqref{sumofcovering}, we have:
\begin{equation}\label{liftonce}
\mathcal{I}=\sum_\Gamma \left|\mathcal{I}_\Gamma\right|^2, \qquad
    |\mathcal{I}_{\Gamma}|^2=\left|f_\Gamma\left\langle \tilde{\mathbb{V}}_{\alpha_1}(0)\tilde{\mathbb{V}}_{\alpha_2}(1)\tilde{\mathbb{V}}_{\alpha_3}(\infty)  \tilde{\Phi}(\eta_1) \tilde{\Phi}(\eta_2)\right\rangle\right|^2
\end{equation}
where $\mathcal{I}_\Gamma$ is the contribution of a specific covering surface (map) $\Gamma$ to the correlator. The function $f_\Gamma$ is the anomalous term \eqref{fGamma} determined by the covering map and the operators with tilde are the ones on the covering surface. As we discussed in section \ref{coveringmaplifted}, if the 5 lifted operators in \eqref{liftonce} are all Virasoro primaries, then we could further collect all the factors coming from the lifting into a function $g_\Gamma$ in \eqref{furtherreduce}. This is in fact true in the bosonic case \cite{Eberhardt:2021vsx}. However, in the SUSY case, while the three lifted operators $\tilde{\mathbb{V}}_{\alpha_i}$ are primaries, the 2 lifted marginal operators $ \tilde{\Phi}(\eta_i)$ $(i=1,2)$ are not. 
Thus,  we need to see in detail the explicit form of  $ \tilde{\Phi}(\eta_i)$, which is
\begin{equation}
\begin{aligned}
     \tilde{\Phi}(\eta_i)=&\epsilon_{\alpha\gamma}\epsilon_{AC}G^{\alpha A}_{-\frac{1}{2}}A_i^{-(\frac{1}{2}-\frac{3k}{8})}\hat{\Psi}^{\gamma C}(\eta_i)\\
     =&\epsilon_{\alpha\gamma}\epsilon_{AC}\oint_{C(\eta_i)}dz_i(\partial \Gamma(z_i))^{-\frac{1}{2}}(\Gamma(z_i)-\xi_i)^{0}\hat{G}^{\alpha A}(z_i)A_i^{-(\frac{1}{2}-\frac{3k}{8})}\hat{\Psi}^{\gamma C}(\eta_i)\\
     =&\epsilon_{\alpha\gamma}\epsilon_{AC}\oint_{C(\eta_i)}dz_i(2A_i(z_i-\eta_i)+3B_i(z_i-\eta_i)^2+...)^{-\frac{1}{2}}\hat{G}^{\alpha A}(z_i)A_i^{-(\frac{1}{2}-\frac{3k}{8})}\hat{\Psi}^{\gamma C}(\eta_i)\\
     =&\epsilon_{\alpha\gamma}\epsilon_{AC}\oint_{C(\eta_i)}dz_i\left(1-\frac{3B_i}{4A_i}(z_i-\eta_i)+...\right)2^{-\frac{1}{2}}A_i^{-(1-\frac{3k}{8})}(z_i-\eta_i)^{-\frac{1}{2}}\hat{G}^{\alpha A}(z_i)\hat{\Psi}^{\gamma C}(\eta_i)
\end{aligned}
\end{equation}
where the factor $A_i^{-(\frac{1}{2}-\frac{3k}{8})}$ in the first line comes from lifting the BPS operator $\Psi^{\gamma C}$, see \eqref{TheformofPsi} (note that the lifted BPS operator $\hat{\Psi}^{\gamma C}$ is a primary on the covering surface). In the last line, the factor $2^{-\frac{1}{2}}=2^{1-\frac{3}{2}}$ is in fact the factor $w^{1-h}$ in the last line of \eqref{liftfractional}, which is of no significance and we always cancel such a factor by rescaling the operator $\tilde{\Phi}\to 2^{\frac{1}{2}}\tilde{\Phi}$.
Since the OPE of the two operators are:
\begin{equation}
    \hat{G}^{\alpha A}(z_i)\hat{\Psi}^{\gamma C}(\eta_i)\sim \frac{\hat{G}^{\alpha A}_0\hat{\Psi}^{\gamma C}(\eta_i)}{(z_i-\eta_i)^{\frac{3}{2}}}+\frac{\hat{G}^{\alpha A}_{-1}\hat{\Psi}^{\gamma C}(\eta_i)}{(z_i-\eta_i)^{\frac{1}{2}}}+...
\end{equation}
One sees that there will be two terms contribute to $\tilde{\Phi}_i(\eta_i)$: 
\begin{equation}\label{2termsmarginal}
    \tilde{\Phi}_i(\eta_i)=\tilde{\Phi}^{(1)}(\eta_i)+\tilde{\Phi}^{(2)}(\eta_i)
\end{equation}
where
\begin{equation}\label{firstmarginal}
\tilde{\Phi}^{(1)}(\eta_i)=
\epsilon_{\alpha\gamma}\epsilon_{AC}A_i^{-(1-\frac{3k}{8})}\hat{G}^{\alpha A}_{-1}\hat{\Psi}^{\gamma C}(\eta_i)
\end{equation}
\begin{equation}\label{secondmarginal}
      \tilde{\Phi}^{(2)}(\eta_i)=\epsilon_{\alpha\gamma}\epsilon_{AC}\left(-\frac{3B_i}{4A_i}\right)A_i^{-(1-\frac{3k}{8})}\hat{G}^{\alpha A}_0\hat{\Psi}^{\gamma C}(\eta_i).
\end{equation}
Now we compare this lifted operator $\tilde{\Phi}_i(\eta_i)$ to the general form \eqref{liftfractional}. 
\begin{itemize}
    \item It is easy to see that the factor $A_i^{-(1-\frac{3k}{8})}$ is just the prefactor $a^{-s}$ in the last line of \eqref{liftfractional}. Note that in our convention, we have canceled the factor $w^{1-h}$ in \eqref{liftfractional} when defining the action of the fractional modes.
    \item  The operator $\tilde{\Phi}^{(1)}(\eta_i)$ is just the lifted operators that we write explicitly in the last line of \eqref{liftfractional}; the operator $\tilde{\Phi}^{(2)}(\eta_i)$ is one  ``...'' term in \eqref{liftfractional}. Thus, it is obvious that the (full) lifted marginal operator $\tilde{\Phi}_i$ is not a Virasoro primary. In fact,  $L_1\tilde{\Phi}^{(1)}\propto \tilde{\Phi}^{(2)}$ so $\tilde{\Phi}^{(1)}$ is not a (quasi-)primary (while $\tilde{\Phi}^{(2)}$ is indeed a primary). 
\end{itemize}
 Since operators like $\tilde{\Phi}^{(1)}(\eta_i)$ always appear regardless of whether the lifted operator is a primary or not. We view such a term as ``regular''. On the other hand, the operator  $\tilde{\Phi}^{(2)}$ reflects the fact that the lifted marginal operators are not a primary, so we view such a term as  ``irregular''. We will use the concepts of regular/irregular to organize the finial results of the five point functions in section \ref{Allterms}.

With these results for the marginal operator $ \tilde{\Phi}(\eta_i)$, 
we can write the correlator as:
\begin{equation}\label{IGamma}
\begin{aligned}
       &\mathcal{I}_{\Gamma}=f_\Gamma\left\langle \tilde{\mathbb{V}}_{\alpha_1}(0)\tilde{\mathbb{V}}_{\alpha_2}(1)\tilde{\mathbb{V}}_{\alpha_3}(\infty)  \left(\tilde{\Phi}_1^{(1)}(\eta_1)+\tilde{\Phi}_1^{(2)}(\eta_1)\right)\left(\tilde{\Phi}_2^{(1)}(\eta_2)+\tilde{\Phi}_2^{(2)}(\eta_2)\right)\right\rangle\\
    &=\epsilon_{\alpha\gamma}\epsilon_{AC}\epsilon_{\beta\delta}\epsilon_{BD}f_\Gamma g_{\Gamma}\\
     &\!\Big\langle \hat{\mathbb{V}}_{\alpha_1}(0)\hat{\mathbb{V}}_{\alpha_2}(1)\hat{\mathbb{V}}_{\alpha_3}(\infty)  \!\left(\hat{G}^{\alpha A}_{-1}\hat{\Psi}^{\gamma C}(\eta_1)\!+\!C_1\hat{G}^{\alpha A}_0\hat{\Psi}^{\gamma C}(\eta_1)\!\right)
       \!\left(\hat{G}^{\beta B}_{-1}\hat{\Psi}^{\delta D}(\eta_2)\!+\!C_2\hat{G}^{\beta B}_0\hat{\Psi}^{\delta D}(\eta_2)\!\right)\!\Big\rangle\\
       &=f_\Gamma g_{\Gamma}\left(\mathcal{I}^{(\hat{G}_{-1},\hat{G}_{-1})}_{\Gamma}+C_1\mathcal{I}^{(\hat{G}_{0},\hat{G}_{-1})}_{\Gamma}++C_2\mathcal{I}^{(\hat{G}_{-1},\hat{G}_{0})}_{\Gamma}+C_1C_2\mathcal{I}^{(\hat{G}_{0},\hat{G}_{0})}_{\Gamma}\right)
\end{aligned}
\end{equation}
where $f_\Gamma$ and $g_\Gamma$ are defined as in \eqref{fGamma} and \eqref{gGamma}:
\begin{equation}\label{fgforms}
\begin{aligned}
    f_\Gamma&=\prod_{i=1}^3 w_i^{-\frac{c(w_i+1)}{24}}a_i^{-\frac{c}{24}\left(1-\frac{1}{w_i}\right)}\prod_{j=1}^2A_j^{-\frac{c}{24}\left(1-\frac{1}{2}\right)}2^{-\frac{c}{4}}\Pi^{-\frac{c}{12}}\\
    g_{\Gamma}&=\prod_{i=1}^3a_i^{-H_i+\frac{c}{24}\left(w_i-\frac{1}{w_i}\right)}\prod_{j=1}^2A_j^{-1+\frac{c}{24}\left(2-\frac{1}{2}\right)},
\end{aligned}
\end{equation}
with $c=6k$. $H_i$ are the weights of the operators $\mathbb{V}^{(w_i)}_{\alpha_i}$, which coincide with the space-time weights of their counterparts  $O^{w_i}_{j_i,h_i}$ on the string side (see \eqref{Thespacetimeweight}). The covering map dependent factors $C_i$ $(i=1,2)$ are defined as: 
\begin{equation}
    C_i\equiv -\frac{3B_i}{4A_i}.
\end{equation}
The 4 terms in \eqref{IGamma} are:
\begin{equation}\label{5pointfunctions}
   \mathcal{I}^{(\hat{G}_{a},\hat{G}_{b})}_{\Gamma}\equiv\epsilon_{\alpha\gamma}\epsilon_{AC}\epsilon_{\beta\delta}\epsilon_{BD}\left\langle \hat{\mathbb{V}}_{\alpha_1}(0)\hat{\mathbb{V}}_{\alpha_2}(1)\hat{\mathbb{V}}_{\alpha_3}(\infty)  \hat{G}^{\alpha A}_{a}\hat{\Psi}^{\gamma C}(\eta_1)
      \hat{G}^{\beta B}_{b}\hat{\Psi}^{\delta D}(\eta_2)\right\rangle. 
\end{equation}

\subsubsection{The computation of all terms}\label{Allterms}
Now we calculate the correlator $\mathcal{I}_\Gamma$. In \eqref{IGamma}, we need to calculate several 5-point correlation functions, where the first 3 operators are simply exponential operators $e^{\sqrt{2}\alpha_i\phi} (i=1,2,3)$ and the remaining 2 (the marginal operators) are of the form   $\hat{G}^{\alpha A}_{-1}\hat{\Psi}^{\gamma C}(\eta_i)$ or
$\hat{G}^{\alpha A}_0\hat{\Psi}^{\gamma C}(\eta_i)$\footnote{In the bosonic case, one needs only calculate one 5-point correlation function where the first three operators have the same form as here and the remaining 2 are also exponential operators $e^{-\sqrt{\frac{k-2}{2}}\phi}$. Thus, one can directly write down the result \cite{Eberhardt:2021vsx}. }. Since the bosonic part of the BPS operator $\hat{\Psi}^{\gamma C}$ is also an  exponential operator $e^{-\sqrt{\frac{k}{2}}\phi}$, We can write $\mathcal{I}_\Gamma$ as:
\begin{equation}\label{remainingfactor}
    \mathcal{I}_\Gamma=f_\Gamma g_{\Gamma}\mathfrak{I}_\Gamma\mathcal{I}_\Gamma^B
\end{equation}
where $\mathcal{I}^B_\Gamma$ is the correlator of the 5 exponential operators. It is in fact identical with the one in the bosonic case (with $k\to k+2$) and has the following form:
\begin{equation}
    \mathcal{I}_\Gamma^B=\prod_{1\leq i<j\leq 3}(z_i-z_j)^{-2\alpha_i\alpha_j}\prod_{i=1}^3\prod_{j=1}^2(z_i-\eta_j)^{\frac{\alpha_i}{b}}(\eta_1-\eta_2)^{-\frac{1}{2b^2}}.
\end{equation}
Since the first three coordinates are fixed  as $(z_1,z_2,z_3)=(0,1,\infty)$, we have\footnote{Since we are ultimately interested in the product of the holomorphic and anti-holomorphic contribution, we have omitted the possible phase in the following expression.}:
\begin{equation}\label{bosonic5point}
    \mathcal{I}_\Gamma^B=\eta_1^{j_1+\frac{k}{2}-1}(\eta_1-1)^{j_2+\frac{k}{2}-1}\eta_2^{j_1+\frac{k}{2}-1}(\eta_2-1)^{j_2+\frac{k}{2}-1}(\eta_1-\eta_2)^{-\frac{k}{2}}
\end{equation}
where we have used \eqref{jalpha}. Note that when viewed as a function of $(\eta_1,\eta_2)$, $\mathcal{I}_\Gamma^B$ in fact does not have any explicit dependence on $\Gamma$. Nevertheless, we add a subscript ``$\Gamma$'' to $\mathcal{I}_\Gamma^B$  to stress that we always need to choose a covering map $\Gamma$ to do the computation (besides, if we view $\mathcal{I}_\Gamma^B$ as a function of $(\xi_1,\xi_2)$, it indeed depends on $\Gamma$).  In the rest of this section, we need to work out the remaining factor $\mathfrak{I}_\Gamma$ in \eqref{remainingfactor}.

 For the computation, we need the explicit form of the supercurrents and the BPS operators. In the following, we write $\hat{G}^{++}_0\hat{\Psi}^{--}$ and $\hat{G}^{++}_{-1}\hat{\Psi}^{--}$ as illustrations. Other terms of the form $\hat{G}^{\alpha A}_{-1}\hat{\Psi}^{\gamma C}$ or
$\hat{G}^{\alpha A}_0\hat{\Psi}^{\gamma C}$ can be written in a similar way. The BPS operator (lifted up to the covering surface) $\hat{\Psi}^{--}$ has the bosonized form:
\begin{equation}
         \hat{\Psi}^{--}=e^{-\sqrt{\frac{k}{2}}\phi}S^{-+--}=e^{-\sqrt{\frac{k}{2}}\phi}e^{\frac{i}{2}(-\hat{H}_1+\hat{H}_2-\hat{H}_3-\hat{H}_4)},
\end{equation}
and the supercurrents  $\hat{G}^{++}$ are (see  Appendix \ref{ThefullN=4algebra}):
\begin{equation}
\begin{aligned}
       \hat{G}^{++}=&\hat{G}^{++}_{\text{LD}}+\hat{G}^{++}_{\mathbb{T}^4}, \\
       \hat{G}^{++}_{\text{LD}}=&i\partial\phi\psi^{++}+i\sqrt{\frac{2}{k}}\left(j^+\psi^{-+}+(j^3-\psi^{+-}\psi^{-+})\psi^{++}-(k-1)\partial\psi^{++}\right),\\
       \hat{G}^{++}_{\mathbb{T}^4}=&\sqrt{2}\lambda^{+0}_{(+)}\partial X^{0+}_{(-)}+\sqrt{2}\lambda^{+0}_{(-)}\partial X^{0+}_{(+)}
\end{aligned}
\end{equation}
where we have used the subscripts ``$LD$'' and ``$\mathbb{T}^4$'' to denote the supercurrents in the $\mathcal{N}=4$ linear dilaton theory and $\mathbb{T}^4$ respectively.
   Then for  $\hat{G}^{++}_{0}\hat{\Psi}^{--}$, it is easy to see that $(\hat{G}^{++}_{\mathbb{T}^4})_{0}\hat{\Psi}^{--}=0$. Thus, $\hat{G}^{++}_{0}\hat{\Psi}^{--}=(\hat{G}^{++}_{\text{LD}})_{0}\hat{\Psi}^{--}$ and its explicit form is:
    \begin{equation}\label{G0action}
    \begin{aligned}
        \hat{G}^{++}_{0}\hat{\Psi}^{--}&=\left((i\partial\phi)_0\psi^{++}_0+i\sqrt{\frac{2}{k}}\left(i\partial H_2+\frac{k-1}{2}\right)\psi^{++}_0\right) e^{-\sqrt{\frac{k}{2}}\phi}e^{\frac{i}{2}(-\hat{H}_1+\hat{H}_2-\hat{H}_3-\hat{H}_4)}\\
      &=\left(i\sqrt{\frac{k}{2}}+i\sqrt{\frac{2}{k}}\left(\frac{1}{2}+\frac{k-1}{2}\right)\right)\psi^{++}_0 e^{-\sqrt{\frac{k}{2}}\phi}e^{\frac{i}{2}(-\hat{H}_1+\hat{H}_2-\hat{H}_3-\hat{H}_4)}\\
      &=i\sqrt{2k} e^{-\sqrt{\frac{k}{2}}\phi}e^{\frac{i}{2}(\hat{H}_1+\hat{H}_2-\hat{H}_3-\hat{H}_4)},
    \end{aligned}
    \end{equation}
    where we have used $\partial\psi^{++}_0=-\frac{1}{2}\psi^{++}_0 $ and the fact that $j^a_0$ $(a=3,\pm)$ annihilate the (lifted) BPS operator  $\hat{\Psi}^{--}$. 
    Note that  $\hat{G}^{++}_{0}\hat{\Psi}^{--}$ is proportional to  $\sqrt{k}$. 
    The form of  $\hat{G}^{++}_{-1}\hat{\Psi}^{--}$ is more complicated. Its explicit form is:
    \begin{equation}\label{G-1action}
        \hat{G}^{++}_{-1}\hat{\Psi}^{--}=(\hat{G}^{++}_{\mathbb{T}^4})_{-1}\hat{\Psi}^{--}+(\hat{G}^{++}_{\text{LD}})_{-1}\hat{\Psi}^{--},
    \end{equation}
    where
       \begin{equation}\label{T4LD}
    \begin{aligned}
     &(\hat{G}^{++}_{\mathbb{T}^4})_{-1}\hat{\Psi}^{--}= \left(\sqrt{2}(\lambda^{+0}_{+})_0(\partial X^{0+}_-)_{-1}
        +\sqrt{2}(\lambda^{+0}_-)_0(\partial X^{0+}_+)_{-1}\right)e^{-\sqrt{\frac{k}{2}}\phi}e^{\frac{i}{2}(-\hat{H}_1+\hat{H}_2-\hat{H}_3-\hat{H}_4)}\\
         &(\hat{G}^{++}_{\text{LD}})_{-1}\hat{\Psi}^{--}
        =\Bigg((i\partial\phi)_0(\psi^{++})_{-1}+(i\partial\phi)_{-1}(\psi^{++})_{0}+i\sqrt{\frac{2}{k}}\Big((j^3-\psi^{+-}\psi^{-+})_{-1}(\psi^{++})_0\\
        &+(j^+)_{-1}(\psi^{-+})_0
        -(\psi^{+-}\psi^{-+})_0(\psi^{++})_{-1}-(k-1)(\partial\psi^{++})_{-1}\Big)\Bigg)e^{-\sqrt{\frac{k}{2}}\phi}e^{\frac{i}{2}(-\hat{H}_1+\hat{H}_2-\hat{H}_3-\hat{H}_4)}
    \end{aligned}
    \end{equation}
    Note that, different from $\hat{G}^{++}_{0}\hat{\Psi}^{--}$, the $k$ dependence of $\hat{G}^{++}_{-1}\hat{\Psi}^{--}$ is complicated: it contains terms proportional to $\sqrt{k}$, $1$, and $1/\sqrt{k}$.

Now we analyze the structure of the five-point functions in \eqref{5pointfunctions}. Since the first 3 operators $\hat{\mathbb{V}}_{\alpha_i}$ $(i=1,2,3)$ are simply vertex operators in the bosonic linear dilaton theory, those terms in \eqref{T4LD} that do not contain $i\partial\phi$ will not change the functional form of the bosonic five-point function $\mathcal{I}_\Gamma^B$. In \eqref{T4LD}, there are   two terms that contain $i\partial\phi$.  For the term $(i\partial\phi)_0(\psi^{++})_{-1}$, the zero mode $(i\partial\phi)_0$ acts on the  bosonic vertex operator $e^{-\sqrt{\frac{k}{2}}\phi}$ by extracting its charge:
\begin{equation}
    (i\partial\phi)_0e^{-\sqrt{\frac{k}{2}}\phi}=i\sqrt{\frac{k}{2}}e^{-\sqrt{\frac{k}{2}}\phi}.
\end{equation}
Thus, this term will  not  change the functional form of $\mathcal{I}_\Gamma^B$ as well. In contrast, 
    the other term $(i\partial\phi)_{-1}(\psi^{++})_{0}$ is different since $(i\partial\phi)_{-1}$ acts on the bosonic vertex operator $e^{-\sqrt{\frac{k}{2}}\phi}$  by taking a derivative:
\begin{equation}\label{takeder}
    (i\partial\phi)_{-1}e^{-\sqrt{\frac{k}{2}}\phi}(\eta_i)=-i\sqrt{\frac{2}{k}}\partial_{\eta_i}\left(e^{-\sqrt{\frac{k}{2}}\phi}(\eta_i)\right).
\end{equation}
For the computation of the five point functions in \eqref{5pointfunctions}, this action is  achieved by taking a derivative of the bosonic five point function $\mathcal{I}_\Gamma^B$ with respect to $\eta_i$ (so it indeed changes the functional form of $\mathcal{I}_\Gamma^B$). Thus,   in the following computation, we divide $\hat{G}^{\alpha A}_{-1}\Psi^{\gamma C}$  into 2 parts:
\begin{equation}
    \hat{G}^{\alpha A}_{-1}\Psi^{\gamma C}=\hat{G}_{-1}^{(1)\alpha A}\Psi^{\gamma C}+\hat{G}_{-1}^{(2)\alpha A}\Psi^{\gamma C},
\end{equation}
where
\begin{equation}
    \hat{G}_{-1}^{(1)\alpha A}\Psi^{\gamma C}\equiv(i\partial\phi)_{-1}(\psi^{\alpha A})_0\Psi^{\gamma C}, \qquad \hat{G}_{-1}^{(2)\alpha A}\Psi^{\gamma C}\equiv \hat{G}^{\alpha A}_{-1}\Psi^{\gamma C}-(i\partial\phi)_{-1}(\psi^{\alpha A})_0\Psi^{\gamma C}.
\end{equation}
Accordingly, in \eqref{IGamma}  we can further divide $\mathcal{I}^{(\hat{G}_{m},\hat{G}_{n})}_{\Gamma}$ $(m,n=0,-1)$ into the following summations:
\begin{equation}
    \mathcal{I}^{(\hat{G}_{-1},\hat{G}_{-1})}_{\Gamma}=\sum_{a,b=1,2}\mathcal{I}^{(\hat{G}^{(a)}_{-1},\hat{G}_{-1}^{(b)})}_{\Gamma}, \quad \mathcal{I}^{(\hat{G}_{-1},\hat{G}_{0})}_{\Gamma}=\sum_{a=1,2}\mathcal{I}^{(\hat{G}^{(a)}_{-1},\hat{G}_{0})}_{\Gamma}, \quad \mathcal{I}^{(\hat{G}_{0},\hat{G}_{-1})}_{\Gamma}=\sum_{b=1,2}\mathcal{I}^{(\hat{G}_{0},\hat{G}_{-1}^{(b)})}_{\Gamma}
\end{equation}
with the terms in the summation defined as:
\begin{equation}
\begin{aligned}
      \mathcal{I}^{(\hat{G}^{(a)}_{-1},\hat{G}_{-1}^{(b)})}_{\Gamma}&\equiv \epsilon_{\alpha\gamma}\epsilon_{AC}\epsilon_{\beta\delta}\epsilon_{BD}\left\langle \hat{\mathbb{V}}_{\alpha_1}(0)\hat{\mathbb{V}}_{\alpha_2}(1)\hat{\mathbb{V}}_{\alpha_3}(\infty)  \hat{G}_{-1}^{(a)\alpha A}\hat{\Psi}^{\gamma C}(\eta_1)
      \hat{G}^{(b)\beta B}_{-1}\hat{\Psi}^{\delta D}(\eta_2)\right\rangle\\
        \mathcal{I}^{(\hat{G}^{(a)}_{-1},\hat{G}_{0})}_{\Gamma}&\equiv \epsilon_{\alpha\gamma}\epsilon_{AC}\epsilon_{\beta\delta}\epsilon_{BD}\left\langle \hat{\mathbb{V}}_{\alpha_1}(0)\hat{\mathbb{V}}_{\alpha_2}(1)\hat{\mathbb{V}}_{\alpha_3}(\infty)  \hat{G}_{-1}^{(a)\alpha A}\hat{\Psi}^{\gamma C}(\eta_1)
      \hat{G}^{\beta B}_{0}\hat{\Psi}^{\delta D}(\eta_2)\right\rangle\\
        \mathcal{I}^{(\hat{G}_{0},\hat{G}_{-1}^{(b)})}_{\Gamma}&\equiv \epsilon_{\alpha\gamma}\epsilon_{AC}\epsilon_{\beta\delta}\epsilon_{BD}\left\langle \hat{\mathbb{V}}_{\alpha_1}(0)\hat{\mathbb{V}}_{\alpha_2}(1)\hat{\mathbb{V}}_{\alpha_3}(\infty)  \hat{G}_{0}^{\alpha A}\hat{\Psi}^{\gamma C}(\eta_1)
      \hat{G}^{(b)\beta B}_{-1}\hat{\Psi}^{\delta D}(\eta_2)\right\rangle
\end{aligned}
\end{equation}
and $ \mathcal{I}^{(\hat{G}_{0},\hat{G}_{0})}_{\Gamma}$ do not need to be  further divided.

Now it is easy to see that the structure of the above terms are as follows: once a  $\hat{G}^{(1)}_{-1}$ appear, one needs to take a derivative of $\mathcal{I}_\Gamma^B$ with respect to $\eta_1$ or $\eta_2$. The contributes of the remaining parts  are some two point functions of the form
\begin{equation}
    \frac{N}{(\eta_1-\eta_2)^M}.
\end{equation}
They come from contracting the 4-th and 5-th operators (with the bosonic vertex operators or their derivatives not included), inserted at $\eta_1$ and $\eta_2$. Note that $M$ is fixed by the weights of the considered 2 operators, so we need to work out the coefficient $N$. 
Making use of the explicit expressions \eqref{G0action} and \eqref{G-1action}, we have carefully worked out all of them\footnote{Note that these $N$'s are not the overall normalization so we need the precise values of them. In particular, the sign of every $N$ is important.}. The results are as follows.
\begin{equation}
\begin{aligned}
    &\mathcal{I}^{(\hat{G}_{-1}^{(1)},\hat{G}_{-1}^{(1)})}_{\Gamma}\\
    =&4i\times \partial_{\eta_1}\partial_{\eta_2}\mathcal{I}_{\Gamma}^B\times\frac{2}{k(\eta_1-\eta_2)}\\
    =&\frac{8i\mathcal{I}_{\Gamma}^B}{k(\eta_1-\eta_2)}\Bigg(\sum_{s,t=1,2}\frac{(j_s+\frac{k}{2}-1)(j_t+\frac{k}{2}-1)}{(\eta_1-\delta_{2,s})(\eta_2-\delta_{2,t})}\!+\!\sum_{u,v=1,2}\frac{k(j_u+\frac{k}{2}-1)}{2(\eta_v-\delta_{2,u})(\eta_1-\eta_2)}-\frac{\frac{k}{2}(\frac{k}{2}+1)}{(\eta_1-\eta_2)^2}\Bigg)
\end{aligned}
\end{equation}
Note that on the second line of the above equation, the  factor 4 appears because the 4 terms in \eqref{The4terms} give identical contributions; the factor $i$ appears due to the cocycles of the bosonized spin fields\footnote{Concretely, it comes from the effects of the cocycles in computing the two point function  $\langle S^{++--}S^{--++}\rangle=\langle S^{--++}S^{++--}\rangle=i$.}. Both of these factors are not crucial and will be  canceled in the calculation of normalizations in  section \ref{Thenormalization}.
\begin{equation}
    \begin{aligned}
    \mathcal{I}^{(\hat{G}_{-1}^{(1)},\hat{G}_{-1}^{(2)})}_{\Gamma}
    &=4i\times \partial_{\eta_1}\mathcal{I}_{\Gamma}^B\times\frac{-2}{k(\eta_1-\eta_2)^2}
    =\frac{-8i\mathcal{I}_{\Gamma}^B}{k(\eta_1-\eta_2)^2}\Bigg(\sum_{s=1,2}\frac{j_s+\frac{k}{2}-1}{\eta_1-\delta_{2,s}}-\frac{k}{2(\eta_1-\eta_2)}\Bigg)\\
    \mathcal{I}^{(\hat{G}_{-1}^{(2)},\hat{G}_{-1}^{(1)})}_{\Gamma}
    &=4i\times \partial_{\eta_2}\mathcal{I}_{\Gamma}^B\times\frac{2}{k(\eta_1-\eta_2)^2}
    =\frac{8i\mathcal{I}_{\Gamma}^B}{k(\eta_1-\eta_2)^2}\Bigg(\sum_{s=1,2}\frac{j_s+\frac{k}{2}-1}{\eta_2-\delta_{2,s}}+\frac{k}{2(\eta_1-\eta_2)}\Bigg)
\end{aligned}
\end{equation}
Then the next term is:
\begin{equation}
    \begin{aligned}
    \mathcal{I}^{(\hat{G}_{-1}^{(2)},\hat{G}_{-1}^{(2)})}_{\Gamma}
    =&4i\times \mathcal{I}_{\Gamma}^B\times\frac{-7}{(\eta_1-\eta_2)^3}.
\end{aligned}
\end{equation}
The  terms involve one $\hat{G}_0$ are:
\begin{equation}
    \begin{aligned}
    \mathcal{I}^{(\hat{G}_{-1}^{(1)},\hat{G}_{0})}_{\Gamma}
    &=4i\times \partial_{\eta_1}\mathcal{I}_{\Gamma}^B\times\frac{-2}{\eta_1-\eta_2}
    =\frac{-8i\mathcal{I}_{\Gamma}^B}{\eta_1-\eta_2}\Bigg(\sum_{s=1,2}\frac{j_s+\frac{k}{2}-1}{\eta_1-\delta_{2,s}}-\frac{k}{2(\eta_1-\eta_2)}\Bigg)\\
    \mathcal{I}^{(\hat{G}_0,\hat{G}_{-1}^{(1)})}_{\Gamma}
    &=4i\times \partial_{\eta_2}\mathcal{I}_{\Gamma}^B\times\frac{-2}{\eta_1-\eta_2}
    =\frac{-8i\mathcal{I}_{\Gamma}^B}{\eta_1-\eta_2}\Bigg(\sum_{s=1,2}\frac{j_s+\frac{k}{2}-1}{\eta_2-\delta_{2,s}}+\frac{k}{2(\eta_1-\eta_2)}\Bigg)
\end{aligned}
\end{equation}
and
\begin{equation}
    \begin{aligned}
    \mathcal{I}^{(\hat{G}_{-1}^{(2)},\hat{G}_{0})}_{\Gamma}
    &=4i\times \mathcal{I}_{\Gamma}^B\times\frac{-2}{(\eta_1-\eta_2)^2}\\
    \mathcal{I}^{(\hat{G}_0,\hat{G}_{-1}^{(2)})}_{\Gamma}
    &=4i\times \mathcal{I}_{\Gamma}^B\times\frac{2}{(\eta_1-\eta_2)^2}.
\end{aligned}
\end{equation}
The last term is:
\begin{equation}\label{Thetestterm}
\begin{aligned}
      \mathcal{I}^{(\hat{G}_{0},\hat{G}_{0})}_{\Gamma}=4i\times\mathcal{I}_{\Gamma}^B\times \frac{2k}{\eta_1-\eta_2}.
\end{aligned}
\end{equation}
With the above expressions, the finial result for the five point function $\mathcal{I}$ can be written as:
\begin{equation}\label{result5pt}
\begin{aligned}
     \mathcal{I}&=\sum_\Gamma |\mathcal{I}_\Gamma|^2, \\
     \mathcal{I}_\Gamma&=f_\Gamma g_\Gamma\left(\sum_{a,b=1,2}\mathcal{I}_\Gamma^{(\hat{G}_{-1}^{(a)},\hat{G}_{-1}^{(b)})}+C_2\sum_{a=1,2}\mathcal{I}_\Gamma^{(\hat{G}_{-1}^{(a)},\hat{G}_{0})}+C_1\sum_{b=1,2}\mathcal{I}_\Gamma^{(\hat{G}_{0},\hat{G}_{-1}^{(b)})}+C_1C_2 \mathcal{I}^{(\hat{G}_{0},\hat{G}_{0})}_{\Gamma}\right)
\end{aligned}
\end{equation}
For our purpose to compare with the string side,  we  write $\mathcal{I}_\Gamma$ as
\begin{equation}\label{CFTsum}
\begin{aligned}
    \mathcal{I}_\Gamma=f_\Gamma g_\Gamma \mathfrak{I}_\Gamma \mathcal{I}_\Gamma^B, \qquad
     \mathfrak{I}_\Gamma=8i\sum_{X\in\mathfrak{S}'} \mathbb{C}_\Gamma^{\text{CFT}}[X]X
\end{aligned}
\end{equation}
where 
\begin{equation}\label{s'}
   \mathfrak{S}'=\left\{ \frac{j_1j_2}{k},\quad  \frac{j_1^2}{k}, \quad \frac{j_2^2}{k},\quad \frac{j_1}{k}, \quad\frac{j_2}{k}, \quad \frac{1}{k}, \quad 1, \quad k, \quad j_1, \quad j_2 \right\}.
\end{equation}
One can see that $\mathfrak{S}'$ in fact coincide with $\mathfrak{S}$ in \eqref{list10} on the string side.

 Now we divide the coefficients $ \mathbb{C}_\Gamma^{\text{CFT}}[X]$ into 2 groups, called regular and irregular respectively.  For each $X\in\mathfrak{S}$, $ \mathbb{C}_\Gamma^{\text{CFT}}[X]$ is called regular if it does not contain any $C_i$ $(i=1,2)$. Otherwise, it is called irregular. Then it is easy to see that the two groups are:
\begin{equation}\label{re/irre}
    \mathfrak{S}^{'reg}=\left\{\frac{j_1j_2}{k},\quad  \frac{j_1^2}{k}, \quad \frac{j_2^2}{k},\quad \frac{j_1}{k}, \quad\frac{j_2}{k}, \quad \frac{1}{k}\right\}, \qquad \mathfrak{S}^{'irreg}=\left\{1, \quad k, \quad j_1, \quad j_2 \right\}.
\end{equation}  
Thus they coincide with the ones we used in the CFT side. In fact, this  notion of  regular/irregular  comes from the comments above  \eqref{IGamma}, where we view $\tilde{\Phi}^{(1)}$ as regular and $\tilde{\Phi}^{(2)}$  as irregular.
Since inserting $\tilde{\Phi}^{(2)}(\eta_i)$ brings $C_i$ $(i=1,2)$, we then have the above notion of  regular/irregular of $ \mathbb{C}_\Gamma^{\text{CFT}}[X]$.  

Finally, we read the explicit form of the coefficients $ \mathbb{C}_\Gamma^{\text{CFT}}[X]$.
For the 6 regular terms, the first 3 terms are: 
\begin{subequations}\label{CFT1}
\begin{align}
   \mathbb{C}_\Gamma^{\text{CFT}}\left[\frac{j_1j_2}{k}\right]&=\left(\frac{1}{\eta_1(\eta_2-1)}+\frac{1}{(\eta_1-1)\eta_2}\right)\times \frac{1}{\eta_1-\eta_2}\\
   \mathbb{C}_\Gamma^{\text{CFT}}\left[\frac{j_1^2}{k}\right]&=\frac{1}{\eta_1\eta_2}\times \frac{1}{\eta_1-\eta_2}\\
   \mathbb{C}_\Gamma^{\text{CFT}}\left[\frac{j_2^2}{k}\right]&=\frac{1}{(\eta_1-1)(\eta_2-1)}\times \frac{1}{\eta_1-\eta_2}.
 \end{align}
\end{subequations}  
The remaining 3 terms are:
\begin{subequations}\label{CFT2}
\begin{align}
   \mathbb{C}_\Gamma^{\text{CFT}}\left[\frac{j_1}{k}\right]
    &=\left(-\frac{1}{\eta_1\eta_2}-\frac{1}{\eta_1(\eta_2-1)}-\frac{1}{(\eta_1-1)\eta_2}\right)\times \frac{1}{\eta_1-\eta_2}\\
     \mathbb{C}_\Gamma^{\text{CFT}}\left[\frac{j_2}{k}\right]
    &=\left(-\frac{1}{(\eta_1-1)(\eta_2-1)}-\frac{1}{\eta_1(\eta_2-1)}-\frac{1}{(\eta_1-1)\eta_2}\right)\times \frac{1}{\eta_1-\eta_2}\\
       \mathbb{C}_\Gamma^{\text{CFT}}\left[\frac{1}{k}\right]
   &=\left(\frac{1}{\eta_1(\eta_2-1)}+\frac{1}{(\eta_1-1)\eta_2}\right)\times\frac{1}{\eta_1-\eta_2}.   
\end{align}
\end{subequations}
For irregular terms, for $ \mathbb{C}_\Gamma^{\text{CFT}}[j_1]$, $ \mathbb{C}_\Gamma^{\text{CFT}}[j_2]$ we have:
\begin{equation}\label{CFT3}
\begin{aligned}
    \mathbb{C}_\Gamma^{\text{CFT}}[j_1]&=\!\left(\frac{1}{2\eta_1\eta_2}+\frac{1}{2\eta_1(\eta_2-1)}+\frac{1}{2\eta_2(\eta_1-1)}-\frac{C_1}{\eta_2}-\frac{C_2}{\eta_1}\!\right)\!\times\! \frac{1}{\eta_1-\eta_2},\\
    \mathbb{C}_\Gamma^{\text{CFT}}[j_2]&=\!\left(\frac{1}{2(\eta_1-1)(\eta_2-1)}+\frac{1}{2\eta_1(\eta_2-1)}+\frac{1}{2\eta_2(\eta_1-1)}-\frac{C_1}{\eta_2-1}-\frac{C_2}{\eta_1-1}\!\right)\!\times \!\frac{1}{\eta_1-\eta_2}.
\end{aligned}
\end{equation}
Then we consider $ \mathbb{C}_\Gamma^{\text{CFT}}[k]$, 
it turns out  that  $ \mathbb{C}_\Gamma^{\text{CFT}}[k]$ can be written into a factorized form, and it is the only coefficient that include the product $C_1C_2$:
\begin{equation}\label{CFT4}
    \mathbb{C}_\Gamma^{\text{CFT}}[k]=\!\left[C_1-\frac{1}{2}\!\left(\frac{1}{\eta_1-1}+\frac{1}{\eta_1}+\frac{-1}{\eta_1-\eta_2}\!\right)\!\right]\!\left[C_2-\frac{1}{2}\!\left(\frac{1}{\eta_2-1}+\frac{1}{\eta_2}+\frac{1}{\eta_1-\eta_2}\!\right)\!\right]\!\times \!\frac{1}{\eta_1-\eta_2}.
\end{equation}
The finial term $ \mathbb{C}_\Gamma^{\text{CFT}}[1]$ is more complicated:
\begin{equation}\label{CFT5}
\begin{aligned}
    \mathbb{C}_\Gamma^{\text{CFT}}&[1]
   =\left(-\frac{3}{(\eta_1-\eta_2)^2}-\frac{1}{\eta_1(\eta_2-1)}-\frac{1}{\eta_1(\eta_2-1)}\right)\times \frac{1}{\eta_1-\eta_2} \\
   &+C_1\left(\frac{1}{\eta_2-1}+\frac{1}{\eta_2}+\frac{1}{\eta_1-\eta_2}\right)\times \frac{1}{\eta_1-\eta_2}+C_2\left(\frac{1}{\eta_1-1}+\frac{1}{\eta_1}+\frac{-1}{\eta_1-\eta_2}\right)\times \frac{1}{\eta_1-\eta_2}.
\end{aligned}
\end{equation}
Note that in these expressions, one can view  $C_1$, $C_2$ as functions of $\eta_1,\eta_2$. They are hard to obtain, as  explicit forms of covering maps with five ramified points are not known. Nevertheless, we will give  (almost) closed formulas for them in section \ref{matchthetwosides} (see \eqref{mainiden1}).

 Finally, we can write the residue of the correlator on the CFT answer as:
\begin{equation}\label{CFTresult1}
\begin{aligned}
    \prod_{i=1}^3\mathcal{N}_{V_i}^{-1}&\left(\mathop{\text{Res}}\limits_{2b(\sum_i\alpha_i-Q)}\mathbb{M}_{VVV}\right)\\
    &\hspace{2cm}=\mathcal{N}_{\Phi}^{-2}\prod_{i=1}^3\mathcal{N}_{V_i}^{-1}
    \frac{(-\mu)^2}{\pi 2!}N^{-\frac{1}{2}}\left(\sqrt{2}\right)^2\prod_{i=1}^3\sqrt{w_i}\int \prod_{l=1}^2d^2 \xi_l \left(\sum_{\Gamma}|\mathcal{I}_{\Gamma}|^2\right),
\end{aligned}
\end{equation}
where $\mathcal{I}_\Gamma$ is expressed in \eqref{CFTsum}. The normalization factor $\mathcal{N}_{\Phi}$ and $\mathcal{N}_{V_i}$ will be calculated in the next subsection, the results are
\begin{equation}\label{resultnormalization}
    \mathcal{N}_{\Phi}=16, \qquad \mathcal{N}_{V_i}=1\quad  (i=1,2,3).
\end{equation}

\subsubsection{The normalization}\label{Thenormalization}
In this subsection, we show the normalization constants \eqref{resultnormalization} of the  operators $V^{(w_i)}_{\alpha_i}$ $(i=1,2,3)$ and the marginal operator $\Phi$. For this purpose, we need to calculate the 2-point functions of them. For  two point functions, we need to find covering maps that have the behavior:
\begin{equation}
\begin{aligned}
   \Gamma_{(2)}(z)&=\lambda_1+\mathfrak{a}_1(z-\theta_1)^{w_1}+\mathfrak{b}_1(z-\theta_1)^{w_1+1}+...,\qquad &\text{near $z=\theta_1$}\\
   \Gamma_{(2)}(z)&=\lambda_2+\mathfrak{a}_{2}(z-\theta_2)^{w_2}+\mathfrak{b}_{2}(z-\theta_2)^{w_2+1}+..., \qquad &\text{near $z=\theta_2$}.
\end{aligned}
\end{equation}
Note that for our purpose, we have not fix the coordinates  $\lambda_i$ $(i=1,2)$ of the two points on the base space  as well as their lifted coordinates  $\theta_i$ $(i=1,2)$ on the covering surface. The covering map that have these 2 ramified points are:
\begin{equation}
    \Gamma_{(2)}(z)=\frac{\lambda_2 (z-\theta_1)^w-\lambda_1 (z-\theta_2)^w}{(z-\theta_1)^w-(z-\theta_2)^w}.
\end{equation}
Then the coefficient in the expansions are:
\begin{equation}
    \mathfrak{a}_1=\frac{\lambda_1-\lambda_2}{(\theta_1-\theta_2)^w},\quad \mathfrak{b}_1=\frac{-w(\lambda_1-\lambda_2)}{(\theta_1-\theta_2)^{w+1}}, \quad \mathfrak{a}_2=\frac{\lambda_2-\lambda_1}{(\theta_2-\theta_1)^w}, \quad \mathfrak{b}_{2}=\frac{-w(\lambda_2-\lambda_1)}{(\theta_2-\theta_1)^{w+1}}.
\end{equation}
In our calculation, for the 3  operators $\mathbb{V}_{\alpha_i}^{(w_i)}$ $(i=1,2,3)$,  the lifted operators are canonically normalized ($\alpha$ could be any one of $\alpha_i$):
\begin{equation}
    \left\langle\hat{\mathbb{V}}_{\alpha}(\theta_1)\hat{\mathbb{V}}_{\beta}(\theta_2)\right\rangle=\delta(\alpha+\beta-Q)\left|(\theta_1-\theta_2)^{-2\alpha\beta}\right|^2=\delta(\alpha+\beta-Q)\left|(\theta_1-\theta_2)^{-2\hat{h}_{\alpha}}\right|^2.
\end{equation}
Then from \eqref{liftfractional}, one find the factor coming from the lifting is:
\begin{equation}
    \left|\mathfrak{a}_1^{-\frac{\hat{h}_\alpha}{w}}\mathfrak{a}_2^{-\frac{\hat{h}_\beta}{w}}\right|^2=\left|(\lambda_1-\lambda_2)^{-\frac{\hat{h}_\alpha+\hat{h}_\beta}{w}}(\theta_1-\theta_2)^{\hat{h}_\alpha+\hat{h}_\beta}\right|^2.
\end{equation}
Thus the 2-point function of $\mathbb{V}_{\alpha}^{(w)}$ is:
\begin{equation}\label{2pointofV}
\begin{aligned}
    \left\langle \mathbb{V}_{\alpha}^{(w)} \mathbb{V}_{\beta}^{(w)}\right\rangle&=\delta(\alpha+\beta-Q)\left|(\lambda_1-\lambda_2)^{-\frac{\hat{h}_\alpha+\hat{h}_\beta}{w}}(\theta_1-\theta_2)^{\hat{h}_\alpha+\hat{h}_\beta-2\hat{h}_\alpha}\right|^2\left|(\lambda_1-\lambda_2)^{-2h_\sigma}\right|^2\\
    &=\delta(\alpha+\beta-Q)\left|(\lambda_1-\lambda_2)^{-2h_\alpha}\right|^2,
\end{aligned}
\end{equation}
with
\begin{equation}
    h_\alpha=h_\sigma+\frac{\hat{h}_\alpha}{w}, \qquad h_\sigma=\frac{c}{24}\left(w-\frac{1}{w}\right).
\end{equation}
In \eqref{2pointofV}, the factor $\left|(\lambda_1-\lambda_2)^{-2h_\sigma}\right|^2$ comes from the 2-point function of the twist fields $\sigma$ and their weights are $h_\sigma$. One can see that $\theta_1, \theta_2$ do not appear in the finial result of the 2-point function, which takes the form of a standard 2-point function in a CFT.
Therefore, $\mathbb{V}_{\alpha}^{(w)}$ is canonically normalized. This is in fact the same as in the bosonic case \cite{Eberhardt:2021vsx}.

Then we consider the  marginal operator. In its form \eqref{Themarginal}, one can see that when lifted up to the covering surface, the marginal operator contains a vertex operator in the linear dilaton theory. Thus, just like  the two point function of $\mathbb{V}_{\alpha}^{(w)}$, the two points function for the marginal operator will also contain a delta function which represents the momentum conservation. However, to determine the normalization factor, we only need to calculate the two point function of the marginal operator and its ``conjugate'', defined by replacing the momentum $\alpha_{\Phi}$ in \eqref{alphaPhi} by $Q-\alpha_{\Phi}$:
\begin{equation}
    \Phi^\dagger(x,\bar{x})\equiv \Phi(x,\bar{x})|_{\alpha_{\Phi}\to Q-\alpha_{\Phi}},
\end{equation}
and drop the divergent term coming from the delta function. Note that the contractions of the 4 terms in  $\Phi$ with their conjugates in $\Phi^\dagger$ give  identical contributions to the 2-point function.  It is easy to see that the calculation of the two point function is similar with the five point function in \eqref{result5pt}, so we can write it as:
\begin{equation}\label{marginal2pt}
\begin{aligned}
    \left\langle \tilde{\Phi}^{\dagger}(\lambda_1,\bar{\lambda}_1) \tilde{\Phi}(\lambda_2,\bar{\lambda}_2)\right\rangle'
    =\left|(\lambda_1-\lambda_2)^{-2h_\sigma}\right|^2\Big|(\mathfrak{a}_1\mathfrak{a}_2)^{\frac{3k}{8}-1}\Big|^2\times \left|\mathcal{I}_{\Gamma_{(2)}}\right|^2.
\end{aligned}    
\end{equation}
In \eqref{marginal2pt}, the  prime on the 2-point function means the divergence term $\delta(0)$ is removed.
The factor   $\left|(\lambda_1-\lambda_2)^{-2h_\sigma}\right|^2$ again comes from the 2-point function of the twist fields and $\Big|(\mathfrak{a}_0\mathfrak{a}_1)^{\left(1-\frac{3k}{8}\right)}\Big|^2$ is the factor that comes from the lifting.  The function $\mathcal{I}_{\Gamma_{(2)}}$ can be calculated as the 5-point function $\mathcal{I}_\Gamma$, except that now we should replace the bosonic 5-point function $\mathcal{I}_\Gamma^B$ by the  the bosonic 2-point function $\mathcal{I}^B_{(2)}$ (again we have removed the divergent delta function):
\begin{equation}
    \mathcal{I}^B_{(2)}=\left\langle e^{\frac{3k-2}{\sqrt{2k}}\phi}(\theta_1)e^{-\sqrt{\frac{k}{2}}\phi}(\theta_2)\right\rangle'=\frac{1}{(\theta_1-\theta_2)^{1-\frac{3k}{2}}}
\end{equation}
 and the covering map $\Gamma$ should be replaced by the covering map $\Gamma_{(2)}$ for the 2-point function. This is realized by replacing $C_1$ by $\mathfrak{c}_1$ and $C_2$ by $\mathfrak{c}_2$, with
 \begin{equation}\label{c1c2}
    \mathfrak{c}_1\equiv-\frac{3 \mathfrak{b}_1}{4 \mathfrak{a}_1}=\frac{3w}{4(\theta_1-\theta_2)}=\frac{3}{2(\theta_1-\theta_2)}, \qquad  \mathfrak{c}_2\equiv-\frac{3 \mathfrak{b}_2}{4 \mathfrak{a}_2}=\frac{3w}{4(\theta_2-\theta_1)}=\frac{3}{2(\theta_2-\theta_1)}
\end{equation}
Thus we have:
\begin{equation}\label{IGamma2}
   \mathcal{I}_{\Gamma_{(2)}}=\sum_{a,b=1,2}\mathcal{I}_{\Gamma_{(2)}}^{(\hat{G}_{-1}^{(a)},\hat{G}_{-1}^{(b)})}+\mathfrak{c}_2\sum_{a=1,2}\mathcal{I}_{\Gamma_{(2}}^{(\hat{G}_{-1}^{(a)},\hat{G}_{0})}+\mathfrak{c}_1\sum_{b=1,2}\mathcal{I}_{\Gamma_{(2)}}^{(\hat{G}_{0},\hat{G}_{-1}^{(b)})}+\mathfrak{c}_1\mathfrak{c}_2 \mathcal{I}^{(\hat{G}_{0},\hat{G}_{0})}_{\Gamma_{(2)}} 
\end{equation}
with
\begin{equation}\label{2ptallterms}
    \begin{aligned}
        \mathcal{I}_{\Gamma_{(2)}}^{(\hat{G}_{-1}^{(1)},\hat{G}_{-1}^{(1)})}&=4i\partial_{\theta_1}\partial_{\theta_2}\mathcal{I}^B_{(2)}\times\frac{2}{k(\theta_1-\theta_2)}=-\frac{2i(3k-2)(3k-4)}{k}(\theta_1-\theta_2)^{\frac{3k}{2}-4}\\
        \mathcal{I}_{\Gamma_{(2)}}^{(\hat{G}_{-1}^{(1)},\hat{G}_{-1}^{(2)})}&=4i\partial_{\theta_1}\mathcal{I}^B_{(2)}\times\frac{-2}{k(\theta_1-\theta_2)^2}=-\frac{4i(3k-2)}{k}(\theta_1-\theta_2)^{\frac{3k}{2}-4}\\
         \mathcal{I}_{\Gamma_{(2)}}^{(\hat{G}_{-1}^{(2)},\hat{G}_{-1}^{(1)})}&=4i\partial_{\theta_2}\mathcal{I}^B_{(2)}\times\frac{2}{k(\theta_1-\theta_2)^2}=-\frac{4i(3k-2)}{k}(\theta_1-\theta_2)^{\frac{3k}{2}-4}\\
           \mathcal{I}_{\Gamma_{(2)}}^{(\hat{G}_{-1}^{(2)},\hat{G}_{-1}^{(2)})}&=4i\mathcal{I}^B_{(2)}\times\frac{-7}{(\theta_1-\theta_2)^3}=-28i(\theta_1-\theta_2)^{\frac{3k}{2}-4}\\
 \mathcal{I}_{\Gamma_{(2)}}^{(\hat{G}_{-1}^{(1)},\hat{G}_{0})}&=4i\partial_{\theta_1}\mathcal{I}^B_{(2)}\times\frac{-2}{\theta_1-\theta_2}=-4i(3k-2)(\theta_1-\theta_2)^{\frac{3k}{2}-3}\\
 \mathcal{I}_{\Gamma_{(2)}}^{(\hat{G}_{0},\hat{G}_{-1}^{(1)})}&=4i\partial_{\theta_2}\mathcal{I}^B_{(2)}\times\frac{-2}{\theta_1-\theta_2}=4i(3k-2)(\theta_1-\theta_2)^{\frac{3k}{2}-3}\\
  \mathcal{I}_{\Gamma_{(2)}}^{(\hat{G}_{-1}^{(2)},\hat{G}_{0})}&=4i\mathcal{I}^B_{(2)}\times\frac{-2}{(\theta_1-\theta_2)^2}=-8i(\theta_1-\theta_2)^{\frac{3k}{2}-3}\\
  \mathcal{I}_{\Gamma_{(2)}}^{(\hat{G}_{0},\hat{G}_{-1}^{(2)})}&=4i\mathcal{I}^B_{(2)}\times\frac{2}{(\theta_1-\theta_2)^2}=8i(\theta_1-\theta_2)^{\frac{3k}{2}-3}\\
  \mathcal{I}_{\Gamma_{(2)}}^{(\hat{G}_{0},\hat{G}_{0})}&=4i\mathcal{I}^B_{(2)}\times\frac{2k}{\theta_1-\theta_2}=8ik(\theta_1-\theta_2)^{\frac{3k}{2}-2}.
    \end{aligned}
\end{equation}
Substituting \eqref{c1c2} and \eqref{2ptallterms} into \eqref{IGamma2}, we find that the finial result for $\mathcal{I}_{\Gamma_{(2)}}$  is simple. In particular, the normalization has no $k$ dependence:
\begin{equation}
    \mathcal{I}_{\Gamma_{(2)}}=-16i(\theta_1-\theta_2)^{\frac{3k}{2}-4}
\end{equation}
Thus
\begin{equation}
\begin{aligned}
    &\left\langle \tilde{\Phi}^{\dagger}(\lambda_1,\bar{\lambda}_1) \tilde{\Phi}(\lambda_2,\bar{\lambda}_2)\right\rangle'\\
    =&256\left|(\lambda_1-\lambda_2)^{-2h_\sigma}\right|^2\Big|(\lambda_1-\lambda_2)^{2\left(\frac{3k}{8}-1\right)}(\theta_1-\theta_2)^{-4\left(\frac{3k}{8}-1\right)}\Big|^2 |(\theta_1-\theta_2)^{\frac{3k}{2}-4}|^2\\
    =&256\left|(\lambda_1-\lambda_2)^{-2}\right|^2.
\end{aligned}    
\end{equation}
Note that there are no $\theta_1$, $\theta_2$ and $k$ dependence in the final answer of the two point function. In addition, the power of $(\lambda_1-\lambda_2)$ is correct since the marginal operator has weight $h=\bar{h}=1$.

\section{Match the two sides}\label{matchthetwosides}
In this section, we match the results of the two sides in the previous two sections. Since both residues of the correlators on the two sides are written as integrals, we will first unify the integration variables. Then we fix the three parameters $\alpha, \beta, \gamma$ in the string answer by comparing 3 regular terms with the CFT answer, and the condition $\alpha+\beta+\gamma=1$ helps to relate the prefactor of normalizations of the two sides (determining the constant $\mu$ in terms of $\nu$). The matching of the remaining 3 regular and 4 irregular terms gives non-trivial checks of the duality. In particular, the matching of the irregular terms involves new mathematical identities of covering maps.

\subsection{Unify the integration variables}

The residues on the string \eqref{stringresult}  and the CFT sides \eqref{CFTresult1} are represented as integrals of $y_i$ $(i=1,2)$ and $\xi_j$ $(j=1,2)$, respectively. We do not  intend to work out these integrals explicitly, which is not easy to do. Instead, we want to show that the integral forms (integrands) of the two sides indeed coincide. For this purpose, we  first unify the integration variables of the two sides by changing the coordinates $\xi_1,\xi_2$ on the CFT side as \cite{Eberhardt:2021vsx}:
\begin{equation}
    (\xi_1,\xi_2)\quad \to \quad (a_1, a_2)
\end{equation}
where $\xi_1, \xi_2, a_1, a_2$ are data in the covering map, defined in \eqref{coveringcondi}. In fact, $(a_1, a_2)$ will finally be identified with $(y_1, y_2)$ on the string side. After this change of  variables, the CFT result \eqref{CFTresult1} can be written as:
\begin{equation}
\begin{aligned}
    \prod_{i=1}^3&\mathcal{N}_{V_i}^{-1}\left(\mathop{\text{Res}}\limits_{2b(\sum_i\alpha_i-Q)=2}\mathbb{M}_{VVV}\right)\\
    =&\frac{1}{16^2}\frac{\left(-\sqrt{2}\mu\right)^2}{\pi 2!}N^{-\frac{1}{2}}\prod_{i=1}^3\sqrt{w_i}\sum_{\Gamma}\int_{D_{\Gamma}} \prod_{l=1}^2d^2 a_l \left|\text{det}\frac{\partial(\eta_1,\eta_2)}{\partial(a_1,a_2)}f_\Gamma(c,a_1,a_2)g_\Gamma(\mathbf{H},a_1,a_2)\mathcal{I}^B_{\Gamma}\mathfrak{I}_{\Gamma}\right|^2
\end{aligned}
\end{equation}
where $D_{\Gamma}$ is the corresponding integration region, which depends on the covering map $\Gamma$.  In the function $f_{\Gamma}, g_{\Gamma}$ (see \eqref{fgforms}), we write the arguments explicitly. The boldface letter $\mathbf{H}$ means the 3 space-time weights of the  vertex operators $\mathbf{H}=(H_1,H_2,H_3)$. It was observed in the bosonic case \cite{Eberhardt:2021vsx}  the following identities relating $X_i$ and some covering map data:
\begin{equation}\label{iden1}
     X_i^4(a_1,a_2,a_3)=\frac{\Pi^2(\eta_1-\eta_2)^2\prod_{j=1}^3(w_ja_j)^{1+w_j+2\delta_{ij}}}{4\prod_{l=1}^2\eta_l^{2+2\delta_{i1}}(\eta_l-1)^{2+2\delta_{i2}}A_l}
\end{equation}
and also the identity about the Jacobi determinant:
\begin{equation}\label{iden2}
     \text{det}\frac{\partial(a_1,a_2)}{\partial(\eta_1,\eta_2)}=\frac{w_1a_1w_2a_2(\eta_1-\eta_2)}{4\prod_{l=1}^2\eta_l^2(\eta_l-1)^2A_l}.
\end{equation} 
In \eqref{iden1}, the  arguments of the functions $X_i$ (defined in \eqref{defXi}) are set to be $(y_1,y_2,y_3)=(a_1,a_2,a_3)$. The right hand sides of \eqref{iden1} and \eqref{iden2}  involve various data in the covering maps (see \eqref{coveringcondi} and \eqref{defPi}).
Making use of these identities, we have:
\begin{equation}
\begin{aligned}
     \text{det}\frac{\partial(\eta_1, \eta_2)}{\partial(a_1, a_2)}f_\Gamma(6(k+2),a_1,a_2)&g_\Gamma(\mathbf{h},a_1,a_2)\mathcal{I}^B_{\Gamma}\\
    =&2^{-2k-1}\prod_{i=1}^3w_i^{j_i-1}\prod_{i=1}^3a_i^{\frac{(k+2)w_i}{2}+j_i-h_i-1}\hat{\mathfrak{B}}(a_1,a_2,a_3)
\end{aligned}
\end{equation}
Note that in the above, the arguments $\mathbf{h}$ in $g_\Gamma$ is $\mathbf{h}=(h_1,h_2,h_3)$, with $h_i$ defined as $h$ in \eqref{3weights}.
Then we can write the CFT answer as:
\begin{equation}
\begin{aligned}
     &\prod_{i=1}^3\mathcal{N}_{V_i}^{-1}\left(\mathop{\text{Res}}\limits_{2b(\sum_i\alpha_i-Q)=2}\mathbb{M}_{VVV}\right)\\ 
=&\frac{1}{16^2}\frac{\left(-\sqrt{2}\mu\right)^2}{\pi 2!N^{\frac{1}{2}}}2^{-4k+4}\prod_{i=1}^3w_i^{2j_i-\frac{3}{2}}\sum_{\Gamma}\int_{D_{\Gamma}} \prod_{l=1}^2d^2 a_l \left|\prod_{i=1}^3a_i^{\frac{(k+2)w_i}{2}+j_i-h_i-1}\hat{\mathfrak{B}}(a_1,a_2,a_3)\mathcal{R}\mathfrak{I}_{\Gamma}\right|^2
\end{aligned}
\end{equation}
where
\begin{equation}\label{R1}
    \mathcal{R}=\frac{1}{8}\frac{f_\Gamma(6k,a_1,a_2)g_\Gamma(\mathbf{H},a_1,a_2)}{f_\Gamma(6(k+2),a_1,a_2)g_\Gamma(\mathbf{h},a_1,a_2)}=\left(\frac{\Pi^2\prod_{j=1}^3(a_jw_j)^{w_j+1}}{A_1A_2}\right)^{\frac{1}{2}}.
\end{equation}
Then from \eqref{iden1}, one has three ways to express the above factor $\mathcal{R}$:
\begin{equation}\label{R2}
\begin{aligned}
  \frac{1}{\mathcal{R}}= \left(\frac{A_1A_2}{\Pi^2\prod_{j=1}^3(a_jw_j)^{w_j+1}}\right)^{\frac{1}{2}}&=\frac{a_1w_1}{2X_1^2}\times \frac{\eta_1-\eta_2}{\eta_1\eta_2(\eta_1-1)(\eta_2-1)}\frac{1}{\eta_1\eta_2}\\
    &=\frac{a_2w_2}{2X_2^2}\times \frac{\eta_1-\eta_2}{\eta_1\eta_2(\eta_1-1)(\eta_2-1)}\frac{1}{(\eta_1-1)(\eta_2-1)}\\
    &=\frac{a_3w_3}{2X_3^2}\times \frac{\eta_1-\eta_2}{\eta_1\eta_2(\eta_1-1)(\eta_2-1)}.
\end{aligned}
\end{equation}
Furthermore, from the above equations, one finds:
\begin{equation}\label{etatoX}
    \left\{ 
    \begin{aligned}
        &\eta_1\eta_2=\frac{X_3^2}{X_1^2}\frac{w_1a_1}{w_3a_3}\\
        &(\eta_1-1)(\eta_2-1)=\frac{X_3^2}{X_2^2}\frac{w_2a_2}{w_3a_3}
    \end{aligned}
    \right.
\end{equation}
 In the above equations, the left hand sides have no explicit dependence on $w_i$ while the right hand sides has explicit $w_i$ dependence. In fact, the 
  equations \eqref{etatoX}  show how to transform the arguments  $(\eta_1,\eta_2)$ to the arguments $(a_1,a_2)$, since $a_3$  is not an independent argument:
 when $(a_1,a_2)$  is given in the covering map \eqref{coveringcondi},    $a_3$ will be fixed as an implicit function of them $a_3=a_3(a_1,a_2)$, satisfying 
 \begin{equation}\label{a_3}
     X_{123}(a_1,a_2,a_3)=0.
 \end{equation}
 This is shown in \cite{Eberhardt:2021vsx} (in the ancillary $\mathtt{Mathematica}$ notebook of that paper). Note that this condition is just the $\delta$-localized condition when we calculate  the residue of the string correlator (see \eqref{delta1} and \eqref{delta2}).  Therefore, when $(a_1,a_2)$ is identified with $(y_1,y_2)$ on the string side, $a_3$ will coincide  with $y_3$ automatically.

Finally, as will be shown in the following subsections,  the integrand becomes 
a single-valued function after the above change of variables.  Thus  we have compensated the necessary
sum over covering maps by this change of variables and accordingly the residue of the CFT side can be written as:
\begin{equation}\label{CFTresult2}
\begin{aligned}
     &\prod_{i=1}^3\mathcal{N}_{V_i}^{-1}\left(\mathop{\text{Res}}\limits_{2b(\sum_i\alpha_i-Q)=2}\mathbb{M}_{VVV}\right)\\ 
=&2\!\times\!\frac{1}{16^2}\frac{\left(-\sqrt{2}\mu\right)^2}{\pi 2!N^{\frac{1}{2}}}2^{-4k+4}\prod_{i=1}^3w_i^{2j_i-\frac{3}{2}}\sum_{\Gamma}\!\int_{D_{\Gamma}} \prod_{l=1}^2d^2 a_l \left|\prod_{i=1}^3a_i^{\frac{(k+2)w_i}{2}+j_i-h_i-1}\hat{\mathfrak{B}}(a_1,a_2,a_3)\mathcal{R}\mathfrak{I}_{\Gamma}\right|^2\\
\equiv& \mathcal{N}_{CFT}\int \prod_{l=1}^2\frac{d^2 a_l}{\pi} \left|\prod_{i=1}^3a_i^{\frac{(k+2)w_i}{2}+j_i-h_i-1}\hat{\mathfrak{B}}(a_1,a_2,a_3)\mathcal{R}\sum_{X\in\mathfrak{S}'} \mathbb{C}_\Gamma^{\text{CFT}}[X]X\right|^2,
\end{aligned}
\end{equation}
where the factor $\mathcal{N}_{CFT}$ is:
\begin{equation}\label{NCFT}
    \mathcal{N}_{CFT}=\frac{2^{-4k+3}\pi\mu^2}{ N^{\frac{1}{2}}}\prod_{i=1}^3w_i^{2j_i-\frac{3}{2}}.
\end{equation}
In the second line of \eqref{CFTresult2}, we have multiplied an extra factor $2$ since both $a_1(\xi_1, \xi_2)$ and $a_2(\xi_1, \xi_2)$ are by construction symmetric functions, so we need to accompany
the change of variables by a factor of $2$ to account for this (similar to the bosonic case \cite{Eberhardt:2021vsx}). In the second line of \eqref{CFTresult2}, the summation over covering maps is removed. 

Now the string result \eqref{stringresult} and the CFT result \eqref{CFTresult2} have  the same form, provided that we identify $y_i$ with $a_i$ $(i=1,2)$ (then $a_3$ coincides with $y_3$ as well, see \eqref{a_3}). In particular, considering the $k,j_1,j_2$ dependence of the two sides, the first non-trivial check of the duality is that the two sides include exactly the same 10 terms ($\mathfrak{S}=\mathfrak{S}'$).  Then we need to check the matching of  every coefficient:
\begin{equation}
     \mathcal{N}_{string}\mathbb{C}^{\text{string}}\left[X\right]\mathop{=}\limits^{?}\mathcal{N}_{CFT}\mathcal{R} \mathbb{C}_\Gamma^{\text{CFT}}\left[X\right] \qquad X\in \mathfrak{S},
\end{equation}
where $\mathcal{R}$ is given in \eqref{R1} and \eqref{R2}; $\mathcal{N}_{string}$ is given in \eqref{Nstring}; $\mathcal{N}_{CFT}$ is given in \eqref{NCFT}; $\mathbb{C}^{\text{string}}[X]$ are given in \eqref{regular} and \eqref{irrgular1}; $\mathbb{C}_\Gamma^{\text{CFT}}[X]$ are given in  \cref{CFT1,CFT2,CFT3,CFT4,CFT5}. 
 In the following, we match these 10 coefficients in order.

\subsection{Match the regular terms}
For the regular terms,  since they do not contain $C_i$ $(i=1,2)$, we can compare the two sides directly.

Firstly, note that there are 3 undetermined  parameters $\alpha, \beta, \gamma$ on the string side.  We fix them by requiring  the  coefficients of $\frac{j_1j_2}{k}, \frac{j_1^2}{k}, \frac{j_2^2}{k}$ of the two sides match:
\begin{equation}
    \left\{ 
    \begin{aligned}
       \mathcal{N}_{string}\mathbb{C}^{\text{string}}\left[\frac{j_1j_2}{k}\right]&=\mathcal{N}_{CFT}\mathcal{R} \mathbb{C}_\Gamma^{\text{CFT}}\left[\frac{j_1j_2}{k}\right]\\
      \mathcal{N}_{string} \mathbb{C}^{\text{string}}\left[\frac{j_1^2}{k}\right]&=\mathcal{N}_{CFT}\mathcal{R} \mathbb{C}_\Gamma^{\text{CFT}}\left[\frac{j_1^2}{k}\right]\\
     \mathcal{N}_{string}  \mathbb{C}^{\text{string}}\left[\frac{j_2^2}{k}\right]&=\mathcal{N}_{CFT}\mathcal{R} \mathbb{C}_\Gamma^{\text{CFT}}\left[\frac{j_2^2}{k}\right].
    \end{aligned}
    \right.
\end{equation}
Together with the constraints:
\begin{equation}
    \alpha+\beta+\gamma=1,
\end{equation}
we obtain the solutions:
\begin{equation}\label{alphabetagamma1}
  \left\{ 
    \begin{aligned}
        &  \gamma= \frac{\eta_1+\eta_2-2\eta_1\eta_2}{(\eta_1-\eta_2)^2}\\
        &\alpha=\frac{\eta_1\eta_2(\eta_1+\eta_2-2)}{(\eta_1-\eta_2)^2}\\
        &\beta=\frac{-(\eta_1-1)(\eta_2-1)(\eta_1+\eta_2)}{(\eta_1-\eta_2)^2},
    \end{aligned}
    \right.
\end{equation}
and
\begin{equation}\label{Nstring=NCFT}
    \mathcal{N}_{string}=\mathcal{N}_{CFT}
\end{equation}
The equation \eqref{Nstring=NCFT} determines the deformation constant:
\begin{equation}
    \mu=\pm \frac{2^{2k-\frac{3}{2}}\nu^{-\frac{k}{2}}}{\pi}.
\end{equation}
 We expect the one with  minus sign should be the correct one, since this is the similar one with the bosonic case\footnote{Note that $\mu$ is looks like but not  the one in the  bosonic  case with $k\to k+2$.}.
The sign can only be fixed by matching  some correlators with $w_1+w_2+w_3\in2\mathbb{Z}$ at the 1-st order (in this work, we do not consider this kind of correlators). 
Notice that by \eqref{etatoX}, $\alpha, \beta, \gamma$ are all single valued function of $a_1,a_2$,  which can be written as:
\begin{equation}\label{alphabetagamma2}
   \left\{ 
    \begin{aligned}
        &  \gamma= \frac{1-Y_1-Y_2}{(Y_1-Y_2+1)^2-4Y_1}\\
        &\alpha=\frac{Y_1(Y_1-Y_2-1)}{(Y_1-Y_2+1)^2-4Y_1}\\
        &\beta=\frac{Y_2(Y_2-Y_1-1)}{(Y_1-Y_2+1)^2-4Y_1}
    \end{aligned}
    \right.
\end{equation}
where $Y_i\equiv \frac{X_3^2}{X_i^2}\frac{w_ia_i}{w_3a_3}$ and $a_3$ is determined by $X_{123}=0$. In \eqref{alphabetagamma2} $\alpha,\beta,\gamma$ are expressed in terms of worldsheet quantities, while in \eqref{alphabetagamma1} they are  expressed in terms of covering map data. 

With $\alpha, \beta, \gamma$ fixed,  we can now check the matching of the remaining  3 terms, proportional to $\frac{j_1}{k}, \frac{j_2}{k}, \frac{1}{k}$ respectively. We observe that on the string side we have the  following relations:  
\begin{equation}\label{relation1}
    \mathbb{C}^{\text{string}}\left[\frac{1}{k}\right]=\mathbb{C}^{\text{string}}\left[\frac{j_1j_2}{k}\right],\qquad\mathbb{C}^{\text{string}}\left[\frac{j_i}{k}\right]=-\mathbb{C}^{\text{string}}\left[\frac{j_i^2}{k}\right]-\mathbb{C}^{\text{string}}\left[\frac{j_1j_2}{k}\right] \quad(i=1,2), 
\end{equation}
and on the CFT side we have exactly the same relations:
\begin{equation}\label{relation2}
     \mathbb{C}_\Gamma^{\text{CFT}}\left[\frac{1}{k}\right]= \mathbb{C}_\Gamma^{\text{CFT}}\left[\frac{j_1j_2}{k}\right],\qquad \mathbb{C}_\Gamma^{\text{CFT}}\left[\frac{j_i}{k}\right]=- \mathbb{C}_\Gamma^{\text{CFT}}\left[\frac{j_i^2}{k}\right]- \mathbb{C}_\Gamma^{\text{CFT}}\left[\frac{j_1j_2}{k}\right] \quad(i=1,2). 
\end{equation}
Thus, given that we have already match the three terms proportional to $\frac{j_1j_2}{k}, \frac{j_1^2}{k}, \frac{j_2^2}{k}$ of the two sides, the remaining three regular terms also match due to the above relations.
Note that the fact that the  two groups of relations \eqref{relation1} and \eqref{relation2} are identical is far from obvious, so this agreement is another non-trivial check of this duality.

\subsection{Match the irregular terms}
Now we match the irregular terms. We have shown above that to obtain the matching of the regular terms, we simply need the interesting identities of covering maps \eqref{iden1}, \eqref{iden2}.  These identities involve the first non-zero  coefficients ($a_1, a_2, a_3, A_1, A_2$) at the ramified points ($0, 1, \infty, \xi_1, \xi_2$). For the matching of the irregular terms, we will need new identities of covering maps that involve not only the first 
but also the second non-trivial Taylor coefficients ($B_1, B_2$) near  the ramified points ($\xi_1, \xi_2$). 

Firstly, we consider the terms proportional to $j_i$ $(i=1,2)$. We need to verify the following equations:
\begin{equation}\label{j_i}
   \mathbb{C}^{\text{string}}\left[j_i\right]\mathop{=}\limits^{?}\mathcal{R} \mathbb{C}_\Gamma^{\text{CFT}}\left[j_i\right], \qquad i=1,2.
\end{equation}
Unlike the regular terms, one can not directly compare the two sides because  on the CFT side the integrand includes the functions $C_j$  $ (j=1,2)$, whose explicit forms are not known  (note that general  covering maps with 5 ramified points have not been constructed in  the literature). At the same time,  on the string side three new functions $\tilde{X}_i$ $ (i=1,2,3)$ have been introduced naturally. We observe the following novel mathematical identities which express $C_j$ $ (j=1,2)$ in terms of $X_i$ and $\tilde{X}_i$  $ (i=1,2)$: 
\begin{equation}\label{mainiden1}
\begin{aligned}
     C_1&=
     \frac{w_1\mathcal{X}_1(\eta_1-1)(\eta_1-\eta_2)+w_2\mathcal{X}_2\eta_1(\eta_1-\eta_2)+\eta_1^2-2\eta_1\eta_2+\eta_2}{2\eta_1(\eta_1-1)(\eta_1-\eta_2)}\\
     &=\frac{w_1\mathcal{X}_1+1}{2\eta_1}+\frac{w_2\mathcal{X}_2+1}{2(\eta_1-1)}-\frac{1}{2(\eta_1-\eta_2)}\\
       C_2&=
     \frac{w_1\mathcal{X}_1(\eta_2-1)(\eta_2-\eta_1)+w_2\mathcal{X}_2\eta_2(\eta_2-\eta_1)+\eta_2^2-2\eta_1\eta_2+\eta_1}{2\eta_2(\eta_2-1)(\eta_2-\eta_1)}\\
     &=\frac{w_1\mathcal{X}_1+1}{2\eta_2}+\frac{w_2\mathcal{X}_2+1}{2(\eta_2-1)}-\frac{1}{2(\eta_2-\eta_1)}
\end{aligned}
\end{equation}
where we defined a new function as the ratio of $X_i$ and its ``conjugate'' 
\begin{equation}
    \mathcal{X}_i\equiv\frac{\tilde{X}_i}{X_i}.
\end{equation} 
Note that $C_1$ and $C_2$ are related   by  $\eta_1\leftrightarrow \eta_2$ ($\mathcal{X}_i$ are invariant  under this exchange). Besides,  if we change the role of the first (inserted at $z=0$) and second  (inserted at $z=1$) operators, namely, make the exchanges  $\eta_1\leftrightarrow1-\eta$ and $w_1\mathcal{X}_1\leftrightarrow w_2\mathcal{X}_2$,  we will find $C_i\leftrightarrow -C_i$ $(i=1,2)$\footnote{The minus sign comes from the fact that  near $1-\eta_i$, while  the coefficient $A_i$ of the covering map $\Gamma(1-z)$ is the same as the original $A_i$ of $\Gamma(z)$, the coefficient $B_i$ of the covering map $\Gamma(1-z)$ is related to the original $B_i$ of $\Gamma(z)$ by a minus sign.}.  With the help of $\mathtt{Mathematica }$, we have checked that the  identities \eqref{mainiden1} indeed hold. From the identities \eqref{mainiden1} and \eqref{etatoX}, one could explicitly write  the  forms of  $C_1$ and $C_2$ as functions of $(\eta_1,\eta_2)$ or $(a_1, a_2)$, but both  will be complicated (in particular, multivalued).  In addition, we  observe that \eqref{mainiden1} are not the only ways to express the quantities $C_i$ $(i=1,2)$ in terms of $\mathcal{X}_i$ $(i=1,2,3)$, since there exists a  linear relation among $\mathcal{X}_i$:
\begin{equation}\label{relamongXi}
    w_1\mathcal{X}_1+ w_2\mathcal{X}_2+ w_3\mathcal{X}_3=1.
\end{equation}
Since $(a_1,a_2,a_3)$ satisfy $X_{123}=0$, we  show that \eqref{relamongXi} indeed holds with $X_{123}=0$ provided in the ancillary $\mathtt{Mathematica }$ notebook. From \eqref{mainiden1} and \eqref{relamongXi}, we have:
\begin{equation}\label{equ1}
\begin{aligned}
      \frac{C_1}{\eta_2}+\frac{C_2}{\eta_1}
     =&\frac{ (w_1\mathcal{X}_1+\frac{1}{2})(\eta_1+\eta_2-2)+ (w_3\mathcal{X}_3-\frac{5}{2})(2\eta_1\eta_2-\eta_1-\eta_2)}{-2(\eta_1-1)(\eta_2-1)\eta_1\eta_2}\\
     =&\frac{ (\eta_1-\eta_2)^2\left(\left(w_1\mathcal{X}_1+\frac{1}{2}\right)\frac{\alpha }{\eta_1\eta_2}-\left(w_3\mathcal{X}_3-\frac{5}{2}\right)\gamma\right)}{-2(\eta_1-1)(\eta_2-1)\eta_1\eta_2}
\end{aligned}
\end{equation}
\begin{equation}\label{equ2}
\begin{aligned}
      \frac{C_1}{\eta_2-1}+\frac{C_2}{\eta_1-1}=&\frac{  -(w_2\mathcal{X}_2+\frac{1}{2})(\eta_1+\eta_2)+ (w_3\mathcal{X}_3-\frac{5}{2})(2\eta_1\eta_2-\eta_1-\eta_2)}{-2(\eta_1-1)(\eta_2-1)\eta_1\eta_2}\\
       =&\frac{ (\eta_1-\eta_2)^2\left(\left(w_2\mathcal{X}_2+\frac{1}{2}\right)\frac{\beta }{\eta_1\eta_2}-\left(w_3\mathcal{X}_3-\frac{5}{2}\right)\gamma\right)}{-2(\eta_1-1)(\eta_2-1)\eta_1\eta_2}
      \end{aligned}
\end{equation}
Making use of these identities, one can easily verify that  equations \eqref{j_i} indeed hold.

Although we have checked the identities \eqref{mainiden1} numerically in the  $\mathtt{Mathematica }$ notebook, here we also verify some special cases of \eqref{mainiden1} analytically. In \cite{Roumpedakis:2018tdb}, a method is developed to construct covering maps. We briefly review the construction in  Appendix \ref{constructcovering}. Generally, one can not write down a closed form of the relevant covering map. Nevertheless, there are special cases where we can explicitly write down the covering map. We call them edge cases, satisfying\footnote{Note that the edge condition \eqref{edgecondi}  coincide with the one that appeared in \cite{Bufalini:2022toj} (see also \cite{Yu:2024kxr}) for a 3-point string correlator whose spectral flow parameters satisfy $w_1+w_2+w_3\in2\mathbb{Z}+1$. There, these edge cases are distinguished because one can not construct covering maps with three ramified points with indices satisfying \eqref{edgecondi}. This means for these edge cases, the contribution from the 0-th order  will vanish (see \cite{Yu:2024kxr}) and  the 2-nd order considered in this work  will give the leading contribution (note that for the 2-nd order, there are 2 extra ramified points with indices $w_4=w_5=2$, so the covering map exists and is in fact unique, see  \eqref{edgecovering} below).}:
\begin{equation}\label{edgecondi}
    w_1+w_2+1=w_3.
\end{equation}
Then
 using the method of  \cite{Roumpedakis:2018tdb}, we can work out the covering map for the edge cases, which is unique (see Appendix \ref{constructcovering}) and  has the following form:
\begin{equation}\label{edgecovering}
    \Gamma(z)=C_{w_1+w_2}^{w_1}\frac{\sum_{n=w_1}^{w_3}(-1)^{w_3-n}(C_{w_2-1}^{n-w_1-2}+C_{w_2-1}^{n-w_1-1}(\eta_1+\eta_2)+C_{w_2-1}^{n-w_1}\eta_1\eta_2)\frac{z^n}{n}}{w_1(w_1+1)-w_1w_3(\eta_1+\eta_2)+w_3(w_3-1)\eta_1\eta_2}\prod_{i=1}^3w_i
\end{equation}
where we use $C_{m}^{n}$ to denote the combinatorial number 
\begin{equation}
    C_{m}^{n}=\frac{\Gamma(n+1)}{\Gamma(m+1)\Gamma(n-m+1)}. 
\end{equation}
From this covering map, one can find  the following identities:
\begin{equation}
\begin{aligned}
     C_1^{\text{edge}}&=\frac{-w_1+1}{2\eta_1}+\frac{-w_2+1}{2(\eta_1-1)}-\frac{1}{2(\eta_1-\eta_2)}\\
      C_2^{\text{edge}}&=\frac{-w_1+1}{2\eta_2}+\frac{-w_2+1}{2(\eta_2-1)}-\frac{1}{2(\eta_2-\eta_1)}.
\end{aligned}
\end{equation}
These identities coincide with \eqref{mainiden1}, since for  edge cases  $P_{w_1,w_2,w_3+1}=P_{w_1-1,w_2,w_3}=P_{w_1,w_2-1,w_3}=0$ \cite{Yu:2024kxr} thus the functions $\mathcal{X}_i$ are simply constants:
\begin{equation}\label{edgecase}
    \mathcal{X}_1=\mathcal{X}_2=-1, \qquad \mathcal{X}_3=1
\end{equation}
With these values of $\mathcal{X}_i$, one can easily see that \eqref{relamongXi} are the same as the edge condition \eqref{edgecondi}. From this fact, one may view the linear relation \eqref{relamongXi} as a generalized ``edge'' condition for the non-edge case (with $X_{123}=0$ provided).   
Besides  edge cases, we also give the simplest example of covering maps which is not edge to verify the  identities \eqref{mainiden1}.  The simplest such case is $w_1=w_2=w_3=1$, for which the covering map is also unique:
\begin{equation}
    \Gamma(z)=\frac{(\eta_1+\eta_2-2)(\eta_1z+\eta_2z-2\eta_1\eta_2)z}{(2z-\eta_1-\eta_2)(2\eta_1\eta_2-\eta_1-\eta_2)}.
\end{equation}
Also in this case  $a_1, a_2, a_3$ can be solved as  single-valued functions of $\eta_1, \eta_2$ (and thus are unique):
\begin{equation}
\begin{aligned}
     a_1&=\frac{2\eta_1\eta_2(\eta_1+\eta_2-2)}{(\eta_1+\eta_2)(2\eta_1\eta_2-\eta_1-\eta_2)}\\
     a_2&=\frac{2(\eta_1-1)(\eta_2-1)(\eta_1+\eta_2)}{(\eta_1+\eta_2-2)(2\eta_1\eta_2-\eta_1-\eta_2)}\\
     a_3&=\frac{2(2\eta_1\eta_2-\eta_1-\eta_2)}{(\eta_1+\eta_2)(\eta_1+\eta_2-2)}
\end{aligned}
\end{equation}
Using these expressions, one can easily verify the above identities \eqref{mainiden1}, as well as the linear relation \eqref{relamongXi}. Note that the case $w_1=w_2=w_3=1$ is special where the covering map $\Gamma(z)$ and $a_i(\eta_1,\eta_2)$ are all unique and  single-valued. These functions immediately become much more complicated (multi-valued and not unique) when we consider other cases.

Now we consider the term proportional to $k$. On the CFT side, this is the only term that contains the factor $C_1C_2$. Again making use of the identities \eqref{mainiden1}, we can translate the (to be verified) equality of the two sides:
\begin{equation}
   \mathbb{C}^{\text{string}}\left[k\right]\mathop{=}\limits^{?}\mathcal{R} \mathbb{C}_\Gamma^{\text{CFT}}\left[k\right]
\end{equation}
to the following (to be verified) mathematical identity:
\begin{equation}\label{idenforC1C2}
\begin{aligned}
-\frac{\alpha w_1^2\mathcal{X}_1^2}{\eta_1\eta_2} -\frac{\beta w_2^2\mathcal{X}_2^2}{(\eta_1-1)(\eta_2-1)}&-\gamma(w_3\mathcal{X}_3-1)^2\mathop{=}\limits^{?}
     \frac{2\eta_1\eta_2(\eta_1-1)(\eta_2-1)}{(\eta_1-\eta_2)^2}\\
     &\times\frac{w_1\mathcal{X}_1(\eta_1-1)(\eta_1-\eta_2)+w_2\mathcal{X}_2\eta_1(\eta_1-\eta_2)}{\eta_1(\eta_1-1)(\eta_1-\eta_2)}\\
     &\times \frac{w_1\mathcal{X}_1(\eta_2-1)(\eta_2-\eta_1)+w_2\mathcal{X}_2\eta_2(\eta_2-\eta_1)}{\eta_2(\eta_2-1)(\eta_2-\eta_1)}
\end{aligned}
\end{equation}
This identity can be easily verified by replacing $w_3\mathcal{X}_3-1$ with $-w_1\mathcal{X}_1-w_2\mathcal{X}_2$ (equation \eqref{relamongXi}).
Note that in the above equation, both sides are single valued functions of $a_1, a_2$ (due to \eqref{etatoX}). 
We can also verify the identity \eqref{idenforC1C2} analytically in the edge cases, since in the edge cases \eqref{relamongXi} holds obviously.

Finally, we deal with the terms proportional to $1$.  Making use of the identities \eqref{mainiden1}, we can simplify the coefficient $ \mathbb{C}_\Gamma^{\text{CFT}}[1]$ on the CFT side to be:
\begin{equation}
    \mathbb{C}_\Gamma^{\text{CFT}}[1]=-\frac{4}{(\eta_1-\eta_2)^2}-\frac{w_3\mathcal{X}_3-1}{2\eta_1(\eta_2-1)}-\frac{w_3\mathcal{X}_3-1}{2\eta_2(\eta_1-1)}
    +\frac{w_1\mathcal{X}_1}{2\eta_1\eta_2}+\frac{w_2\mathcal{X}_2}{2(\eta_1-1)(\eta_2-1)}.
\end{equation}
On the string side, one needs to know the terms $\partial_{a_1}\alpha$ and $\partial_{a_2}\beta$ in the last line of \eqref{irrgular1} (note that $(a_1,a_2)$ is identified with $(y_1,y_2)$), which come from  integration by parts. We observe the following identities for them:
\begin{equation}\label{partialiden}
\begin{aligned}
    \partial_{a_1}\alpha&=\frac{\eta_1\eta_2}{w_1a_1(\eta_1-\eta_2)^2}\Bigg(w_1\mathcal{X}_1(2-\eta_1-\eta_2)\left(\frac{4\eta_1\eta_2(\eta_1-1)(\eta_2-1)}{(\eta_1-\eta_2)^2}+1\right)\\
    &\hspace{1cm}+(1-w_2\mathcal{X}_2)\left(\frac{4\eta_1\eta_2(\eta_1-1)(\eta_2-1)(\eta_1+\eta_2)}{(\eta_1-\eta_2)^2}+\eta_1\eta_2\right)\Bigg)\\
    \partial_{a_2}\beta&=\frac{(\eta_1-1)(\eta_2-1)}{w_2a_2(\eta_1-\eta_2)^2}\Bigg(w_2\mathcal{X}_2 (\eta_1+\eta_2)\left(\frac{4\eta_1\eta_2(\eta_1-1)(\eta_2-1)}{(\eta_1-\eta_2)^2}+1\right)\\
    &\hspace{1cm}+(1-w_1\mathcal{X}_1)\left(\frac{4\eta_1\eta_2(\eta_1-1)(\eta_2-1)(2-\eta_1-\eta_2)}{(\eta_1-\eta_2)^2}+(\eta_1-1)(\eta_2-1)\right)\Bigg)
\end{aligned}
\end{equation}
We have verified these identities in the ancillary $\mathtt{Mathematica }$ notebook.  Note that the above ways to express $\partial_{a_1}\alpha$ and $\partial_{a_2}\beta$ are also not unique due to the linear relation \eqref{relamongXi}. In addition, we find that  the above identities only hold when all $w_i$ $(i=1,2,3)$ are odd, while other identities observed in this work  generally hold (as long as $\sum_{i}w_i$ is odd).  It is interesting that one can understand this mathematical fact naturally from the (supersymmetric) AdS$_3/$CFT$_2$ model considered here: on both sides the parity (odd) of $w_i$ comes from the physical condition of the considered operators. On the string side, $w$ must be odd for $O_{j,h}^w$ due to the GSO projection; on the CFT side, $w$ must be odd since the lifted operator $\mathbb{V}_\alpha$ is in the NS sector on the covering surface \cite{Lunin:2001pw}.  For $w$ even, one needs to consider other operators and their correlators will be different from the one studied in this work (see \cite{Yu:2024kxr}).    Note that the two sides of the  equation \eqref{partialiden} are also single valued functions of $a_1,a_2$. Then, by \eqref{etatoX} one can write \eqref{partialiden} in a form that only involves  $X_i$ and $\tilde{X}_i$  but no covering map is needed to construct.  Thus, 
 this identity could be understood as a mathematical relation among $X_i$ and $\tilde{X}_i$, with the condition $X_{123}=0$ provided.

 Making use of the identities \eqref{mainiden1},
 the (to be verified) equality of the two sides:
\begin{equation}
   \mathbb{C}^{\text{string}}\left[1\right]\mathop{=}\limits^{?}\mathcal{R} \mathbb{C}_\Gamma^{\text{CFT}}\left[1\right]
\end{equation}
predicts the following (to be verified) mathematical identity:
\begin{equation}
\begin{aligned}
    &-\frac{w_1a_1\partial_{a_1}\alpha}{\eta_1\eta_2}-\frac{w_2a_2\partial_{a_2}\beta}{(\eta_1-1)(\eta_2-1)}-\frac{\alpha w_1\mathcal{X}_1}{\eta_1\eta_2}-\frac{\beta w_2\mathcal{X}_2}{(\eta_1-1)(\eta_2-1)}
    -\frac{\alpha }{ 2\eta_1\eta_2}-\frac{ \beta }{2(\eta_1-1)(\eta_2-1)}\\
    &+\gamma w_3\mathcal{X}_3-2\gamma
     \xlongequal{?}\frac{2\eta_1\eta_2(\eta_1-1)(\eta_2-1)}{(\eta_1-\eta_2)^2}\\
     &\hspace{5em} \times \left(-\frac{4}{(\eta_1-\eta_2)^2}-\frac{w_3\mathcal{X}_3-1}{2\eta_1(\eta_2-1)}-\frac{w_3\mathcal{X}_3-1}{2\eta_2(\eta_1-1)}
    +\frac{w_1\mathcal{X}_1}{2\eta_1\eta_2}+\frac{w_2\mathcal{X}_2}{2(\eta_1-1)(\eta_2-1)}\right)
\end{aligned}
\end{equation}
From \eqref{partialiden} one can easily verify the above identity (in the above identity, the terms proportional to $w_3\mathcal{X}_3$ on the two sides happen to  cancel, so we do not need to use the linear relation \eqref{relamongXi}).

\section{Comments on the matching}\label{more Comments}

In this section, we give  some  comments on the matching.

\subsection{Uniqueness of the deforming operator}
From our conformal perturbation computation on the CFT side, we have found that the 4 terms in \eqref{The4terms} give identical contributions. Thus, it seems that to reproduce the string result, the marginal operator could have different forms. For example, it could only contain 2 terms:
\begin{equation}\label{fake}
    \Phi(x)=G^{++}_{-\frac{1}{2}}\Psi^{--}(x)+G^{--}_{-\frac{1}{2}}\Psi^{++}(x)
\end{equation}
The matching at the second order $(m=2)$ studied in this work can not rule out such possibilities. 

One can see that the  operator \eqref{fake} is no longer a singlet of $SU(2)_{\text{R}}\oplus SU(2)_{\text{outer}}$, so intuitively it should not be the correct marginal operator\footnote{This means the global symmetries should fix the ambiguity in identifying the deforming operator. In particular, if the operator was not a singlet with respect to $SU(2)_{\text{R}}\oplus SU(2)_{\text{outer}}$, it would affect the correlators of states that are not singlets either in different ways --- and this is probably not consistent with what one knows from the string background. We thank Matthias Gaberdiel for a discussion on this point.}. Note that in the twist 2 sector, there are in total 16 degenerate ground states. With respect to  $SU(2)_{\text{R}}\oplus SU(2)_{\text{outer}}$, they form:
\begin{equation}
\left[\textbf{(2,1)}\oplus\textbf{(1,2)}\right]\otimes\left[\textbf{(2,1)}\oplus2\textbf{(1,1)}\right]=\textbf{(3,1)}\oplus\textbf{(1,1)}\oplus\textbf{(2,2)}\oplus 2\textbf{(2,1)}\oplus2\textbf{(1,2)}
\end{equation}
where the representations in the first and second square brackets comes from the  ground states of Ramond fermions  in the $\mathcal{N}=4$ linear dilaton and the $\mathbb{T}^4$ theory respectively (note that all  states in the  second square brackets are singlets of $SU(2)_{\text{outer}}$, since  $SU(2)_{\text{outer}}$ is constructed by the bosons in the  theory $\mathbb{T}^4$). Then since the 4 supercurrents form a $\textbf{(2,2)}$,  their action on the above ground states will  give only one singlet, which is just the  marginal operator proposed in \cite{Eberhardt:2021vsx}.
Quantitatively, one could also try to rule out  operators like \eqref{fake} as the marginal operator by considering the matching of other correlators (for example, the ones calculated in \cite{Yu:2024kxr}).

\subsection{General backgrounds AdS$_3\times X$ }\label{generalbackground}

Our calculation could be generalized to the CFT duals of superstrings on  general backgrounds AdS$_3\times X$ with few modifications. In fact, our calculation on the string side is universal (that is,  it does not depend on  $X$), since the string correlator (and its residues) studied in the present work (see section \ref{Thestringcorrelator}) only depends on the supersymmetric AdS$_3$ part. However, the calculation on the CFT side in section \ref{TheCFTside} is based on the specific seed theory \eqref{undeformed} and the marginal operator \eqref{Themarginal} so is not universal. Concretely, while the operator $\mathbb{V}^{(w)}_\alpha$ in \eqref{theoperator} depends only on the linear dilaton theory when lifted up to the covering surface (so is universal), the form of the marginal operator does depend on the other part ($X$) of the seed theory\footnote{For the calculation at the 0-th order in \cite{Yu:2024kxr}, the CFT result is also universal since no marginal operator is inserted in the conformal perturbation computation. Thus, the matching found at the 0-th order in \cite{Yu:2024kxr} holds for all superstrings on  general backgrounds AdS$_3\times X$. }.  Nevertheless,  our calculation in fact shows the key for the matching for  general backgrounds AdS$_3\times X$. One can see this as follows.

 Since the residue \eqref{stringresult} of the string correlator is universal, it should be reproduced by the second order conformal perturbation computation of the CFT dual of any superstring on  AdS$_3\times X$.  In practice, this is not very helpful for finding (or verifying) the marginal operator on the CFT side, since \eqref{stringresult} is far from being a result of conformal perturbation.  Thanks to the mathematical identities observed in the present work and \cite{Eberhardt:2021vsx}, we have shown  in section \ref{matchthetwosides}   that \eqref{stringresult} could indeed be translated into a form \eqref{CFTresult1}, which has the correct structure of  the conformal perturbation computation of a symmetric orbifold theory\footnote{Of course, the calculation we did in section \ref{TheCFTside} shows that it is indeed the answer of conformal perturbation  for the CFT dual of the superstring on AdS$_3\times$S$^3\times\mathbb{T}^4$. Besides, when rewriting \eqref{stringresult} as   an equivalent form \eqref{CFTresult1}, we need to fix the quantities $\alpha,\beta,\gamma$ as in \eqref{alphabetagamma1}, which is expected to hold in the  cases of superstrings on general backgrounds  AdS$_3\times X$.  }.  This means that in the undeformed symmetric orbifold theory, the five point function $\mathcal{I}$ in \eqref{result5pt} should be universal. Then we have the following requirements for the marginal operator of the CFT dual of superstring theory on AdS$_3\times X$:
    \begin{itemize}
        \item Assume $\Phi_{X}$ is the deforming operator that defines the CFT dual of the superstring on  AdS$_3\times X$ with pure NS-NS flux. Then in the undeformed theory, the five point function 
        \begin{equation}
           \mathcal{I}_X\equiv \left\langle  V^{(w_1)}_{\alpha_1}(0) V^{(w_2)}_{\alpha_2}(1) V^{(w_3)}_{\alpha_3}(\infty)\Phi_{X}(\xi_1)\Phi_X(\xi_2)\right\rangle
        \end{equation}
        should equal to $\mathcal{I}$ in \eqref{result5pt} (up to normalization) for all $X$.  In other words, if we view the covering map related function $C_i$ $(i=1,2)$ as variables in  $\mathcal{I}$, then the coefficients of $1$, $C_1$, $C_2$, $C_1C_2$ in \eqref{result5pt}  should be reproduced by  lifted correlators of $\mathcal{I}_X$, which are simply some 5-point functions in the seed theory.
    \end{itemize}
    The above requirement in particular implies that the 
      marginal operator should not be a primary when lifted up to the covering surface, since only in this way $C_1$ $C_2$ could appear. From our calculation in section \ref{TheCFTside}, for general $X$ 
 the marginal operator presumably could also be a linear combination of super-descendents of some states in the twist 2 sector, which contains a bosonic vertex operator $e^{-\sqrt{\frac{k}{2}}\phi}$ when lifted up to the covering surface. For superstrings that have at least $\mathcal{N}=2$ space-time superconformal symmetry, they could be some linear combinations of super-descendents of  BPS operators in the twist 2 sector. In fact, in \cite{Balthazar:2021xeh,Chakraborty:2025nlb}, such  marginal operators had been proposed for (a large class of) general $\mathcal{N}=2$ superconformal string backgrounds \cite{Giveon:1999jg,Berenstein:1999gj,Giveon:2003ku}. We believe that our calculation  could be generalized to the $\mathcal{N}=2$ dualities in \cite{Balthazar:2021xeh,Chakraborty:2025nlb} without many difficulties. In particular,   generalizing our calculation to the (small and large) $\mathcal{N}=4$  dualities of AdS$_3\times$S$^3\times$K3 and AdS$_3\times$S$^3\times$S$^3\times$S$^1$ could be more straightforward. In the following, we give a simple computation for the more general $\mathcal{N}=2$ cases as an illustration. 

We first review the general procedure for constructing superstring theory on AdS$_3\times X$ with  (at least) $\mathcal{N}=2$ space-time supersymmetry, following \cite{Chakraborty:2025nlb}.  
Consider a string theory on
\begin{equation}
    \text{AdS}_3\times\mathbb{S}^1\times\mathcal{M}
\end{equation}
where $\mathcal{M}$ is an $\mathcal{N}=2$ SCFT. Then the requirement of a space-time $\mathcal{N}=2$ superconformal
algebra leads to a chiral GSO projection, which acts on $\mathbb{S}^1\times\mathcal{M}$ as a $\Gamma$-orbifold ($\Gamma$ is a discrete group). Thus we take  $X=(\mathbb{S}^1\times\mathcal{M})/\Gamma$ and the superstring theory on AdS$_3\times X$ has $\mathcal{N}=2$ space-time supersymmetry. Then the CFT dual is a deformed symmetric orbifold theory, with the seed theory being
\begin{equation}
    \mathbb{R}_Q^{(1)}\times X,
\end{equation}
where $\mathbb{R}_Q^{(1)}$ is an $\mathcal{N}=1$ linear dilaton with background charge $Q$. Furthermore, one can combine the $U(1)$ current (denoted as $i\partial Y$) and its associated fermion  in $X$ with $\mathbb{R}_Q^{(1)}$ to obtain an $\mathcal{N}=2$ linear dilaton, which has the following supercurrents\footnote{Compared to the convention in \cite{Balthazar:2021xeh}, we have $Q_{\text{their}}=-\sqrt{2}Q_{\text{ours}}$ (our convention is the same as in the bosonic analysis \cite{Eberhardt:2021vsx}). Besides, since our convention for the R-charge $q$ of a chiral operator is $q=h$,  our  $J_{LD}$ is half of the one in \cite{Balthazar:2021xeh}.  }:
\begin{equation}
\begin{aligned}
    G^+_{LD}&=\psi^+(\partial\phi-i\partial Y)-\sqrt{2}Q\partial\psi^+\\
    G^-_{LD}&=\psi^-(-\partial\phi-i\partial Y)+\sqrt{2}Q\partial\psi^-,
\end{aligned}
\end{equation}
and the R-current:
\begin{equation}
    J_{LD}=\frac{1}{2}\psi^+\psi^-+\frac{\sqrt{2}}{2}Qi\partial Y,
\end{equation}
where we have recombined the 2 fermions as a complex fermion $\psi^\pm$.
The deforming operator proposed in \cite{Balthazar:2021xeh,Chakraborty:2025nlb} is  a linear combination of  super-descendents of the BPS states in the twist 2 sector (we only write the the left-moving part)\footnote{Note that if the string theory has more supersymmetries (supercurrents) than $\mathcal{N}=2$, one needs to further consider the global symmetries to construct the marginal operator (as a linear combination of super-descendents of BPS operators). We have seen such a construction in the case of  superstring on AdS$_3\times$S$^3\times\mathbb{T}^4$.}:
\begin{equation}\label{marginalforN=2}
    \Phi'\equiv G^+_{-\frac{1}{2}}\Sigma^-+G^-_{-\frac{1}{2}}\Sigma^+, 
\end{equation}
where $\Sigma^\pm$ are two BPS states with $h=\frac{1}{2}$, and their R-charges are $q=\pm\frac{1}{2}$. There are in fact two ways to describe these BPS states \cite{Chakraborty:2025nlb}: directly write them in the $\mathbb{Z}_2$ twisted sector or write them by their lifted operator on the covering surface. The former form is obtained in  \cite{Balthazar:2021xeh,Chakraborty:2025nlb}:
\begin{equation}\label{BPSforN=2}
    \Sigma^{\pm}=e^{-\frac{1}{2\sqrt{k}}(\phi_S\mp iY_S)}\left(\sigma_{\phi_A}\sigma_{Y_A}\sigma_{\psi_A}^\pm\right)\Sigma^\pm_\mathcal{M}
\end{equation}
where the subscripts ``S'' and  `` A'' denote the symmetric and antisymmetric combinations of the two copies of operators respectively.  The operators $\sigma$ in the brackets  are the  $\mathbb{Z}_2$ twist operators for various free fields, and $\Sigma^\pm_\mathcal{M}$ is the $\mathbb{Z}_2$ twist operator that
creates the BPS twisted ground states in $\mathcal{M}$.
Here we need to write \eqref{BPSforN=2} in the latter form, since we always use covering maps to calculate  correlators.  Thus $\Sigma^{\pm}$ could be write in terms of its lifted operator $\hat{\Sigma}^{\pm}$ as:
\begin{equation}\label{BPSforN=2 2}
    \Sigma^{\pm}=\hat{\Sigma}^\pm \sigma_2 \qquad \text{with} \quad \hat{\Sigma}^\pm=e^{-\sqrt{\frac{k}{2}}\phi}e^{\pm\frac{i}{\sqrt{2k}}Y}\mathbb{S}^\pm\hat{\Sigma}^\pm_\mathcal{M},
\end{equation}
where $\mathbb{S}^\pm$ are the 2 spin fields for the complex fermions $\psi^\pm$ in the Ramond sector and $\hat{\Sigma}^\pm_\mathcal{M}$  denote the lifted operators of $\Sigma^{\pm}_{\mathcal{M}}$.  One can verify \eqref{BPSforN=2 2} by checking that the weights and the R-charges of the operators in the free theories  $\phi,Y,\psi^\pm$ in \eqref{BPSforN=2} and \eqref{BPSforN=2 2} indeed coincide.  
Note the bosonic part of the lifted vertex operator in the linear dilaton theory is $e^{-\sqrt{\frac{k}{2}}\phi}$, the same as in the CFT dual of AdS$_3\times$S$^3\times\mathbb{T}^4$.

  Now we do the conformal perturbation computation. We can lift the five-point function up to the covering surface, just as in the computation we did in section \ref{TheCFTside}. Then we get the same form as \eqref{result5pt}, but now the 4 coefficients of $1$, $C_1$, $C_2$, $C_1C_2$ should be calculated using $\Phi'$ in \eqref{marginalforN=2}. We do not intend to do the full calculation but only calculate the coefficient of $C_1C_2$ as an example. This coefficient should be the following lifted correlator:
 \begin{equation}
     \left\langle\hat{\mathbb{V}}_{\alpha_1}(0)\hat{\mathbb{V}}_{\alpha_2}(1)\hat{\mathbb{V}}_{\alpha_3}(\infty) \left(G^+_{0}\hat{\Sigma}^-+G^-_0\hat{\Sigma}^+\right)(\eta_1)\left(G^+_{0}\hat{\Sigma}^-+G^-_0\hat{\Sigma}^+\right)(\eta_2)\right\rangle.
 \end{equation}
 The result should be the same as  \eqref{Thetestterm} (up to normalization), which is proportional to $k$. Thus, we need to show that $G_0^{\pm}\hat{\Sigma}^{\mp}$ is proportional to $\sqrt{k}$.  We stress that this is  not a matter of normalization because $\sqrt{k}$ will appear as the coefficient of $C_1C_2$ in \eqref{result5pt} so is not an overall factor. Since the full supercurrents are summations of the ones in the linear dilaton theory and $\mathcal{M}$: $G^{\pm}_0=(G_{LD}^\pm)_0+(G_{\mathcal{M}}^\pm)_0$, we calculate the action of these two parts in order. For $(G_{LD}^\pm)_0\hat{\Sigma}^{\mp}$ we have:
 \begin{equation}
 \begin{aligned}
      (G_{LD}^\pm)_0\hat{\Sigma}^{\mp}&=\left[\big(\pm(\partial\phi)_0-(i\partial Y)_0\big)\psi^\pm_0\mp\sqrt{2}Q(\partial\psi^\pm)_0\right]\hat{\Sigma}^{\mp}\\
     & =\left[\pm\sqrt{\frac{k}{2}}\pm\frac{1}{\sqrt{2k}}\mp\sqrt{2}\left(\sqrt{k}-\frac{1}{\sqrt{k}}\right)\frac{-1}{2}\right]\psi^\pm_0\hat{\Sigma}^{\mp}\\
     &=\pm\sqrt{2k}\psi^\pm_0\hat{\Sigma}^{\mp}.
 \end{aligned}
 \end{equation}
Thus, this part is already proportional to $\sqrt{k}$. For $(G_{\mathcal{M}}^\pm)_0\hat{\Sigma}^{\mp}$, we will show that it is in fact null. 
The norm of  $(G_{\mathcal{M}}^\pm)_0\hat{\Sigma}^{\mp}$ is:
\begin{equation}\label{Thennorm}
\begin{aligned}
    \big|\big|(G_{\mathcal{M}}^\pm)_0\hat{\Sigma}^{\mp}\big|\big|^2= &\Big\langle\hat{\Sigma}^{\pm}\Big|\left\{\left(G_{\mathcal{M}}^\mp\right)_0,\left(G_{\mathcal{M}}^\pm\right)_0\right\}\Big|\hat{\Sigma}^{\mp}\Big\rangle\\
     =&\Big\langle\hat{\Sigma}^{\pm}\Big|\left(2L_{\mathcal{M}}-\frac{c_\mathcal{\mathcal{M}}}{12}\right)\Big|\hat{\Sigma}^{\mp}\Big\rangle\\
     =&\Big\langle\hat{\Sigma}^{\pm}\Big|\left(2h(\hat{\Sigma}^\pm_\mathcal{M})-\frac{c_\mathcal{\mathcal{M}}}{12}\right)\Big|\hat{\Sigma}^{\mp}\Big\rangle=0
\end{aligned}
\end{equation}
where in the second line we used the $\mathcal{N}=2$ superconformal algebra in $\mathcal{M}$. The reason for the finial equality in \eqref{Thennorm} is as follows. The weight of  $\Sigma^\pm_\mathcal{M}$ is \cite{Klemm:1990df,Fuchs:1991vu}:
\begin{equation}
   h(\Sigma^\pm_\mathcal{M})=\frac{c_\mathcal{M}}{12}.
\end{equation}
On the other hand,
it could also be calculated as:
\begin{equation}
    h(\Sigma^\pm_\mathcal{M})=\frac{c_\mathcal{M}}{24}\left(2-\frac{1}{2}\right)+\frac{h(\hat{\Sigma}^\pm_\mathcal{M})}{2}.
\end{equation}
Then we immediately get $h(\hat{\Sigma}^\pm_\mathcal{M})=\frac{c_\mathcal{M}}{24}$, so the last equality in \eqref{Thennorm} holds.

\section{Discussion}\label{discussion}

In this work, we study the  CFT dual of superstring theory on AdS$_3\times$ S$^3\times \mathbb{T}^4$ with pure NS-NS flux. We have verified the proposed marginal operator in \cite{Eberhardt:2021vsx} (which deforms the symmetric orbifold theory on the CFT side) by matching the residues of correlators of the two sides. On the string side, the residue could be read from the  the superstring 3-point correlator obtained in \cite{Yu:2024kxr}, and the result is written as an integral over the $y$-variables introduced in \cite{Dei:2021xgh}; on the CFT side, we need to do a conformal perturbation computation and the result can be written as an integral over the coordinates of the deforming operators.  Similar to the bosonic case,   we have not tried to work out these integrals  explicitly. Instead, we directly match the integral form of these residues of the two sides, i.e. the integration region and integrands of the two integrals are shown to be the same. 

The main differences between the bosonic case  \cite{Eberhardt:2021vsx} and the SUSY case studied here are as follows. Firstly, one can not naturally obtain a unique integral form of the  residue of the string correlator, since  different picture choices will give different integrands.   It turns out that the correct integrand is not  one that comes from any specific picture choice but is rather a linear combination of them. Secondly, to match with the CFT result, we need to use the mass shell condition, which does not play a role in the matching for the bosonic case. Finally, we observe that the marginal operator is no longer a primary when lifted up to the covering surface. Then the correlator involves both the first and the second non-trivial Taylor coefficients of the covering maps at their ramified points. The matching of the two sides then predicts novel mathematical identities that relate these data with the functions $X_i$ (defined in \cite{Dei:2021xgh}) and their ``conjugate'' $\tilde{X}_i$ (defined  in this work).

We discuss several possible future research directions. Since in this work we focus on the matching of residues at the second order, it would  be desirable to generalize the result to all orders or even to the higher point case.  This is in fact achieved in the bosonic case, where the near-boundary approximation \cite{Knighton:2023mhq,Knighton:2024qxd} was employed to show the matching at all orders. Thus, for the SUSY case one can either extract the residues from the exact string correlators (just like what we do in the present work) or use the near-boundary approximation. To complete the matching of arbitrary $n$-point correlators at all orders, it seems that the near-boundary approximation is more practicable since exact formulas for higher point string correlators are hard to work out.  The discussion in the present work gives several  lessons that may be crucial for the generalization. First,  the correct integral form of the residue on the string side is not simply the one calculated from a specific picture choice, which is probably also true if one uses the near-boundary approximation to get the residues\footnote{One may bypass the complexity of linearly combining different picture choices by working in superspace on both sides, which probably would make the matching a bit more natural. We thank Lorenz Eberhardt for bringing us this point of view. }. Second, it is very likely that the mass-shell condition  also plays a role in the method of near-boundary approximation (while it plays no role in the bosonic case). Third, the conformal perturbation computation at the second order we did in this work is a good illustration for  performing the computation  at arbitrary orders. In particular, it will be interesting to see whether the near-boundary approximation could match the two integrals of the two sides without involving the complexity that comes from covering maps\footnote{In the bosonic case, the answer is yes. In particular, in the method of  near-boundary approximation, one does not need to know the mathematical identities of covering maps observed in \cite{Eberhardt:2021vsx}. In the SUSY case, this may be subtle, since the  fermionic correlator on the worldsheet involves the function $P_{w_1w_2w_3}$ \cite{Yu:2024kxr}, which are closely related to the data of the relevant covering maps. }.

As we have discussed in section \ref{generalbackground}, our result could be generalized to  the CFT duals of superstrings on general backgrounds AdS$_3\times X$. In particular, it will not be difficult to generalize our calculation to the backgrounds AdS$_3\times$S$^3\times$K3 and AdS$_3\times$S$^3\times$S$^3\times$S$^1$. The marginal operators for these cases could be found in \cite{Eberhardt:2021vsx,Chakraborty:2025nlb}, see also \cite{Sriprachyakul:2024gyl,Gaberdiel:2024dva} for related discussions.  One  could also try to generalize our result to the more general  $\mathcal{N}=2$ AdS$_3/$CFT$_2$ dualities proposed in \cite{Chakraborty:2025nlb}, for which we have successfully matched the terms proportional to $C_1C_2$ of the two sides in section \ref{generalbackground}. For general superstring backgrounds AdS$_3\times X$, unlike in the bosonic case, the marginal operator on the CFT side will involve the internal CFT $X$. Nevertheless, one probably could perform  the conformal perturbation computation in such a way that  the result is valid for all $X$   (similar to our computation in section \ref{generalbackground}, which is valid for all $\mathcal{M}$). In addition, it will be interesting to  verify these dualities by matching the (residues of) higher genus correlators, which are only well defined for the superstring case.

Finally, notice that our work, together with \cite{Eberhardt:2021vsx} (see also \cite{Dei:2022pkr,Yu:2024kxr}), uncover  surprising connections between  covering maps and the special functions $X_I$ ($\tilde{X}_I$) that appear as building blocks in the exact formula of correlators in the $SL(2,\mathbb{R})$ WZW model \cite{Dei:2021xgh,Dei:2021yom}. Thus, it would also be interesting to further explore these  connections, which could play a role in matching the residues of other correlators  of  the two sides.  These connections may also be helpful for the study of general symmetric orbifold CFTs, since covering maps are crucial for computing correlators in these theories \cite{Baggio:2015jxa}.

\section*{Acknowledgments}
I would like to thank Cheng Peng for initial collaboration. I also thank Matthias Gaberdiel, Sergio Iguri, Nicolas Kovensky, Cheng Peng, Juli\'{a}n H. Toro for useful discussions.   I am  grateful to Lorenz Eberhardt, Nicolas Kovensky and Matthias Gaberdiel for their valuable comments on a draft of this paper. This work is supported by the Postdoctoral Fellowship Program of CPSF under Grant Number GZC20241685 and  funds from the UCAS program of Special Research Assistant.

\appendix

\section{Conventions for the seed theory}\label{seedthoryconvention}

For a bosonic   linear dilaton $\phi$ with background charge $Q$, the defining OPE of the $U(1)$ current $i\partial\phi$ is:
\begin{equation}
    i\partial\phi(z)i\partial\phi(w)\sim \frac{1}{(z-w)^2}. 
\end{equation}
In the presence of the background charge $Q$, the stress-energy tensor is modified to be:
\begin{equation}
    T(z)=-\frac{1}{2}:\partial\phi\partial\phi:(z)+\frac{1}{\sqrt{2}}Q\partial^2\phi(z).
\end{equation}
Consequently, the central charge is modified as:
\begin{equation}
    c=1+6Q^2
\end{equation}
Then a vertex operator $e^{\sqrt{2}\alpha\phi}$ has conformal weight:
\begin{equation}
    h(e^{\sqrt{2}\alpha\phi})=\alpha(Q-\alpha).
\end{equation}
An  $\mathcal{N}=4$  linear dilaton theory  with background charge $Q$ has 
 generating fields: $i\partial\phi$,   $j^i$ $(i=3,\pm)$, $\psi^a$ $(a=0,1,2,3)$,  the OPEs among them are:
\begin{equation}
\begin{aligned}
       i\partial\phi(z)i\partial\phi(w)&\sim \frac{1}{(z-w)^2}, \\
       \psi^{a}(z)\psi^{b}(w)&\sim \frac{\delta^{ab}}{z-w},\\
       j^a(z)j^b(w)&\sim \frac{\frac{k-2}{2}\delta^{ab}}{(z-w)^2}+\frac{i{\epsilon^{ab}}_cj^c(w)}{z-w},\\
\end{aligned}
\end{equation}
For the theory $\mathbb{T}^4$, the  generating fields are two complex bosons $X^a, X^{a\dagger}$ $(a=1,2)$ and two complex fermions $\lambda^b, \lambda^{b\dagger}$ $(b=1,2)$. The OPEs among them are:
\begin{equation}
\begin{aligned}
    X^a(z)X^{b\dagger}(w)&\sim \delta^{ab}\text{log}(z-w),\\
    \lambda^a(z)\lambda^{b\dagger}(w)&\sim \frac{\delta^{ab}}{z-w}.
\end{aligned}
\end{equation}

\section{The $\mathcal{N}=4$ superconformal currents}\label{ThefullN=4algebra}
In this section, we give the construction of the $\mathcal{N}=4$ superconformal currents in the seed theory. Since the seed theory is a direct sum of an $\mathcal{N}=4$ linear dilaton and $\mathbb{T}^4$, we write the superconformal currents in these 2 theories in order.

\subsection{The $\mathcal{N}=4$   currents  in the $\mathcal{N}=4$ linear dilaton}\label{TheN=4algebra}

 In fact, the construction of the $\mathcal{N}=4$ currents is known in the literature (e.g. in \cite{Eguchi:2016cyh}). The chiral algebra of a general $\mathcal{N}=4$ linear dilaton theory is a large $\mathcal{N}=4$ superconformal algebra generated by $\{T, G^a, J^{\pm,i}, Q^a, U\}$ ($a=0,1,2,3$, $i=1,2,3$), defined as:
\begin{equation}
\begin{aligned}
    T&=-\frac{1}{2}\partial\phi\partial\phi+\frac{\sqrt{2}Q}{2}\partial^2\phi-\frac{1}{2}\psi^a\partial\psi^a+\frac{1}{k}j^ij^i,  \\
    G^0&=i\psi^0\partial\phi+\sqrt{\frac{2}{k}}\left(\psi^ij^i-i\psi^1\psi^2\psi^3\right)-i\sqrt{2}Q\partial\psi^0, \\
    G^i&=i\psi^i\partial\phi-\sqrt{\frac{2}{k}}\left(\psi^0j^i+\epsilon_{ijk}\phi^j\psi^k-\frac{i}{2}\epsilon_{ijk}\psi^j\psi^k\psi^0\right)-i\sqrt{2}Q\partial\psi^i,\\
    J^{+,i}&=j^{+,i}_f+j^i, \quad J^{-.i}=j^{+,i}_f, \\
    \Big(j^{+,i}_f&=-\frac{i}{2}\psi^i\psi^0-\frac{i}{4}\epsilon_{ijk}\psi^j\psi^k+j^i,\quad j^{-,i}_f=\frac{i}{2}\psi^i\psi^0-\frac{i}{4}\epsilon_{ijk}\psi^j\psi^k\Big),\\
    U&=\sqrt{\frac{k}{w}}i\partial\phi,\quad Q^a=\sqrt{\frac{k}{w}}\psi^a.
\end{aligned}
\end{equation}
Note that for the background charge,  compared with the convention in \cite{Eguchi:2016cyh} we have $Q_{\text{their}}=-\sqrt{2}Q_{\text{ours}}$. The case at hand  is special: for $Q=\frac{k-1}{\sqrt{k}}$ one can check that $J^{-,i}, Q^a, U$  decouple and  the large $\mathcal{N}=4$ superconformal algebra reduces to the small $\mathcal{N}=4$ superconformal algebra, generated by $\{T, G^a, J^{+,i}\}$. As for the generators $J^{-,i}=j^{+,i}_f$, their zero modes in fact generate the outer automorphism $SU(2)_{\text{outer}}$ of the small $\mathcal{N}=4$ algebra.

Now we write the  small $\mathcal{N}=4$  currents in a basis where  supercurrents  are labeled by 2 spinor indices, that is,  $G^a\to G^{\alpha\beta}$, where $\alpha,\beta$ are spinor indices with respect to $SU(2)_{\text{R}}\oplus SU(2)_{\text{outer}}$ (the automorphism $SU(2)_{\text{R}}$ and $SU(2)_{\text{outer}}$ are  generated by the zero modes of $J^{+,i}$ and $J^{-,i}$ respectively). For this purpose, we recombine the 4 Majorana
fermions as 2 complex fermions:
\begin{equation}
\begin{aligned}
    \Psi^{(1)}&=\frac{1}{\sqrt{2}}(\psi^1+i\psi^2),\quad \Psi^{(1)\dagger}=\frac{1}{\sqrt{2}}(\psi^1-i\psi^2),\\
    \Psi^{(2)}&=\frac{1}{\sqrt{2}}(\psi^3+i\psi^0),\quad \Psi^{(2)\dagger}=\frac{1}{\sqrt{2}}(\psi^3-i\psi^0).\\
\end{aligned}
\end{equation}
Then we rename them as:
\begin{equation}
    \psi^{++}\equiv\Psi^{(1)}, \quad \psi^{--}\equiv\Psi^{(1)\dagger},\quad \psi^{+-}\equiv\Psi^{(2)}, \quad \psi^{-+}\equiv-\Psi^{(2)\dagger}
\end{equation}
Note that to obtain the correct OPE of $\psi^{+-}\psi^{-+}$, in the above we add  a extra sign in the definition of $\psi^{-+}$. Accordingly, we recombine the current $J^{\pm,i}$ as $ J^{\pm,\pm}=J^{\pm,1}\pm iJ^{\pm,2}$ (with $J^{\pm,3}$ unchanged) and  the fermionic currents then have the following form:
\begin{equation}
\begin{aligned}
     j^{+,+}_f&=\psi^{+-}\psi^{++},\quad j^{+,-}_f=\psi^{-+}\psi^{--}, \quad j^{+,3}_f=\frac{1}{2}(\psi^{++}\psi^{--}+\psi^{-+}\psi^{+-})\\
      j^{-,+}_f&=\psi^{-+}\psi^{++},\quad j^{-,-}_f=\psi^{+-}\psi^{--}, \quad j^{-,3}_f=\frac{1}{2}(\psi^{++}\psi^{--}-\psi^{-+}\psi^{+-})
\end{aligned}
\end{equation}
Then one can easily check $\psi^{\alpha\beta}$ ($\alpha, \beta=\pm$) form a $(\textbf{2},\textbf{2})$ of $SU(2)_{\text{R}}\oplus SU(2)_{\text{outer}}$. 
They can be compactly written as:
\begin{equation}
     j^{+,a}_f=\frac{1}{4}(\sigma^a)_{\alpha\gamma}\epsilon_{\beta\delta}(\psi^{\alpha\beta}\psi^{\gamma\delta}), \qquad j^{-,a}_f=\frac{1}{4}\epsilon_{\alpha\gamma}(\sigma^a)_{\beta\delta}(\psi^{\alpha\beta}\psi^{\gamma\delta}),
\end{equation}
where $\sigma^a (a=\pm,3)$ are the Pauli matrices with non-zero elements:
\begin{equation}
    (\sigma^-)_{--}=2, \quad  (\sigma^3)_{+-}=1, \quad (\sigma^3)_{-+}=1, \quad (\sigma^+)_{++}=-2
\end{equation}
Then we defined $G^{\alpha\beta}$ as
\begin{equation}
\begin{aligned}
    G^{++}&=\frac{1}{\sqrt{2}}(G^1+iG^2), \quad  G^{--}=\frac{1}{\sqrt{2}}(G^1-iG^2),\\
    G^{++}&=\frac{1}{\sqrt{2}}(G^3+iG^0), \quad  G^{--}=-\frac{1}{\sqrt{2}}(G^3-iG^0).
\end{aligned}
\end{equation}
So their expressions are:
\begin{equation}
\begin{aligned}
    G^{++}&=i\partial\phi\psi^{++}+i\sqrt{\frac{2}{k}}\left(j^+\psi^{-+}+(j^3-\psi^{+-}\psi^{-+})\psi^{++}-(k-1)\partial\psi^{++}\right),\\
    G^{--}&=i\partial\phi\psi^{--}+i\sqrt{\frac{2}{k}}\left(j^-\psi^{+-}-(j^3-\psi^{+-}\psi^{-+})\psi^{--}-(k-1)\partial\psi^{--}\right),\\
      G^{+-}&=i\partial\phi\psi^{+-}+i\sqrt{\frac{2}{k}}\left(j^+\psi^{--}+(j^3+\psi^{++}\psi^{--})\psi^{+-}-(k-1)\partial\psi^{+-}\right),\\
    G^{-+}&=i\partial\phi\psi^{-+}+i\sqrt{\frac{2}{k}}\left(j^-\psi^{++}-(j^3+\psi^{++}\psi^{--})\psi^{-+}-(k-1)\partial\psi^{--}\right).
\end{aligned}
\end{equation}
They can be compactly written as:
\begin{equation}
     G^{\alpha\beta}=i\partial\phi\psi^{\alpha\beta}+i\sqrt{\frac{2}{k}}\left( {(\sigma_a)^\alpha}_{\gamma}\left(j^a+J^{+,a}_f\right)\psi^{\gamma\beta}-{(\sigma_a)^\beta}_\gamma J^{-,a}_f\psi^{\alpha\gamma}-(k-1)\partial\psi^{\alpha\beta}\right)
\end{equation}
where 
\begin{equation}
    {(\sigma_+)^+}_{-}=1, \quad  {(\sigma_3)^-}_{-}=-1, \quad {(\sigma_3)^+}_{+}=1, \quad {(\sigma_-)^-}_{+}=1.
\end{equation}

\subsection{The $\mathcal{N}=4$  currents in $\mathbb{T}^4$}

This theory has a small $\mathcal{N}=4$ superconformal symmetry with $c=6$, whose generators are $T, G^a, G^{a\dagger}, J^i$ $(a=1,2, \hspace{2mm} i=1,2,3)$. They have the following form \cite{David:2002wn}:
\begin{equation}\label{generatorT4}
\begin{aligned}
     T&=\sum_{i=1,2}\partial X^{i}\partial X^{i\dagger}+\frac{1}{2}\sum_{a=1,2}(\lambda^a\partial \lambda^{a\dagger}-\partial \lambda^a\lambda^{a\dagger})\\
     G^a&=\left[
     \begin{aligned}
         G^1\\
        G^2
     \end{aligned}
     \right]=\sqrt{2}\left[
     \begin{aligned}
         \lambda^1\\
         \lambda^{2}
     \end{aligned}
     \right]\partial X^{2}+
     \sqrt{2}\left[
     \begin{aligned}
         -\lambda^{2\dagger}\\
         \lambda^{1\dagger}
     \end{aligned}
     \right]\partial X^{1}\\
     J^i&=\frac{1}{2}\left[\lambda_1,\lambda_2\right]\sigma^i\left[
     \begin{aligned}
         \lambda^{1\dagger}\\
         \lambda^{2\dagger}
     \end{aligned}
     \right].
\end{aligned}
\end{equation}
The 4 supercurrents are $G^a$ $(a=1,2)$ and their conjugates $G^{a\dagger}$ $(a=1,2)$.
The algebra $SU(2)_\text{R}$ is generated by the zero modes of above R-currents $J^i$. Besides,
there are 2 global $SU(2)$ symmetries which correspond to the $SO(4)$ rotations of the 4 bosons, whose generators are denoted as $I^i_1$ and $I^i_2$ respectively in \cite{David:2002wn} (so the bosons form a $(\textbf{2},\textbf{2})$ under these 2 $SU(2)$'s). In particular, the algebra generated by  $I^i_2$ could play the role of the outer automorphism $SU(2)_{\text{outer}}$, under which the 4 supercurrents from 2 $\textbf{2}$'s. Thus,  the generators $I^i_2$  of $SU(2)_{\text{outer}}$  are:
\begin{equation}
    I^i=\frac{1}{4}\oint \frac{dz}{2\pi i}[X^1, -X^{2\dagger}]\sigma^i\left[
     \begin{aligned}
         \partial X^{1\dagger}\\
         -\partial X^{2}
     \end{aligned}\right]-\frac{1}{4}\oint \frac{dz}{2\pi i}[\partial X^1, -\partial X^{2\dagger}]\sigma^i\left[
     \begin{aligned}
         X^{1\dagger}\\
         - X^{2}
     \end{aligned}\right]
\end{equation}
Now we label the fermions and bosons by two spinor indices (two superscripts) of $SU(2)_{\text{R}}\oplus SU(2)_{\text{outer}}$.  Let:
\begin{equation}
\begin{aligned}
      \lambda^{+0}_{(+)}\equiv&\lambda^1, \quad \lambda^{-0}_{(-)}\equiv\lambda^{1\dagger}, \quad \lambda^{+0}_{(-)}\equiv -\lambda^{2\dagger}, \quad \lambda^{-0}_{(+)}\equiv \lambda^{2}\\
     X^{0+}_{(+)} \equiv& X^1, \quad X^{0-}_{(-)} \equiv -X^{1\dagger}, \quad X^{0+}_{(-)} \equiv X^2, \quad X^{0-}_{(+)} \equiv X^{2\dagger},\\
\end{aligned}
\end{equation}
where ``$0$'' means singlet. 
Note that we also have added a subscript to distinguish  fermions (bosons) that have the same superscripts.  In fact, these subscripts are just the spinor indices of the $SU(2)$ algebra generated by $I_1^i$ in  \cite{David:2002wn} (note that the construction of this $SU(2)$ algebra involves both the bosons and fermions). 
Then we have:
\begin{equation}
\begin{aligned}
    X^{0\alpha}_{(\gamma)}(z)X^{0\beta}_{(\delta)}(z)&\sim \epsilon^{\alpha\beta}\epsilon^{\gamma\delta}\text{log}(z-w)\\
    \lambda^{\alpha 0}_{(\gamma)}(z)\lambda^{\beta 0}_{(\delta)}(w)&\sim \frac{\epsilon^{\alpha\beta}\epsilon^{\gamma\delta}}{z-w}
\end{aligned}
\end{equation}
Now the supercurrents are:
\begin{equation}
\begin{aligned}
    G^1&=\sqrt{2}\lambda^{+0}_{(+)}\partial X^{0+}_{(-)}+\sqrt{2}\lambda^{+0}_{(-)}\partial X^{0+}_{(+)}, \quad G^2=\sqrt{2}\lambda^{-0}_{(+)}\partial X^{0+}_{(-)}+\sqrt{2}\lambda^{-0}_{(-)}\partial X^{0+}_{(+)}\\
    G^{1\dagger}&=\sqrt{2}\lambda^{-0}_{(-)}\partial X^{0-}_{(+)}+\sqrt{2}\lambda^{-0}_{(+)}\partial X^{0-}_{(-)}, \quad G^{2\dagger}=-\left(\sqrt{2}\lambda^{+0}_{(-)}\partial X^{0-}_{(+)}+\sqrt{2}\lambda^{+0}_{(+)}\partial X^{0-}_{(-)}\right)
\end{aligned}
\end{equation}
Finally, we can rename the supercurrents as:
\begin{equation}
      G^{++}\equiv G^1, \quad G^{--}\equiv G^{1\dagger}, \quad G^{-+}\equiv G^2, \quad G^{+-}\equiv -G^{2\dagger}.
\end{equation}
Now one can check that the superscript $\alpha, \beta$ of $G^{\alpha\beta}$ are just the two spinor indices of $SU(2)_{\text{R}}\oplus SU(2)_{\text{outer}}$. 
Note that the situation here is different from the $\mathcal{N}=4$ linear dilaton, where the algebra $SU(2)_{\text{outer}}$ is constructed by fermions.

\section{The construction of covering maps}\label{constructcovering}
In this section, we review the procedure to construct covering maps in \cite{Roumpedakis:2018tdb}. Firstly, note that closed forms of general covering maps with 2 and 3 ramified points had been constructed in \cite{Lunin:2000yv}. So here let's consider covering maps with $n$ ramified points $(n\geq 4)$ (though the author of  \cite{Roumpedakis:2018tdb} focus on the case of $n=4$, the method developed there could be applied to the cases of general $n$). The covering map $\Gamma: z\to x$ satisfies: 
\begin{equation}
    \Gamma(z)\sim x_i+a_i(z-z_i)^{w_i}+..., \qquad \text{near} \quad z=z_i
\end{equation}
We use the $SL(2,\mathbb{R})$ symmetry to fix the first three point as:
\begin{equation}
    x_1=z_1=0, \quad x_2=z_2=1, \quad x_3=z_3=\infty.
\end{equation}
Recall that in \eqref{RH}, for $g=0$ we have:
\begin{equation}
    d=1+\frac{1}{2}\sum_{i=1}^n(w_i-1),
\end{equation}
Then we consider the derivative of the covering map, which will have the following form:
\begin{equation}\label{dGamma}
    \Gamma'(z)=N\frac{z^{w_1-1}(z-1)^{w_2-1}\prod_{m=1}^{n-3}(z-z_m)^{w_m-1}}{\prod_{j=1}^{d-w_3}(z-l_j)^2},
\end{equation}
where $N$ is the normalization factor, which is a function of $z_m$ $(m=1,2,...,n-3)$. $l_j$ $(j=1,2,...,d-w_3)$ are also functions of $z_m$ $(m=1,2,...,n-3)$,  determined by the condition that $\Gamma'(z)$ does not have simple poles at $z=l_i$:
\begin{equation}\label{Rescondition}
   \mathop{\text{Res}}\limits_{z=l_i} \Gamma'(z)=0, \qquad \forall i=1,2,...,d-w_3
\end{equation}
From \eqref{dGamma} and \eqref{Rescondition}, we get:
\begin{equation}
    \frac{w_1-1}{l_i}+\frac{w_2-1}{l_i-1}+\sum_{m=1}^{d-w_3}\frac{w_m-1}{l_i-z_m}=\sum_{j\neq i}^{d-w_3}\frac{2}{l_i-l_j}
\end{equation}
We need to find the solutions of the above equation and then substitute them into \eqref{dGamma}. Next, we determine the normalization factor $N$ by requiring:
\begin{equation}
    \Gamma(1)=1, \qquad \text{or} \qquad \int_0^1dz\Gamma'(z)=1.
\end{equation}
Finally, we can integrate \eqref{dGamma} to obtain the covering maps:
\begin{equation}
    \Gamma(z)=\int_{0}^zdt\Gamma'(z).
\end{equation}
For the calculation in the main body, we let $n=5$. In particular, in the so-called edge cases, we have $w_3=w_1+w_2+1$, $w_4=w_5=2$, so $d=w_3$.  There is no $l_j$ in \eqref{dGamma} and $\Gamma'(z)$ is simply a polynomial. Consequently, the covering map is unique. This fact makes it  much simpler to work out   the (unique) covering map in these edge cases (their explicit form could be found in  \eqref{edgecovering}).

\bibliographystyle{JHEP}
\bibliography{refs}

\end{document}